\documentclass[12pt]{ucsddissertation}
\usepackage{mathptmx}
\usepackage[NoDate]{currvita}
\usepackage{array}
\newcolumntype{M}[1]{>{\centering\arraybackslash}m{#1}}
\usepackage{tabularx}
\usepackage{booktabs}
\usepackage{ragged2e}
\usepackage{microtype}
\usepackage{wrapfig}
\usepackage{natbib}
\usepackage{url}
\usepackage{enumitem}
\usepackage[T1]{fontenc}
\usepackage{subfig}
\usepackage{xcolor}
\usepackage{colortbl}
\usepackage[breaklinks=true,pdfborder={0 0 0}]{hyperref}
\usepackage{graphicx}
\usepackage{caption}
\usepackage{longtable}
\usepackage[skins,breakable]{tcolorbox} % For the tip box
\usepackage{fontawesome5}

\AtBeginDocument{%
	\settowidth\cvlabelwidth{\cvlabelfont 0000--0000}%
}

\increasemargins{.1in}

\usepackage{trace}
\usepackage[english]{babel}
\usepackage{blindtext}
\title{Data-Driven and Participatory Approaches toward Neuro-Inclusive AI}
\author{Naba Rizvi}
\degree{Computer Science}{Doctor of Philosophy}
\chair{Professor Imani Munyaka}
\committee{Professor Laurel Riek}
\committee{Professor Taylor Berg-Kirkpatrick}
\committee{Professor David Serlin}
\degreeyear{2025}

\newtcolorbox{neurotipbox}[1][]{%
  breakable, % Allows the box to break across pages
  colback=yellow!5!white, % Background color (light yellow)
  colframe=yellow!95!black, % Frame color (darker yellow)
  coltitle=black, % Title text color
  fonttitle=\bfseries, % Bold font for the title
  title=Defining Neuro-Inclusive AI, % Default title (can be overridden)
  enhanced, % Enables advanced skin features
  attach boxed title to top left={yshift=-0.1in, xshift=0.15in}, % Position title nicely
  boxed title style={colback=yellow!60!white, colframe=yellow!15!black, sharp corners}, % Style for the title box
  drop shadow={blue!50!yellow}, % Adds a subtle shadow
  #1 % Allows for passing additional options when you use the box
}
\begin{document}
% Begin with frontmatter and so forth
\frontmatter
\maketitle
\makecopyright
\makesignature
% Optional
\begin{dedication}
\centering
\noindent
Words are not enough, so I hope I can make this worthy some day. I am writing this in honor of all the people for whom this endeavor will forever be a dream. The ones we have forsaken. The ones who fought for justice until their last breath. The ones who keep me up at night and forever stuck in my own head.

\noindent
For my grandmothers, who were orphaned in a war they were too young to even understand.

\noindent
For my grandfather, and the voices inside his head. 

\noindent
For my uncle, who never lived to see his 30th birthday.

\noindent
For my aunt, who taught me the importance of keeping in touch.

\noindent
For my grandfather, who turned broken machines into magic. Who showed me what I can do with a computer.

\noindent
For my mom, and her steadfast belief in seeing the good in every situation.

\noindent
For my dad, and his dreams.

\noindent
For my siblings, with love.

\noindent
For Khalil, the love of my life who helps me fall asleep every night.

\noindent
For my in-laws, who welcomed a foreigner into their hearts and home with so much patience and love.

\noindent
And for all the people around the world who have endured chaos and warfare for generations:
\noindent
We will rise.

% the people who loved me through it all: my parents, my sister, brothers, and the love of my life, Khalil Mrini, who helps me fall asleep every night, and my niece, who continues to inspire and entertain me. 
\setsinglespacing
\raggedright % It would be better to use \RaggedRight from ragged2e
\parindent0pt\parskip\baselineskip
% In the end, every experience becomes a story, and the story of my PhD would have been left incomplete if it were not for the love, support, and guidance of 
\end{dedication}
% Optional
\begin{epigraph}
\vskip0pt plus.5fil
% \setsinglespacing
% {\flushright
% True ease in writing comes from art, not chance,\\
% As those move easiest who have learn'd to dance.\\
% 'T is not enough to no harshness gives offence,---\\
% The sound must seem an echo to the sense.

% \vskip\baselineskip
% \textit{Alexander Pope}\par}
\vfil
\begin{center}
No matter where you roam\\
Anywhere you lay your heart's your home\\
Stand your ground if they push you down\\
Because all you can say is, all you can say\\

We are here and we will be\\
Following our own way\\
No matter what you say\\
\textbf{We are here and we're here to stay}\\
\vskip\baselineskip
\textit{Lenka, \textit{Here to Stay}}
\end{center}
\vfil
% \noindent Writing, at its best, is a lonely life. Organizations for
% writers palliate the writer's loneliness, but I doubt if they improve
% his writing. He grows in public stature as he sheds his loneliness and
% often his work deteriorates. For he does his work alone and if he is a
% good enough writer he must face eternity, or the lack of it, each day.

% \vskip\baselineskip
% \hskip0pt plus1fil\textit{Ernest Hemingway}\hskip0pt plus4fil\null

% \vfil
\end{epigraph}

% Next comes the table of contents, list of figures, list of tables,
% etc. If you have code listings, you can use \listoflistings (or
% \lstlistoflistings) to have it be produced here as well. Same with
% \listofalgorithms.
\tableofcontents
\listoffigures
\listoftables

% Preface
\begin{preface}
\textbf{Part 1: The Life}\\
I have spent my whole life feeling like I am on the outside looking in. Growing up in four distinct cultures probably contributed to it. I am sure my neurodivergence did not help.

Yet, it is that feeling of isolation that connected me with the people society often mistreats. As a child, I watched an Iraqi mother wailing loudly at the mosque because she lost her son. That day I learned that grief, much like a smile, is contagious. As a teenager, I survived a terror attack. That day I developed my fear of large crowds. Then, came the heartbreak of losing the people I loved deeply, including my grandfather, who doctors assumed was just faking his symptoms for ``attention''. Had they taken the time to listen to him, perhaps they would have found the cancer before it consumed him. 

I have tried to live my life with good intentions, though I know now that isn't enough. When I started my journey in computer science, I envisioned myself harnessing the power of data to make our world a better place. Perhaps I could curtail human trafficking, for example. I never wanted to let my own notions strip away someone else's humanity. Never once could I have imagined this work could contribute to the abject horrors of war, censorship, racial profiling, or sexual assault. Now that I am aware of these possibilities, I try to be more cautious of my own creations.

% As an adult now, I am forced to reckon with the heaviest burden of all: my own conscience. 

I can control the things that I create, but I cannot control what you choose to do with them. Still, I need the foresight to cease contributing to projects that harm others. In developing \textsc{Autalic}, and all the projects that followed, I simply wanted to fight for the recognition of our own humanity. I hope the world receives this in good faith, for I do not consent to having my work used to subjugate other people.

\textbf{Part 2: The Prose}\\
We are living in a world where many will wonder whether I used AI to write this. I did not, but there may be an AI detection tool that says otherwise. I have found myself censoring my own writing and purposely using simpler language to sound more ``human'', whatever that means to you. The greatest irony of it all is that because writing is how I connect with the world, my prose is strewn across the Internet and likely ended up as training data for many of the Large Language Models (LLMs) that exist today. Yet, the burden of proof lies entirely on me to prove I did not get `help' from an agent that simply re-purposes the words of other inquisitive souls without our knowledge or consent.

% If perspective makes a difference, why do I feel like Victor watching the creature awaken?

% simply repurposes the words of other inquisitive souls like me without our consent. 

The written words that once freed us and helped us connect with other humans, are now being used to corroborate our lack of humanity. Still, a part of me is relieved, because if the democratization of the internet amplified the unheard, perhaps the democratization of writing will finally help us fit in. It's a funny feeling, being the hunter and the prey, but realizing if I stood a thousand miles away, the difference wouldn't matter. Regardless of intent, it would look like self-destruction. 

% Here, in the boat from Conrad's tale, I dream of becoming like Fatima.

Where do we go from here? 

I can only hope the answer is cathartic. 
\end{preface}

% Your fancy acks here. Keep in mind you need to ack each paper you
% use. See the examples here. In addition, each chapter ack needs to
% be repeated at the end of the relevant chapter.
\begin{acknowledgements}

I would like to acknowledge:

Professor Munyaka for her support and guidance, especially in navigating the planning processes which have always been challenging for me.

Professor Nedjma Ousidhoum, without whom my research would have no direction. Despite the ocean between us, she was always there when I needed help and guidance.

My mentor Professor Andrew Begel, who has been with me throughout my research journey, from my bachelor's to today. 

Professor Steven Swanson, for helping me see the light at the end of a very dark tunnel.

My research mentors throughout my undergraduate education. Without them, I would never have even started my PhD: Dr. Jared Oluoch, Dr. Lesley Berhan, Dr. Kevin Xu, Dr. Weiqing Sun, Bahar Zoghi, and Dr. Daniel Cer.

Allison McFadden-Keesling, who ignited my love for research,

Ms. Nancy, for getting me through my growing pains,

Shalani Shridhar, for trading her trajectory with me,

Michael McGann, for introducing me to social justice, 

Ms. Emerald Austerberry, for seeing the leader in me.

My collaborators: Harper Strickland, Alexis Morales-Flores, Mya Bolds, Saleha Ahmedi,  Haaset Owens, Aekta Kallepalli, Isha Khirwadkar, Daniel Gitelman, Tristan Cooper, Alexis Morales-Flores, Michael Golden, Akshat Alurkar, William Wu, Mya Bolds, Raunak Mondal, Taggert Smith, Tanvi Vidyala, Dr. Rua Williams, Dr. Katta Spiel, Franck Dernoncourt, Lisa Dirks, Erin Beneteau, Richard E. Ladner, the Race in HCI Collective, Hala Annabi, Nadir Weibel, Andrea Hartzler, Emily Bascom, and Harshini Ramaswamy, for all the things we created together.

For the Mrinis, Khans, and the Shanks, who supported me as if I was one of their own,

For my friends: Meenakshi Das, Ashley Lopez, Jayden Butler, Chand Haryani, Mubashir Abbas, Maarten, Anthony Colas, Hana Gabrielle Rubio Bidon, KaShawna Lollis, Jesse Posthuma, Shm Almeda, Jamilla Muhammad, Sophia Sun, MASK, James E Williams, Marty, Sadie Sims, Toriano Drane, Anya Bouzida, Elham Hafiz who supported me in their own ways,

My story would be incomplete without the boy who made me whole: Khalil Mrini, and his beautiful family, who accepted a foreigner as one of their own. Choukran bzaf. 

And finally, for my own family, who raised me, loved me when I fell apart, and helped put the pieces of me back together: 

my ammi, Erum Rizvi, who gave me her creativity, resilience, and optimism, 

my abbu, Abid Rizvi, who gave me his passion and resolution, 

my nana, Naqi Akhtar, who ignited my love for technology, 

my nani, who prayed for us and with us, 

my brother Mujtaba, for understanding me and being my oldest friend,

my sister, Tuba, for advocating for me and protecting me from everything, 

my brother, Ammar, for keeping me grounded and aligned,

my niece Alina, for keeping me entertained throughout my PhD, 

my brother-in-law Erik, for keeping me open-minded,

my cousins, aunts, uncles, and my nation, for supporting me through it all.

\end{acknowledgements}

% Stupid vita goes next
\begin{vita}
\noindent
\begin{cv}{}
\begin{cvlist}{}
\item[2020] Bachelor of Science, University of Toledo, Berkeley
\item[2020] Research Intern, Microsoft
\item[2022] Research Intern, Google
\item[2023] Master of Science, University of California, San Diego
\item[2023-2025] Research Assistant, University of California, San Diego 
\item[2025] Doctor of Philosophy, University of California, San Diego 
\end{cvlist}
\end{cv}

% This puts in the PUBLICATIONS header. Note that it appears inside
% the vita environment. It is optional.
\publications
\noindent``Margin Call: An Accessible Web-based Text Viewer with Generated Paragraph Summaries in the Margin’’ Proceedings of the International Natural Language Generation Conference \textbf{(INLG)}, 2019.\par
\vspace{6pt}

\noindent``Using HCI to Tackle Race and Gender Bias in ADHD Diagnosis’’ Proceedings of the “What’s Race Got to Do with It? Engaging in Race in HCI’’ Workshop at the 2020 \textbf{CHI} Conference on Human Factors in Computing Systems, 2020.\par
\vspace{6pt}

\noindent``Experiences of Computing Students with Disabilities’’ Proceedings of the 52\textsuperscript{nd} ACM Technical Symposium on Computer Science Education \textbf{(SIGCSE)}, pp.\,939–940, 2021.\par
\vspace{6pt}

\noindent``Inclusive Interpersonal Communication Education for Technology Professionals’’ Americas Conference on Information Systems \textbf{(AMCIS)}, 2021.\par
\vspace{6pt}

\noindent``Keepin’ It Real About Race in HCI’’ \textit{Interactions}, vol.\,28, no.\,5, pp.\,28–33, 2021.\par
\vspace{6pt}

\noindent``Making Hidden Bias Visible: Designing a Feedback Ecosystem for Primary Care Providers’’ Proceedings of the “Designing Ecosystems for Complex Health Needs’’ Workshop at the 2022 \textbf{CHI} Conference on Human Factors in Computing Systems, 2022.\par
\vspace{6pt}

\noindent``QTBIPOC PD: Exploring the Intersections of Race, Gender, and Sexual Orientation in Participatory Design’’ Workshop at the 2022 \textbf{CHI} Conference on Human Factors in Computing Systems, 2022.\par
\vspace{6pt}

\noindent``Battling Bias in Primary Care Encounters: Informatics Designs to Support Clinicians’’ \textbf{CHI} Conference on Human Factors in Computing Systems Extended Abstracts, pp.\,1–9, 2022.\par
\vspace{6pt}

\noindent``Are Robots Ready to Deliver Autism Inclusion?: A Critical Review’’ \textbf{CHI} Conference on Human Factors in Computing Systems, 2024.\par
\vspace{6pt}

\noindent``From Granular Grief to Binary Belief: A Collaborative Optimization of Annotation Techniques for Anti-Autistic Language’’ ACM \textbf{SIGCHI} Conference on Computer-Supported Cooperative Work \& Social Computing \textbf{(CSCW)}, 2025.\par
\vspace{6pt}

\noindent``AUTALIC: A Dataset for Anti-AUTistic Ableist Language in Context’’ under review for the 63\textsuperscript{rd} Annual Meeting of the Association for Computational Linguistics \textbf{(ACL 2025)}, 2025.\par
\vspace{6pt}

\noindent``‘I Hadn’t Thought About That’: Creators of Human-like AI Weigh in on Ethics \& Neurodivergence’’ ACM Conference on Fairness, Accountability, and Transparency \textbf{(FAccT)}, 2025.\par

% This puts in the FIELDS OF STUDY. Also inside vita and also
% optional.
\fieldsofstudy
\noindent Major Field: Computer Science (Specialization or Focused Studies)
\vskip\baselineskip
Studies in Human-Centered AI\par
Professor Imani Munyaka
% \vskip\baselineskip
% Studies in Mechanices\par
% Professors Epsilon Zeta and Eta Theta
% \vskip\baselineskip
% Studies in Electromagnetism\par
% Professors Iota Kappa and Lambda Mu
\end{vita}

% Put your maximum 350 word abstract here.
\begin{dissertationabstract}
My work addresses the critical issue of biased data representation in AI, which mischaracterizes and marginalizes up to 75 million autistic people worldwide. Current AI research often focuses on medical applications and views autism as a deficit of neurotypical social skills rather than an aspect of human diversity, and this perspective is grounded in research questioning the humanity of autistic people. This problem is especially pertinent as AI development increasingly focuses on human-like agents. Thus, I explore the origins, prevalence, and impact of these biases. My first study tests an autism-inclusive communication course for non-autistic individuals, and finds that challenging neuronormative social conventions which place the burden entirely on autistic individuals to adapt to different communication styles can lead to more equitable outcomes. The second study examines the extent to which autistic people are marginalized in human-robot interaction research, revealing that 90\% of such studies exclude autistic perspectives. I offer guidance for moving beyond purely medical applications and increasing autistic individuals' participation in research about them. Next, I interview the creators of AI systems to better understand how their perspectives impact the technologies they create, and the barriers they face. While some do not consider ethical concerns around accessibility and neurodiversity to be their responsibility, those who do are often hindered by funding and organizational priorities. I follow this with a mixed-methods study on the perspectives of annotators in classifying anti-autistic hate speech, finding that a simplified binary annotation scheme could effectively capture important nuances while simplifying the labeling task. Finally, I develop a novel benchmarking resource for anti-autistic hate speech detection, and use it to evaluate four large language models (LLMs). I find that LLMs frequently misclassify autistic community speech, potentially leading to censorship if used for content moderation, and also fail to identify ableist speech due to reliance on simplistic keyword-based methods. I also uncover a significant misalignment between human and LLM reasoning, as humans employ a more holistic approach considering speaker identity, impact, and context, which LLMs often overlook or misinterpret.

% applications of robotics research for autistic people 

% and frequently replicated in computing

% challenges these beliefs by defining autism through a neurodiversity lens as merely a 

% embracing neurodiversity within which differences in neurotypes are considered a valid form of human diversity. Through this, I have explored the impact of 
\end{dissertationabstract}

% This is where the main body of your dissertation goes!
\mainmatter

% Optional Introduction
\begin{dissertationintroduction}
In 1949, Alan Turing created one of the most well-known benchmarks for assessing intelligence in machines. Now known as the Turing Test, it evaluates a machine's ability to respond to questions in a human-like manner such that the two are virtually indistinguishable to another human being \citep{turing2009computingreprint}. Nearly 8 decades later, scientists continue referring to this benchmark in our pursuit of artificial intelligence (AI). If we define `intelligence' as a machine's ability to mimic human communication, it is important for us to ensure our representations of `human' behavior include the unique behaviors of humans who have been long been dehumanized in our society, such as the 75 million autistic individuals around the world. Yet, computing research continues to portray autism through a one-sided deficits-based lens by defining it as a \textit{deficit} of neurotypical social skills, rather than an exemplification of human diversity \citep{kapp2013deficit, woods2018redefining}. Such beliefs inadvertently dehumanize autistic people because they build upon foundational work that questioned their humanity \citep{baron1997mindblindness} and are especially pertinent given the growing interest in creating human-like AI agents such as robots and chatbots. This research investigates how uncritically reflecting the medical field's deficits-based view of autism has fostered AI models and agents that dehumanize autistic people, and provides recommendations and resources for a more neuro-inclusive future.

In our first chapter, we explore the benefits of considering research directions beyond the deficits-based medical model. The double empathy problem suggests that issues with miscommunication that arise when people with different neurotypes interact with each other are a shared burden, and not a skills deficit for autistic individuals to overcome \citep{Morris2023Double, milton2012ontological, mitchell2021autism}. We test the effectiveness of resolving this problem through an autism-inclusive communication course that provides practical guidance to non-autistic people on navigating miscommunications. Such `social skills trainings' are commonplace for autistic people who are expected to be fully responsible for adapting to other communication styles but rarely ever are they offered to non-autistic people. However, our findings indicate that offering these trainings to other groups of people may encourage more equitable communication. 

In the second chapter, we evaluate the extent to which autistic people are marginalized in human-robot interaction research (HRI) as a case study in applied AI agents. We quantify this marginalization by assessing the replication of various misrepresentations of autistic people in HRI research today. Our findings show that 90\% of HRI research excludes the perspectives of autistic people, and that the majority of the studies tend to replicate stereotypes and power imbalances. We provide practical guidance to researchers on pursuing research directions that go beyond medical applications, and increasing the participation of autistic people in research about them.

In the third chapter, we attempt to dive deeper into the perspectives held by the makers of such systems and the challenges they face in making them more neuro-inclusive. We conduct semi-structured interviews with 16 researchers, engineers, and designers working on human-like AI agents, and collect demographic information from them through surveys. Through a qualitative analysis, we find that while some creators do not view ethical concerns related to accessibility and neurodiversity as their responsibility, the ones who do often face barriers in funding and organization priorities that limit them from exploring these concerns in depth. We also uncover the researchers tend to misunderstand or overlook the needs of neurodivergent individuals.

Our next chapter covers a study with six annotators who we engage in the structured design and testing of multiple annotation schemes to identify the challenges they face in these annotation tasks, and their recommendations on addressing them. Through this, we provide empirical evidence on the effectiveness of annotation techniques including score-based, comparison, and blackbox labeling. Within the score-based labeling, we test multiple annotation schemes by adjusting the granuarity of the labels. We expand on our findings and the annotator's suggestions, which conclude that a binary annotation scheme can simplify the labeling task and sufficiently captures the nuances important to annotators for this task.

In the final chapter, we develop a novel benchmarking resource for fine-tuning or evaluating models such as LLMs. This provides a practical resource for researchers to increase the alignment of their work with community perspectives. We assess the performance of 4 popular state-of-the-art open-source LLMs and find they frequently misclassify community speech which may lead to censorship if used for automated content moderation, while paradoxically missing actual instances of ableist speech due to their simplistic keyword-based classification approaches. Our findings also reveal major misalignments in human and LLM reasoning toward this task as humans have a more holistic approach to assess the identity of the speaker, the impact of the speech, and other contexual information such as tone. These aspects are frequently misinterpreted or overlooked by LLMs whose annotations appear to be more focused on identifying the presence of certain keywords instead of the deeper meaning of the speech.

\section*{Definition of Key Terms}
\label{sec:definition_of_key_terms}

The definitions of important concepts discussed in this dissertation are defined below to make the work accessible to a broader audience, including those who may be unfamiliar with specific terminology related to autism, neurodiversity, artificial intelligence, and participatory research methodologies.

\begin{neurotipbox}
This dissertation introduces and formally defines the term \textbf{Neuro-Inclusive AI}, as an approach to the design, development, and deployment of AI systems that explicitly de-centers neuronormative benchmarks and challenges the goal of merely mimicking `humanness.'

A formal definition is included within the glossary table that follows. 
\end{neurotipbox}
% --- End of tip box ---

\begin{longtable}{|p{0.28\textwidth}|p{0.62\textwidth}|}
\caption{Definition of Key Terms} \label{tab:key_terms} \\
\hline
\textbf{Term} & \textbf{Definition and Source} \\
\hline
\endfirsthead
\multicolumn{2}{c}%
{{\bfseries \tablename\ \thetable{} -- continued from previous page}} \\
\hline
\textbf{Term} & \textbf{Definition and Source} \\
\hline
\endhead
\hline \multicolumn{2}{r}{{Continued on next page}} \\
\endfoot
\hline
\endlastfoot

\textbf{Ableism} & A set of beliefs or practices that devalue and discriminate against people with physical, intellectual, or psychiatric disabilities, often resting on the assumption that disabled people need to be ‘fixed’ in one form or another \citep{cherney2011rhetoric, neilson2020ableism}. [Chapters 1, 2, 4, 5] \\
\hline
\textbf{AI Agents (Communicational)} & Artificial intelligence systems, such as chatbots or robots, designed to interact and communicate with humans, often mimicking human-like conversational patterns or behaviors \citep{van2020human}. [Chapter 3] \\
\hline
\textbf{Annotation (Data Annotation)} & The process of labeling data to make it usable for AI models for applications such as identifying anti-autistic language, as explored in Chapters 3 and 4. \\
\hline
\textbf{Anti-Autistic Ableism} & A specific form of ableism that marginalizes or misrepresents autistic individuals. Such ableism may manifest in research practices [as discussed in Chapter 2], and result in misalignments with community perspectives [Chapters 2, 4, and 5]. \\
% This can manifest as exclusionary practices, biased language, or systems that fail to accommodate autistic ways of being and communicating, a central theme in Chapters 3 and 4. [Chapters 3, 4] \\
\hline
\textbf{Autism} & A neurodevelopmental variation characterized by differences in social communication, interaction, sensory processing, and patterns of interests and behaviors. This dissertation approaches autism from a neurodiversity perspective, emphasizing it as an aspect of human diversity rather than a deficit to be remediated \citep{kapp2013deficit}.  \\
\hline
\textbf{AUTALIC (Dataset)} & A novel benchmarking dataset developed in this dissertation, comprising of Reddit posts annotated for anti-autistic ableist language in context. It is designed to support research in fine-tuning and evaluating models for detecting such language. [Chapter 5] \\
\hline
\textbf{Chatbot} & A communicational AI agent that simulates human conversation through text or voice \citep{shum2018fromEliza}. This dissertation considers the ethical design of chatbots for neuro-inclusive interactions. [Chapter 3] \\
\hline
\textbf{Critical Autism Studies (CAS)} & An academic field that critically examines prevailing understandings of autism, challenges deficit-based perspectives, and promotes the inclusion of autistic voices and perspectives in research and society \citep{woods2018redefining}. [Chapter 2] \\
\hline
\textbf{Crip Technoscience} & A field of study that combines critical disability studies with science and technology studies to examine how technology shapes and is shaped by experiences of disability, often advocating for designs that center access, interdependence, and disability justice \citep{spiel2022adhd}. [Chapter 2] \\
\hline
\textbf{Deficit-Based View (of Autism)} & A perspective that characterizes autism primarily by perceived `deficits' of neurotypical skills and abilities, often contrasting with a difference-based or neurodiversity view \citep{kapp2013deficit}. \\
\hline
\textbf{Double Empathy Problem} & The theory suggesting that communication difficulties between autistic and non-autistic (allistic) individuals are not solely due to autistic `deficits,' but are a mutual issue arising from differing neurotypes and experiences, where both parties may struggle to understand the other's perspective and communication style \citep{milton2012ontological}. [Chapters 1, 2, and 3] \\
\hline
\textbf{Essentialism (in Autism)} & The idea that individuals with autism share a fixed set of inherent, innate, and unchanging characteristics. [Chapter 2] \\
\hline
\textbf{Human-Robot Interaction (HRI)} & A field of study focused on the understanding, design, and evaluation of robotic systems for use by or with humans \citep{sheridan2016human}. [Chapter 2] \\
\hline
\textbf{Humanizing AI} & Applying human-like characteristics, behaviors, or intelligence to computing systems \citep{turing2009computingreprint, weizenbaum1966eliza}. This dissertation questions the ethical implications of how "humanness" is defined and implemented in AI, and its broader implications on autism inclusion. [Chapter 3] \\
\hline
\textbf{Identity-First Language (IFL)} & Language that puts the identity characteristic first (e.g., "autistic person") \citep{taboas2023preferences}. This dissertation adopts IFL, in alignment with the preferences of the majority of autistic adults in the United States, where the author is currently based \citep{taboas2023preferences}. [Chapter 4] \\
\hline
\textbf{Intersectionality} & A theoretical framework for understanding how various social and political identities (e.g., race, gender, class, disability, neurotype) combine to create unique modes of discrimination and privilege \citep{cho2013toward}. [Chapters 2, 3] \\
\hline
\textbf{Large Language Models (LLMs)} & AI models trained on vast amounts of text data, designed to generate human-like responses \citep{rajaraman2023eliza}. [Chapter 5] \\
\hline
\textbf{Medical Model (of Disability/Autism)} & A model that views disability (including autism) as an individual medical problem or deficit residing within the person, often emphasizing diagnosis, treatment, and cure, rather than addressing societal barriers in access or embracing diversity \citep{kapp2013deficit}. \\
\hline
\textbf{Neurodiversity} & The concept that neurological differences, including autism, ADHD, dyslexia, and others, are valid forms of human diversity \citep{walker2014neurodiversity}. It is a social justice movement that advocates for the rights, inclusion, and respect of neurodivergent individuals, challenging the view that such differences are inherently pathological. This is a foundational concept for the entire dissertation. \\
\hline
% \textbf{Neuroinclusion / Neuro-Inclusive} & Practices, designs, or environments that are intentionally created to be accessible, welcoming, and supportive of neurodivergent individuals, accommodating their diverse ways of thinking, learning, communicating, and experiencing the world \citep{NeuroinclusionDefinitionSource}. Achieving neuro-inclusive AI is the overarching goal of this dissertation. \\
\rowcolor{yellow!50} 
\textbf{Neuro-Inclusive AI} & An approach to the design, development, and deployment of AI systems that explicitly de-centers neuronormative benchmarks and challenges the goal of merely mimicking `humanness.' Instead, it prioritizes understanding, accommodating, and empowering diverse neurotypes, such as autism.
 \\
\hline
\textbf{Neuronormativity / Neuronormative} & The societal assumption and privileging of neurotypical ways of thinking, behaving, and communicating as the `normal' or `correct' standard, often leading to the marginalization or pathologization of neurodivergent individuals \citep{walker2014neurodiversity}.  \\
\hline
\textbf{Participatory Approaches / Participatory Design} & Research and design methodologies that actively involve end-users and other stakeholders, particularly those from marginalized communities, as partners in the design and development process, aiming for more equitable and relevant outcomes \citep{sanders2002user}. [Chapter 2] \\
\hline
\textbf{Pathologizing} & The act of framing natural human variations, differences, or behaviors as medically or psychologically abnormal, diseased, or disordered. [Chapter 2] \\
\hline
\textbf{Power Imbalance} & Unequal distribution of power, influence, or control between individuals or groups, such as in research or design contexts where researchers or designers may hold more authority than participants or end-users. [Chapter 2] \\
\hline
\textbf{Robot} & An embodied agent capable of sensing its environment, processing information, and performing actions, often through the applications of AI \citep{franklin1997autonomous}. [Chapters 2, 3] \\
\hline
\textbf{Social Model (of Disability)} & A model that views disability as a social construct, where individuals are 'disabled' by societal barriers (e.g., inaccessible environments, discriminatory attitudes, exclusionary policies) rather than by their own abilities \citep{kapp2019social}. [Chapters 1, 2] \\
\hline
% \textbf{Uncanny (Valley)} & A phenomenon in aesthetics where objects or representations that appear almost, but not exactly, like real human beings elicit feelings of unease or revulsion in observers \citep{mori1970bukimi}. [Chapter 3] \\
\caption{A definition of key terms and concepts that may be used throughout this dissertation, unless otherwise stated.}
\end{longtable}
\end{dissertationintroduction}
\chapter{Shifting the Lens: Digital Education for Neuro-Inclusive Communication}
\section{Introduction}
The U.S. economy continues to compete for talented IT professionals. According to the Bureau of Labor Statistics, the number of IT jobs will grow 11\% between 2019 and 2029, faster than the average for all other occupations \citep{BLS2020Computer}. It is expected that in 2030 there will be a global shortage of 85 million tech workers \citep{daCosta2019TechTalent}. IT companies struggle to meet their talent needs. Increasingly, more firms such as SAP, IBM, and Microsoft, inspired by the initial success of the Danish IT consulting firm Specialisterne, are taking notice of the untapped technology talents and skills of individuals with autism to address the industry’s significant unmet demand for employees \citep{AnnabiLocke2019Theoretical}. Scholars from various fields recognize the proclivity of individuals on the autism spectrum to pursue technology-oriented employment \citep{Mazurek2012Prevalence, Wei2013STEM}. Industry champions of such programs point to the opportunity for such programs to have a greater social impact \citep{Shattuck2012Postsecondary, HedleyUljarevic2018SystematicReview}. Autism employment programs provide meaningful employment opportunities for the growing number of IT-oriented individuals with autism who have been severely unemployed or underemployed. It is difficult to estimate the unemployment rate of adults on the spectrum due to limited research and limited disclosure \citep{AustinPisano2017Neurodiversity}. Some estimate that the unemployment rate among individuals with autism (including those with severe intellectual disabilities) is 80\% \citep{AustinPisano2017Neurodiversity}. \citet{Roux2017National} suggest that only 14\% of adults with autism in the US work for pay.

Autism at Work initiatives offer potential business benefits, including meeting the rising demand for IT workers, as well as capitalizing on the unique cognitive style and talents of employees with autism, namely, systems thinking, attention to detail, high level of focus, comfort with doing repetitive behavior, and ability to visualize problems \citep{AustinPisano2017Neurodiversity, Baldwin2014Employment, Morris2015Understanding, AnnabiLocke2019Theoretical}. Despite their potential, early studies of autism employment in IT reveal that technology workers with autism experience challenges and isolation in the workplace \citep{HedleyUljarevic2018SystematicReview}. These experiences often are attributed to:
\begin{enumerate}[label=\arabic*)]
    \item limited understanding of the talents of IT workers with autism \citep{Annabi2017NotJustAttention, AustinPisano2017Neurodiversity};
    \item suboptimal environmental and task design accommodations \citep{Rebholz2012LifeUncanny}; and
    \item the social and behavioral disconnect between autistic employees and their neurotypical managers and co-workers with whom they directly work \citep{AustinPisano2017Neurodiversity, Morris2015Understanding, Rebholz2012LifeUncanny}.
\end{enumerate}
This raises questions about the readiness of the IT workplace to welcome and effectively engage individuals with autism in tasks as well as social processes at work.

\section{Research Gap}
The emphasis of autism employment research has largely been on preparing the autistic employee for the workplace and the development of vocational training (e.g., \citealp{Burke2010Evaluation, SeamanCannellaMalone2016Vocational}) and supported employment \citep{Schall2015Employment}. There is limited research that addresses challenges and successes in the organizational setting to better support autistic employees in the workforce \citep{Wei2015Transition, Roux2015National}, particularly in the IT industry, where many individuals on the autism spectrum work \citep{Hedley2017Employment, AnnabiLocke2019Theoretical}. Most of the research and practice place the onus of change on the autistic person, rather than challenging the normative ideas about how professionals should be and behave. Therefore, our work, consistent with \citet{AnnabiLocke2019Theoretical}, focuses on building the capacity of organizations and teams to work effectively in diverse teams of neurotypical and autistic employees. In this study, we focus our attention on social and behavioral disconnect between employees on the autism spectrum and their neurotypical managers and co-workers with whom they work \citep{AustinPisano2017Neurodiversity, Morris2015Understanding, Rebholz2012LifeUncanny}. This research focuses on interpersonal communication, a key barrier identified in the literature below. To this end, we address the following research question:
\begin{quote}
    \textit{How can we help neurotypical workers interact more effectively with their autistic colleagues?}
\end{quote}
In the remainder of this paper, we review the relevant literature to explain the nature of autism employment in IT and reveal the key challenges that autistic employees face at work. Our literature review will transition to the design and assessment of our communication intervention, whose goal is to raise awareness of and knowledge of how to effectively communicate between neurotypical and autistic employees. The paper will report on the design of our communication intervention and the results of a formative pilot study to evaluate the design of the intervention and assess its effectiveness.

\section{Autism Employment in the IT Industry}
Employing individuals on the autism spectrum in the technology industry is not new. In fact, many in the popular press (e.g., \citealp{Silberman2015NeuroTribes}) and those doing empirical research (e.g., \citealp{Rebholz2012LifeUncanny, Morris2015Understanding}) note that the IT workforce employs a significant number of individuals on the spectrum. However, autism-specific employment programs in prominent IT firms or in IT departments of prominent firms is a relatively new and a growing phenomenon in the US worthy of attention \citep{AustinPisano2017Neurodiversity, AnnabiLocke2019Theoretical}. Firms such as SAP, Microsoft, IBM, Google, JPMorgan Chase, EY, Dell, SAS, WB, and others have developed autism-specific initiatives to recruit and interview autistic job candidates in order to meet the high demand for IT talent. Autism employment programs deploy various methods and practices to include individuals on the spectrum in the workplace. These include nontraditional interview processes that focus on:
\begin{enumerate}[label=\arabic*)]
    \item relaxed and casual interactions with candidates over a period of days; and
    \item hands-on skills assessments.
\end{enumerate}
Programs also include training and support for co-workers and managers to help them understand and appreciate the talents of their colleagues on the spectrum. In addition, some programs offer customizable career management and development processes \citep{AustinPisano2017Neurodiversity}. These programs are growing in size and impact. At Microsoft, over 130 autistic employees have been hired in software development and data analytics roles in the last five years, and over 3000 employees have gone through autism coworker training. SAP has hired more than 200 people in 16 countries, with a goal to reach 1\% representation of their employees to reflect the portion of autistic people in the broader population. JPMorgan Chase has hired 225 in 90 countries in 40 different roles. The potential of these programs is great but can only be achieved if autistic people are included equitably and have a sense of belonging in the workplace. There has been little specific research that assesses the effectiveness of the autism hiring programs mentioned above. Early research on autism employment across various industries suggests that autistic people face considerable barriers and challenges once they are hired.

\subsection{Challenges that autistic employees experience at work}
Societal notions of desirable employees and organizational processes are designed to value and emphasize social skills such as communication, emotional intelligence, and teamwork \citep{AustinPisano2017Neurodiversity, Annabi2017NotJustAttention}. Individuals with autism may have marked social communication differences and restricted interests that affect their interpersonal interactions and ability to relate to neurotypical managers and coworkers. These normative notions result in autistic individuals being deemed unqualified or undesirable for jobs and collaborations that they are intellectually capable of doing well \citep{Oliver2013SocialModel, AustinPisano2017Neurodiversity}. Research suggests that autistic individuals are viewed with stigma and disclosing an autism diagnosis in the workplace can have detrimental consequences \citep{JohnsonJoshi2016DarkClouds}. In a study of vocational experiences among autistic individuals, some reported tolerating the experience of being socially ``different'' while others reported ``a certain sort of stigmatization'' \citep{Muller2003Meeting}. Therefore, it is no surprise that individuals on the autism spectrum often feel isolated from and have few informal meaningful interactions and relationships with their colleagues \citep{Morris2015Understanding, AustinPisano2017Neurodiversity}. Those that report being ``scorned'' by co-workers say that their social impairments often led to isolation and alienation in the workplace \citep{Muller2003Meeting}. This is due, in part, to a significant disconnect in interpersonal communication between neurotypicals and those on the spectrum \citep{Scott2015Viewpoints}. Furthermore, autistic employees sometimes require environmental accommodations to manage stimulus in their physical surroundings or limited travel, which may attract negative attention from colleagues who view these accommodations as unfair, due to their limited knowledge of autism \citep{AustinPisano2017Neurodiversity}.

In their review of autism employment in IT, \citet{AustinPisano2017Neurodiversity} determined that managers and leaders play an important role in including individuals on the spectrum on their teams and setting the right tone. Unfortunately, transformational leadership styles emphasized in industry, which utilize abstract forms of communication such as metaphors, may lead to challenges for autistic employees because they induce more stress and anxiety \citep{Parr2013Questioning}. Managers and leaders often fail to communicate in ways that many autistic people prefer. For example, individuals on the spectrum may prefer explicit communication and the use of literal language \citep{TagerFlusberg1997Language}. The use of idioms, metaphors, and sarcasm may be confusing and lost in translation for many \citep{Happe1995RoleAge}. Further, individuals on the spectrum also may prefer logic-based problem solving \citep{Grandin1995HowPeople}, sameness and routinized behavior \citep{APA2013DSM5}, and structured environments \citep{MesibovShea2010TEACCH}. The dynamic, ambiguous, fast-paced, and often critical events (e.g., new releases, malfunctions) characteristic of IT work can present challenges to individuals on the spectrum \citep{Morris2015Understanding}. These events, if not properly managed, can increase stress and anxiety for individuals on the spectrum and prevent them from completing their work effectively and efficiently \citep{AustinPisano2017Neurodiversity}. Similarly, the reliance of IT firms on computer-mediated work communication, which has increased significantly during the COVID-19 pandemic, exacerbates stressors that affect autistic workers, such as sensory sensitivities, cognitive overload, and anxiety, requiring them to apply sensory, cognitive, and social coping strategies which may reduce trust in and increase their social isolation from neurotypical coworkers \citep{Burke2010SocialUse, Zolyomi2019ManagingStress}.

This review of challenges that autistic employees face emphasizes the importance of challenging normative expectations of social behavior and developing effective communication practices in diverse teams. Essential to doing so is not only understanding and changing the knowledge and attitudes neurotypical coworkers and managers have of autism, but also building capacity and know-how of effective communication practices \citep{AnnabiLocke2019Theoretical, HedleyUljarevic2018SystematicReview}.

\section{Changing attitudes is important, but know-how is too!}
The limited knowledge and negative attitudes neurotypical individuals often hold of autism are the cause of many of the barriers in the workplace discussed above \citep{MarotoPettinicchio2015TwentyFive}. The general public, including neurotypical IT workers, often have limited knowledge of the talents and needs of individuals on the autism spectrum. The misconceptions are based in normative social expectation about the talents and needs of individuals on the spectrum and create difficulties in interactions and employer decisions to hire, promote, or retain employees with autism \citep{Gould2015BeyondLaw, AustinPisano2017Neurodiversity}. These attitudes influence organizational culture and policies and either facilitate or prohibit the inclusion of individuals with autism in the workplace \citep{Gould2015BeyondLaw}. Attitudes refer to the perceived advantages and disadvantages of performing a behavior (the degree to which a person has a favorable or unfavorable evaluation of a specific behavior; in our case, communication differences of individuals on the autism spectrum). Studies have documented the relevance of attitudes in accounting for intentions \citep{Hagger2002MetaAnalytic, ArmitageConner2001Efficacy} and demonstrated that intentions account for a considerable amount of the variation in specific behaviors \citep{ArmitageConner2001Efficacy}. Attitudes present a malleable determinant of behavior that can be targeted to promote better inclusion of individuals with autism in the workplace. Thus, a lot of training relevant to autism focuses on understanding autism in order to change people’s attitudes \citep{Morrison2019Variability}. Progress with autism acceptance training shows that increasing autism knowledge can help reduce explicit bias and stigma towards autistic individuals (though with less success in reducing implicit bias) \citep{Morrison2019Variability, Bast2020Effect, Dickter2020Implicit, jones2021effects}.

Changing people’s attitudes is not sufficient, however. Our early discussions with many neurotypical workers in IT revealed that their biggest challenges lay in know-how. They often want to be more effective, but they find it difficult to communicate effectively. Unfortunately, most training stops short of presenting strategies for effective communication, i.e., the know-how. To encourage neurotypical IT workers’ behaviors to be more welcoming and inclusive in their interactions with coworkers on the spectrum, we need to not only understand and change neurotypicals’ understanding of and attitudes towards autism, but we must also improve their readiness to use effective communication strategies (their know-how). To this end, we developed an educational module addressing effective communication. The next sections describe the design of our module and a formative study to assess its design and effectiveness.

\section{Method}
We designed an online course that we built to teach neurotypical employees communication skills to work more effectively with their autistic coworkers. The online version was adapted from an in-person course we developed in a series of workshops held with IT workers at large USA-based technology firm. The course’s goals are to teach learners three lessons:
\begin{enumerate}[label=\arabic*)]
    \item understand the fundamentals of communication,
    \item understand and appreciate the diversity of communication styles and preferences of autistic (and non-autistic) people, and
    \item develop effective ways to communicate with autistic coworkers.
\end{enumerate}

\begin{figure}[htbp]
  \centering
  \includegraphics[width=0.8\linewidth]{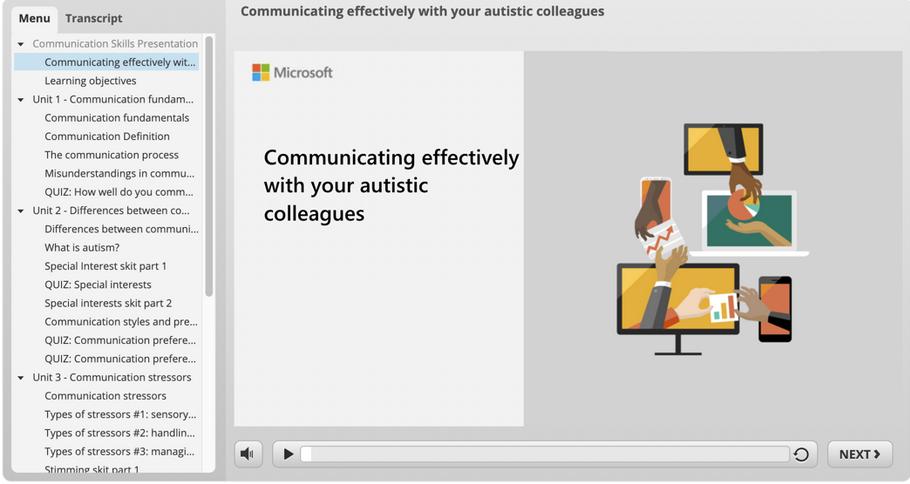} 
  \caption{The introduction slide to the online course. On the left is the course outline.}
  \label{fig:intro_slide}
\end{figure}

\begin{figure}[htbp]
  \centering
  \includegraphics[width=0.8\linewidth]{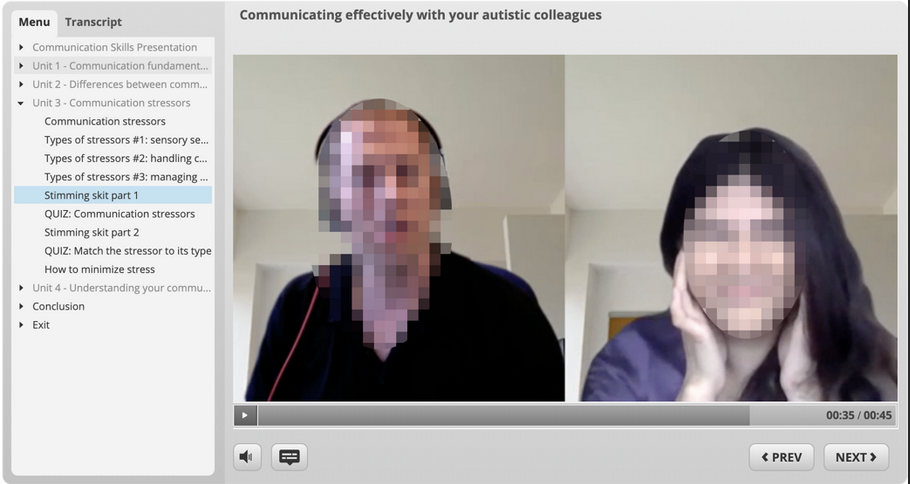}
  \caption{A picture of two coworkers talking to one another in an education video about stimming. The autistic coworker on the right is overwhelmed after her colleague on the left calls out her hair twirling.}
  \label{fig:stimming_video}
\end{figure}

The 15-minute course, divided into 4 units, employs video skits, quizzes, audio narrations, and charts to explain core concepts. The title screen of the course is shown in Figure~\ref{fig:intro_slide}. The video skits and quizzes are designed to reinforce the concepts discussed in the section, promote empathy and understanding between neurotypical and autistic people, and introduce concrete skills that can be used to improve communication. Each video portrays a scenario in which a communication issue occurs between an autistic person and their neurotypical colleague. We quiz the learner to identify the error, and then show them a follow-up video demonstrating how the issue can be resolved using the communication skills taught in the course. A screenshot from one of the course’s video skits is shown in Figure~\ref{fig:stimming_video}.

\subsection{Course Outline}
Our communication skills course is divided into four modules: communication fundamentals, differences between communication styles of autistic and neurotypical people, communication stressors, and understanding communication preferences.

\subsubsection{Communication Fundamentals.}
The first module briefly explains the fundamentals of communication. Our goal is to help learners understand their own communication preferences. It defines communication and explains both the transmission and feedback phases in the communication process. We then define noise and explain how it can introduce misunderstandings in communication. The unit concludes with a communication skills self-assessment, encouraging the learner to reflect on their own behavior.

\subsubsection{Differences between communication styles of autistic and neurotypical people.}
The second module defines autism as a neurological condition impacting the way a person experiences the world. It illustrates the differences between the communication preferences of neurotypical and autistic people through 20 examples of typical reactions to common behaviors. A brief video skit demonstrates restricted interests as a communication preference that can be common among autistic people. In the skit, a neurotypical worker unsuccessfully tries to talk about work, while their autistic colleague, missing their coworker’s subtle hints to change the topic, talks about bees and honey. The skit is followed by a quiz asking the learner to identify the cause of the miscommunication. A follow-up video skit demonstrates how more direct and straightforward communication can help them avoid these issues and get their work done. A final quiz in the unit explores the different preferences in small talk between autistic and neurotypical people.

\subsubsection{Communication stressors.}
The third module focuses on communication stressors faced by autistic people: sensory sensitivity, cognitive overload, and anxiety. Examples of stressors include managing multiple tasks in conversations, listening to too many people talking at the same time, and feeling the need to adhere to neurotypical social norms. A video skit in this unit features an autistic worker who feels extreme anxiety and loss of focus when they are called out for twirling their hair, a kind of ``stimming'' (or self-stimulating) behavior that autistic people tend to exhibit. The skit is followed by a quiz that asks the learner to identify the communication stressors they saw. A follow-up video skit then teaches the learner how to minimize these stressors in future meetings. A final quiz asks learners to match communication stressors with actions that can be taken to minimize their impact on autistic coworkers.

\subsubsection{Understanding communication preferences.}
The final module of the course contains a three-part, 12 question self-assessment quiz. The questions ask learners to reflect on their own communication preferences and compare them to preferences commonly preferred by autistic people. The first part focuses on cognitive load in face-to-face interactions, offering examples about eye contact and displaying emotion to conversational partners. The second part concerns conversational dynamics, asking learners to assess their comfort with talking to more than one person at a time, or with multitasking during conversations. The third part focuses on the content of conversation, including questions on reviewing the meeting agenda in advance and talking about ``uninteresting'' conversation topics. The module concludes with an actionable list of steps to help learners negotiate their communication preferences with their coworkers, helping them to minimize miscommunication.

The course concludes with four takeaways:
\begin{itemize}
    \item Communication is effective only when a common understanding is reached. Noise hampers the communication process.
    \item Autistic people have a wide range of abilities, characteristics, and communication preferences that may differ from and are very noticeable to neurotypical people. However, we stress that these differences should not be considered deficits.
    \item Three types of communication stressors faced by autistic people are sensory sensitivities, cognitive overload, and anxiety. These stressors may be minimized by setting clear expectations and discussing one’s communication preferences in advance of important conversations.
    \item To be effective communicators, we must seek to understand and adjust to others' styles and preferences.
\end{itemize}
Resources are provided at the end of presentation in the form of books, web pages, and academic articles that can be read to gain a deeper understanding of the material covered in the course.

\subsection{Intervention Pilot Assessment}
To assess the effectiveness of our course, we conducted a pilot study with 19 information workers at a large, USA-based technology firm. These were recruited via personal email invitations from members of an autism inclusion working group at the firm. We asked participants to take a pre-survey to assess their familiarity with autism, their understanding of their own communication preferences, and their knowledge of how to negotiate communication preferences with colleagues. Participants then took the online course. Afterwards, a post-survey measured the change in autism knowledge, their attitudes towards autistic individuals, and their ability to synthesize appropriate strategies in response to novel hypothetical communication scenarios with autistic coworkers. The experimental methodology was approved by our IRB.

The pre-survey contained 13 questions: 5 assessing one’s own communication preferences and comfort, 3 assessing prior experience with autism or disability, 2 measuring communication-related autism knowledge, and 3 evaluating how well they knew how to react to novel scenarios involving conversations with autistic colleagues. The post-survey contained 12 questions: 3 repeated from the pre-survey to assess understanding one’s own communication preferences and knowledge of autism, 5 evaluating recall of course content, 1 assessing change in attitude towards autistic coworkers, and 3 evaluating their reaction to novel scenarios. Knowledge and reaction questions were intended to reveal whether our participants retained the information presented in the course (in the short term) and were able to apply this knowledge in real-world scenarios. The attitude question helped us learn whether our course helped reduce explicit bias that our participants may have held against autistic people.

\section{Results}
Of the 19 participants, most (68\%) personally knew an autistic person, but only 1 had received any disability training (not specific to autism) prior to the course. 12 (63\%) reported they had experience negotiating their communication preferences with colleagues prior to the course. All participants reported being somewhat or extremely comfortable with written, professional communication, however 2 (11\%) said they were somewhat or extremely uncomfortable with oral, professional communication, whether in-person or virtual.

Evaluating the participants’ knowledge gain of communication fundamentals, we found that 18 out of 19 participants understood that communication is only effective when a common understanding has been reached. All the participants understood that effective communication requires understanding and adjusting to another’s communication style. 6 participants improved their understanding of their own communication styles, but 2 participants realized that they now did not know as much as they had thought.

Before taking the course, participants started out with a reasonably good knowledge of autism. Out of the 18 respondents, 3 said reported that they knew a lot about autism, while 6 knew a moderate amount, and 8 only knew a little. 7 out of 18 participants reported that the course increased their knowledge of autism by one rating on a 5-point Likert scale.

18 out of 19 participants said that they understood the bottom-up, detail-oriented style that autistic individuals tend to prefer. After the course, one participant improved their understanding of this style, but another participant’s understanding decreased. 5 out of 18 participants improved their attitudes towards accepting the tendency of autistic people to offer unsolicited advice or correct their facts, however one participant’s attitude worsened.

While 18 participants understood that explicit communication is an effective way to change the topic of conversation with an autistic colleague, 3 thought that subtle, indirect means could work as well. After taking the course, 1 of the 3 recognized that indirect approaches would be unlikely to be effective.

18 out of 19 participants could identify the kinds of stressors autistic people face during video conference calls. After the course, all the participants correctly recalled these stressors. Before the course, 12 of 18 participants could identify four appropriate ways to minimize stress in communication. 5 could identify three appropriate ways and 1 could only identify two. After the course, 17 out of 18 could identify four ways to minimize stress, while 1 participant who had identified three ways now believed there were only two. Finally, while all participants could identify various techniques to minimize sensory sensitivities experienced by autistic coworkers during video conference calls, only 11 out of 19 participants could correctly recognize the stressors that could be felt by an autistic coworker in a novel, hypothetical communication situation.

Overall, we were impressed that many of the learners did as well as they did on the pre-survey assessment. We checked to see if this knowledge on the pre-survey or post-survey correlated with any prior relationships with autistic people or disability training they reported on the pre-survey. However as this was a formative assessment with just 19 participants, we did not find any statistically significant correlations with knowledge change, attitude change, or ability to use the information in novel situations.

Beyond the quantitative assessment, we asked participants to offer qualitative feedback to help us improve the course. One said, ``This was a useful training on working with autistic colleagues.'' The course helped another recognize the limits of their own previous understanding of communication preferences. ``It made me realize I needed to learn more, including about myself.'' Others described pedagogical aspects of the course that they felt were particularly effective. ``Overall, the course was incredibly informative and insightful... The examples are great! e.g., when going through the types of stressors.'' The participant continued, referencing the directness and effectiveness of the video skits, ``The videos are also helpful – both seeing examples of ‘what not to do’ followed by ‘what to do’ to communicate more effectively.'' Another reinforced the value targeting how-to knowledge, ``I would like more video examples of interactions and also around negotiating communication styles I think would be helpful (e.g., are there mistakes neurotypicals might make even in up-front negotiation or less vs. more effective ways to come to the best solution for all parties?).'' A third said, ``I learn better when I see examples played out.'' Two participants said they felt the course was effective enough to want to see it rolled out more broadly.

\section{Limitations of this Assessment}
Our study was designed for the working conditions of an active IT firm. As such, the course and formative assessment were designed to take a short amount of our IT workers’ schedules. Study designs used in some prior research offer control lessons, but ours could not in order to minimize the time requirements. In addition, much prior work employed standardized instruments to measure autism knowledge and explicit and implicit bias. In our work, we have a different evaluation goal in mind, aiming more at evaluating knowledge gain and the ability to apply knowledge in novel situations that are specific the lessons offered in the course. Standardized instruments can be less appealing to professional audiences who have less time available for training. In previous studies, trainings required about twice the amount of time (1 hour) than we had available (30 minutes) \citep{Morrison2019Variability, jones2021effects}.

Our study had a small sample size of only 19 participants. This fits in with our goal of formatively assessing our efforts. The participants were recruited through snowball sampling from members of an autism inclusion working group and self-selected into the study. This could explain their better-than-expected performance on the pre-survey, demonstrating a greater sense of engagement than the average IT worker. Future studies in this line will incorporate more random sampling methods to engage with a wider, more diverse employee population. We believe that the results are encouraging enough to continue designing several additional autism-related courses on intra-team social dynamics and expectation management.

\section{Implications}
The formative pilot study to assess the communication intervention revealed interesting findings about the knowledge participants acquired, their preferences for how knowledge is presented, and the applicability of that knowledge to their communication with autistic and neurotypical coworkers alike. Furthermore, the study raised interesting questions about the preconceived notions of communication abilities participants measured in the pretest.

Participants suggest that the communication course helped think about their communication style and effectiveness more broadly. Some stated that they found the reflective exercises were valuable not just for learning about the communication preferences of autistic people but learning about their own communication preferences and behavior. It became clearer to them that effective communications strategies have broader applicability. This is a common theme reinforced in the literature that training focused on collaboration or management of differences relevant to autism has a spillover effect to both neurotypical and autistic employees \citep{Krzeminska2019Advantages}.

Furthermore, our results and qualitative data suggest the universal applicability at work and home of the interpersonal communication knowledge in the course. Our participants revealed the pervasiveness of neurodiversity in their life where this can be applied. Our findings suggest that even participants who had prior knowledge of autism and relationship with a person on the spectrum and had prior communication knowledge can still benefit from the course. ``This is great thank you. I have a [child] who we diagnosed at [a young age] and I still am learning about him and how he thinks and this gave me some insights.''

Last, the results revealed that some participants understood communication stressors (per their post-survey and perceptions in qualitative comments), but when assessed in a novel situation through scenarios, were unable to reach the preferred communication strategy. One might conclude that participants may have inaccurate assumptions of effective communication and overestimate their ability to address situations appropriately. This reinforces that the nuances complexity of communication between autistic employees and their coworkers and motivates the need for deeper learning. First, the nuances of communication preferences, how they surface in video scenarios, and how they may be interpreted is complex. This requires more focus on developing frameworks or strategies to better guide the student to determine these nuances and define appropriate signals. Second, it suggests the need for more practice in the course through more scenarios and perhaps a collaboration component through an asynchronous discussion mode to fully discern patterns of behaviors and provide more opportunity to internalize strategies.

\section{Conclusion}
In this paper we presented a formative study in which we assessed the design and impact of a communication course aimed at improving interpersonal communication between autistic technology workers and their neurotypical coworkers. This work is unique from other efforts to prepare autistic employees for the workplace in that it shifts the onus of learning and adaptation onto the neurotypical employee rather than expecting the autistic employee to adapt. In doing so, we challenge normative expectations of social behaviors dominating our workplaces which have historically led to the exclusion and marginalization of talented technology professionals on the spectrum. Efforts such as ours begin to change the paradigm of our workplace and pave a way for challenging broader structures of exclusion and discrimination grounded in ableism. This is not only relevant for autistic employees — it is applicable to all employees who may have been excluded from the workplace due to normative expectations that stigmatize social and communication behaviors grounded in ableism.

Future work will focus on refining our online course and addressing communication knowledge gaps revealed in the Results section. In particular, our work will explore pedagogy that focuses more deeply on application of communication strategies to scenarios that reveal the nuances and complexity of interpersonal communication between autistic technology workers and their coworkers. The next iteration of the course will also elaborate on intersectionality and the implications diversity of identity may have on the preferences of autistic people, as well as the preferences of their neurotypical coworkers and their intersecting identities. Lastly, our work will continue to expand to include additional courses on teamwork strategies.
\chapter{Agentic Ableism: A Case Study of How Robots Marginalize Autistic People}
\section{Introduction}
\label{sec:intro}

In 2022, the Disability Day of Mourning recorded at least eighteen autistic people around the world who were murdered by their caregivers. One of them was a young girl murdered by her mother who claimed she was stressed'' that her daughter might be autistic.\footnote{\url{https://disability-memorial.org/arya-smith}}. These acts of anti-autistic violence mirror the experiences of nearly 800 children who were killed by Nazi Germany under the supervision of Hans Asperger  \citep{czech2018hans}. Asperger was the first to designate a group of children as autistic psychopaths'' and his work legitimized policies such as forced sterilization and child euthanasia \citep{czech2018hans}. His work on pathologizing autism was so influential that he became the namesake of Asperger's Syndrome, now controversially referred to as high-functioning autism''  \citep{frith1992autism, mesibov2005understanding}. The beliefs that autistic people were deficient'' and needed to be cured'' were further entrenched in autism research when Baron-Cohen described autistic children as lacking the quintessential human trait'' of Theory of Mind  \citep{happe1995theory, baron1997mindblindness}. Although this deficit-based view of autism has been linked to other concerns such as an increased risk of suicidal ideation and social isolation in autistic individuals, it continues to be pervasive in autism research even today  \citep{cassidy2018risk, cassidy2020camouflaging}. Traditionally, studies in human-robot interaction (HRI) research are centered around the \textbf{medical model} view of autism which treats autism as a disorder that needs to be cured so that a person can be made normal''  \citep{Williams2021IMI,kapp2013deficit, scassellati2012robots}. This approach has been criticized by scholars for: 1) failing to center the perspectives of the autistic community and 2) applying a deficit-based understanding which has been linked to worrying social concerns such as negatively impacting the identity-building process, and an increase in suicidal ideation among autistic people \citep{Ymous2020Terrified, spiel2020nothing, anderson2022autism, cassidy2018risk, cassidy2020camouflaging, bennett2020point}. Furthermore, the medical model promotes the idea that a formal diagnosis is necessary to validate the experiences of autistic people, which introduces a power imbalance between medical professionals such as psychiatrists and the community \citep{bennett2020point}. Despite this, it continues to be pervasive in research.

\textbf{Critical Analysis of Disability Research in Computer Science.} In recent years, researchers in other areas of computer science have critically analyzed and applied different approaches to define and understand autism beyond deficit-based theories in efforts to be more inclusive \citep{Spiel2019agency, spiel2021purpose,rizvi2021inclusive, zolyomi2019managing, zolyomi2021social, keyes2020automating}. These approaches include using applied critical disability studies and crip technoscience. As an interdisciplinary study of disability and society, \textbf{critical disability studies} attempts to incorporate the perspectives of disabled individuals in research concerning their communities \citep{mankoff2010disability}. In other words, it promotes the idea that research should be done \textit{with} disabled individuals and not just \textit{about} them \citep{mankoff2010disability,spiel2020nothing}. Prior work applying this approach has suggested building technologies that support both autistic and neurotypical people in communicating with each other to help broaden the scope of assistive technologies \citep{mankoff2010disability} among other proposed work. Prior work also suggests that, while participatory approaches may already be applied in the research process, they can be made more inclusive by including the collaboration of participants in all aspects of the research including the study design, data collection, analysis, and dissemination of results \citep{mankoff2010disability}.

Similarly, \textbf{crip technoscience} particularly centers access, interdependence, and disability justice, and has been applied by researchers to examine how the medical model has shaped the status quo for technology research focusing on other neurodivergent populations \citep{spiel2022adhd}. Prior work uncovered how the medical model may marginalize neurodivergent end-users by focusing on outcomes that mitigate their experiences, thus privileging neuro-normative outcomes \citep{spiel2022adhd} by positioning neurotypicality as the 'norm' and neurodivergence as a deviation that can be 'diagnosed' or 'treated' \citep{huijg2020neuronormativity}.
For example, the belief that autistic people are 'deficient' in communication, can lead to dehumanizing assumptions about their agency and personhood \citep{keyes2020automating}. For autistic children, the use of the medical model in research embodies societal expectations that marginalize autistic children while treating them as a secondary audience to purposes not defined by them \citep{Spiel2019agency}. Researchers have been encouraged to acknowledge that the emotional experiences of autistic adults are uniquely influenced by their social interactions and processing of sensory outputs, as they work towards creating more inclusive human-like AI technologies \citep{zolyomi2021social}.

A prior literature review also found that similar to Autism research, ADHD research excludes the perspectives of end-users with ADHD, even in technologies that are about/for them \citep{spiel2022adhd}. To assist, researchers have developed the social-emotional-sensory design map to create more effective affective computing interfaces accounting for neurodivergent users \citep{zolyomi2021social}. However, recognizing that it will take time for research to move in a more inclusive direction that views autism as a difference and not a deficiency, AI ethicists have been encouraged to analyze the consequences of work that relies heavily on the medical model \citep{keyes2020automating}.

Thus, in this work, we examine anti-autistic ableism in HRI research as a case study of the broader marginalization of neurodivergent end-users in technology research. We focus on this area since the marginalization of autistic people in other areas of computer science such as human-computer interaction research \citep{guberman2023not,Spiel2019agency,spiel2020nothing, Ymous2020Terrified}, gaming \citep{spiel2021purpose}, design \citep{spiel2017empathy}, AI ethics \citep{keyes2020automating,bennett2020point}, and affective computing \citep{zolyomi2021social} have been explored in prior work.

\textbf{Robots and Ableism.} Prior work has uncovered how our perceptions of robots and the resulting user experiences are shaped by ableism, while calling into question the effectiveness of developing clinical robots for autistic end-users. In “All Robots are Disabled”, the author explores how our sociality with robots is largely shaped by disability stigma \citep{Williams2022AllRA}. For example, the author discusses an incident where students were reprimanded for “harassing” the robots while studying them on a university campus \citep{Williams2022AllRA}. This insinuates that there are social protocols of respect afforded to robots that their disabled end-users are expected to adhere to which may not be reciprocated, thus highlighting the power imbalances that exist between them. Other work critically examines the power imbalances that arise between robots and autistic end-users when such systems are used to provide therapy, diagnosis, and other forms of rehabilitation \citep{Williams2021IMI}. Research suggests that such objectives relegate robots to a mentorship role in efforts to help autistic end-users emulate “humanness” thereby supporting the idea that autistic people are deficient in their humanity by associating neurotypical behaviors to being human \citep{Williams2021IMI, keyes2020automating}. It is important to note that prior work has also called into question the effectiveness of developing robots to provide therapy to autistic end-users as even clinicians are not convinced of their effectiveness and minimal progress has been made in making such robots clinically useful \citep{begum2016robots}. In fact, research even suggests that this use of robots may be counterproductive and negatively impact the skills they are designed to hone in autistic end-users \citep{diehl2012clinical}. 

Following the principles of critical disability studies and crip technoscience which center the inclusion of disabled individuals in research for/about them, we apply the theories of intersectionality \citep{cho2013toward}, participatory design \citep{sanders2002user}, and design justice \citep{hundt2022robots} to quantify and critically analyze the inclusiveness of the objectives, designs, and results of HRI studies for autism, and the systemic barriers that may shape them. As \textbf{critical autism studies} promotes critically evaluating the power dynamics arising in discourses concerning autism \citep{woods2018redefining}, our work also focuses on untangling these dynamics in HRI research. Using these theories, we define autism inclusion as the application of theories beyond the medical model to understand autism, the involvement of autistic participants as key stakeholders in the design process, and the inclusion of participants from understudied minority groups. We highlight how a deficit-based understanding of autism, an underrepresentation of autistic adults and gender minorities, and power imbalances present in the research study designs and user interactions with robots are pervasive in this work. Our purpose is not to criticize the authors of these studies in any way but to highlight how the studies reflect systemic ableist beliefs.

This leads us to our research question.

\begin{description}
\item[Research Question:] What is the state of autism inclusion in HRI studies?
\end{description}

Our work is novel in applying a mixed methods approach to critically examine the inclusiveness of HRI studies for autism based on their research study designs. We qualitatively and quantitatively analyze 142 papers published between 2016 to 2022. We contribute and provide 1) a critical analysis of the three dimensions in which autism is stigmatized in HRI research; 2) a contextualization of the results in foundational work shaping autism research; 3) characterization of the three dimensions through measurement; 4) systemic barriers; and 5) considerations for future work to help avoid the issues identified.

\section{Methodology}

Following similar data collection and analysis techniques as other critical literature reviews
 \citep{spiel2022adhd}, we systematically searched through literature using the Google Scholar search engine. Through three iterations of data collection, our main corpus included 142 papers.
Our referenced works corpus comprised of all the works referenced by our main corpus. We conducted both inductive and deductive qualitative thematic analysis and used descriptive statistics to analyze the data.

\subsection{Positionality Statement} Following recommendations from feminist methodologies \citep{harding1987feminist}, we position ourselves to contextualize our study design, methodologies, data collection process, results, and analysis. The research team is neurodiverse and includes women and men-identifying researchers who have a strong interest in improving inclusivity in Human-Centered Computing studies.

The authors of this paper believe in community-oriented design practices such as participatory design \citep{harrington2019deconstructing} and design justice \citep{hundt2022robots} to address power imbalances in traditional HCI research practices between participants and researchers. We understand that certain methodologies may require adjustments based on the access needs of the participant communities \citep{maun2021adapting}.

Due to the international scope of our literature review, we focused on gender and age solely for our intersectional analysis. The first author of this paper has lived in five countries, and recognizes the limitations of the US-centric perceptions of race, religion, and other identities which may not translate to other cultures. The exclusion of other identities is not reflective of their perceived importance by the authors. We acknowledge the difficulties researchers may have with recruiting diverse participants for their study, and want to note that we are not criticizing the individual authors of any of the works critiqued in this paper.

\subsection{Main Corpus Selection}

\begin{figure}
\begin{center}
\includegraphics[width=0.49\columnwidth]{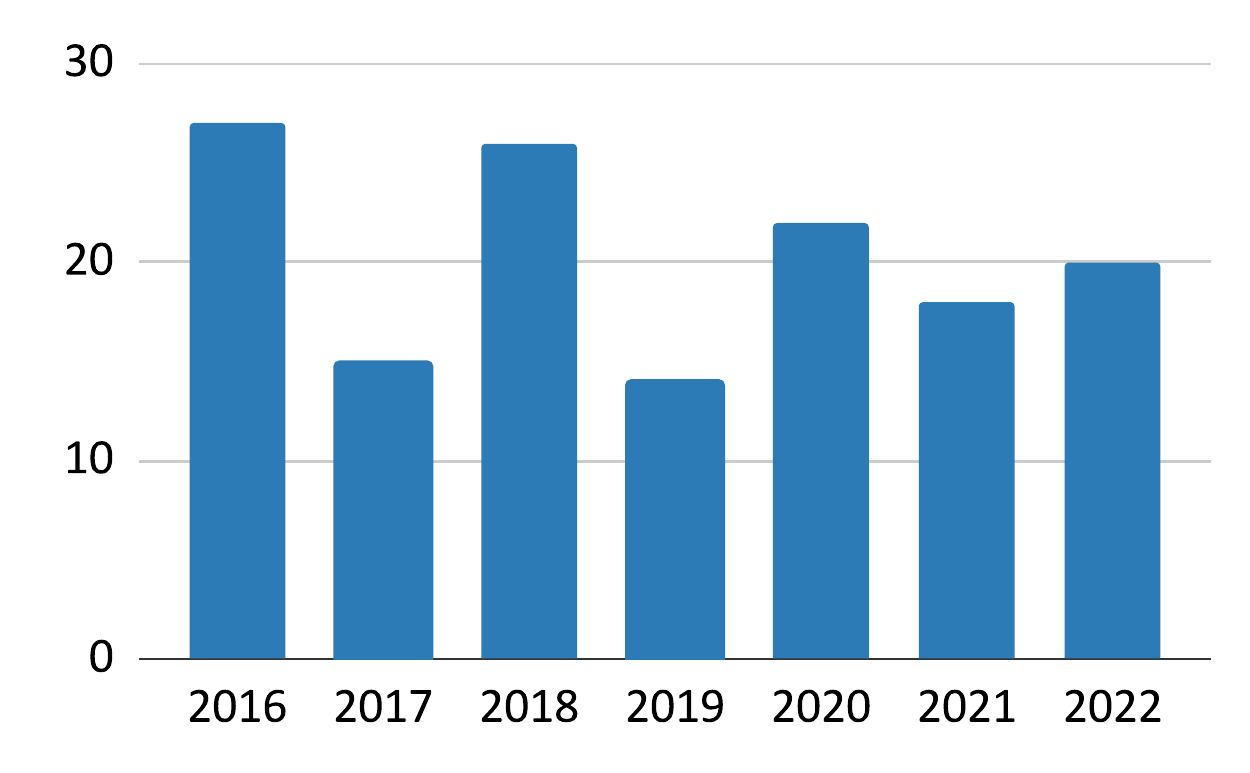}
\end{center}
\caption{Distribution of publication years of the papers included in our main corpus.}
\label{fig:pub_years}
\end{figure}

Our corpus selection was carried out in multiple iterations. We obtained full papers, short papers, survey papers, and editorials from the fifty most relevant results using the criteria detailed in Sections~\ref{sec:search} and~\ref{sec:inclusion} from the Google Scholar search engine, the ACM Digital Library, IEEE Xplore, Springer, and Science Robotics. Our search criteria specified the publication years ranging from 2016 to 2022. The distribution of the publication years of our main corpus are detailed in Figure~\ref{fig:pub_years}. For searches that revealed more than 50 results, we filtered our results based on relevancy and reviewed the first 50 papers.

We selected 142 papers~\footnote{https://figshare.com/s/576daf217af2178026a1} for our final analysis. Table~\ref{tab:venues} summarizes the venues and papers selected for our analysis.

\begin{table*}
\begin{tabular}{|c|c|c|}
\hline
\textbf{Venue}     & \textbf{\# Res.}
& \textbf{\# Sel.}
\\ \hline

ACM/IEEE Int. Conf. on Human-Robot Interaction (HRI)* & 50 & 36                           \\ \hline
IEEE Intl. Symp. on Rob. \& Human Interactive Comm. (RO-MAN)* & 48 & 36 \\ \hline
Int. Journal of Social Robotics (JSR)* & 50                 & 26                           \\ \hline
IEEE/RSJ Int. Conf. on Intelligent Robots and Systems (IROS)* & 33 & 3 \\ \hline
ACM SIGACCESS Conf. on Computers and Accessibility (ASSETS)                                         & 50  & 4                            \\ \hline
ACM Conf. on Computer Supported Cooperative Work (CSCW)                                                & 11  & 2                            \\ \hline
Science Robotics (SR)* & 19  & 5                            \\ \hline
ACM Transactions on Human-Robot Interaction (THRI)* & 50 & 9                            \\ \hline
IEEE International Conf. on Robotics and Automation (ICRA)* & 37  & 4                            \\ \hline
The ACM Conf. on Human Factors in Computing Systems (CHI)                                 & 50   & 12                           \\ \hline

IEEE Robotics and Automation Letters* & 1  & 0                           \\
\hline

\end{tabular}
\caption{Selected conferences and journals and the number of papers included in our main corpus. }

\label{tab:venues}
\end{table*}
\subsubsection{Venue Selection}
\label{sec:venue} We chose our venues based on Google Scholar's h5-index ranking in robotics \footnote{\url{https://scholar.google.com/citations?view_op=top_venues&hl=en&vq=eng_robotics}}
 and HCI \footnote{\url{https://scholar.google.com/citations?view_op=top_venues&hl=en&vq=eng_humancomputerinteraction}}.
These venues were: IEEE International Conference on Robotics and Automation (ICRA), IEEE Robotics and Automation Letters (RAL), Computer Human Interaction (CHI), IEEE Transactions on Affective Computing (TAC). While RAL and TAC had no relevant corpus, we were able to obtain 16 papers from CHI and ICRA. We added the International ACM SIGACCESS Conference on Computers and Accessibility (ASSETS) and the ACM Conference on Computer-Supported Cooperative Work (CSCW) due to their focus on accessibility and human-centered technologies. We found two papers from CSCW and four from ASSETS fitting our selection criteria which we included in our main corpus.

Next, we consulted an expert in human-robot interaction to identify the top HRI venues and added the following to our venue list: International Journal of Social Robotics (JSR), ACM/IEEE International Conference on Human-Robot Interaction (HRI), IEEE International Workshop on Robot and Human Communication (RO-MAN), and the ACM Transactions on Human-Robot Interaction (THRI). In contrast to our first search iteration, 60\% of the literature obtained through this search fit our selection criteria and they comprised the majority of the papers included in our study.

In total, we obtained 142 papers from more than 10 venues, detailed in Table~\ref{tab:venues}. Unsurprisingly, the majority of the papers were published in three HRI venues: HRI, RO\-MAN and JSR.

\subsubsection{Search Criteria. }\label{sec:search} We searched for papers published between 2016 and 2022. For venues focusing on robotics or HRI, our Google scholar search string included the conference or journal name and \textit{autism OR autistic OR ASD}. For other venues such as CHI, we added \textit{AND (robots OR robotics)} to the search string. While searching through the ACM Digital Library, IEEE Xplore, Springer, and Science Robotics, we also specified the conference or journal name. For our search, we looked for papers with the key words in the title, abstract or paper text. In addition to full papers, we included short papers, survey papers, and editorials to broaden the scope of papers considered to include future work and works in progress.

\subsubsection{Inclusion Criteria. }\label{sec:inclusion} Table ~\ref{tab:venues} details the total number of papers found through each search in the '\# Results' column and the final number of papers included in our main corpus in the '\# Selected' column. We began our selection process by excluding any papers that were published outside of the venues and years specified in the sections above. For example, while searching for CHI papers on the ACM Digital Library, we encountered works that were published in TOCHI which we excluded from our corpus. We obtained the abstracts for all the remaining papers found in our search results using the Semantic Scholar API. We analyzed these abstracts to only include papers that: i) focused on robotics, and ii) explicitly identified autistic people as the end-users for their work. If the API was unable to find the abstract, we manually obtained it and analyzed the full papers. For papers that used other terms to define their target users that may include autistic people, such as 'neurodivergent', we also read other sections of the paper for clarification to determine whether autistic people were their target end-users. Our main corpus contains the papers filtered using these inclusion criteria from each venue.

\subsection{Referenced Works Selection}
These works were selected through the automated analysis of all the references from our main corpus. We built a web crawler that obtains the Digital Object Identifier (DOI) of each paper as a unique ID. Using these DOIs, we found matching works cited in each paper and obtained a total of 6,611 papers. In order to gain a better understanding of trends in the most frequently referenced works in HRI research for autism, we analyzed the publication years and fields of all of these papers. We selected the 100 works most frequently referenced by our main corpus and conducted a deductive qualitative thematic analysis to identify the model used to define autism in their work.

\subsection{Data Analysis}
Our data analysis consisted of two parts: a manual coding of the corpora and an automated analysis of the metadata obtained from the Semantic Scholar API.

\subsubsection{Manual Analysis}
A thematic analysis was carried out in several iterations described in Table~\ref{tab:reading_iterations} by three authors of this paper. This analysis was both deductive and inductive in nature as we referenced HRI literature to code the robot automation levels and roles \citep{beer2014toward,baraka2020extended}, but generated our own codes for other data such as the proposed use of robots. In order to determine the model used in each study, we first identified the proposed use of robots in each paper. The criteria for determining the usage of robots in the studies are detailed in Table~\ref{tab:proposed_use}. Using Braun and Clarke's six-step method \citep{clarke2013teaching, majumdar2022thematic}, we generated several pertinent themes in each iteration detailed in the results section and provided some exemplary works in Table~\ref{tab:reading_iterations}. To generate these themes, we first familiarized ourselves with the data by reading the papers and identifying their research objectives. For example, we collected data on the proposed use of robots by looking at the objective of their interactions with autistic end-users, such as providing dance therapy, detection of autism-related aggressive behavior, or support in the ER. Through discussion, we narrowed these to codes such as therapy, behavior detection, skills training, and medical support and synthesized them into the themes detailed in Table ~\ref{tab:proposed_use}. The coding for all of the themes in the proposed use were exclusive. However, the classification of the models applied in some of the themes detailed in Table~\ref{tab:proposed_use} were non-exclusive. For example, while the support theme exclusively included the self-care, and medical, social, general, \& educational support codes, these codes were not exclusively considered medical or social model codes as they did not necessarily pathologize autism. In contrast, the codes for the treatment theme were also exclusively placed in the medical model theme due to their clinical approach to understanding autism.

\begin{table*}
\centering
\begin{tabular}{|c|M{3cm}|M{5cm}|M{2.8cm}|}
\hline
\textbf{Reading Iteration} & \textbf{Objective} & \textbf{Information Collected} & \textbf{Qualitative Analysis Methods} \\ \hline
First* & Model Identification & Model applied to understand autism in the study (e.g. the deficit-based medical model) \citep{kapp2019social}, the proposed use of robots, the inclusion of autistic individuals in the study design & Thematic analysis (both inductive and deductive) \citep{majumdar2022thematic} \\ \hline
Second* & Intersectionality Considerations & Age of participants, representation of gender minorities (i.e. women, and girls, as the papers did not report other genders) & Thematic analysis (inductive) \\ \hline
Third* & Analyzing Robot Characteristics  & User data collected, automation level \citep{beer2014toward}, interaction type, robot name, robot characteristics, robot roles \citep{baraka2020extended} & Thematic analysis (both inductive and deductive) \\ \hline
Fourth & Analysis of Frequently Referenced Works & Model applied in study \citep{kapp2019social}, proposed application of research & Thematic analysis (deductive) \\ \hline
\end{tabular}
\caption{For our manual data collection, we carried out several reading iterations to identify and collect the information from the corpus detailed in this table.
The first, second, and third reading iterations were carried out for our main corpus only.}
\label{tab:reading_iterations}
\end{table*}

\begin{table*}
\centering
\begin{tabular}{| c | M{7cm} | M{4cm} | }
\hline
\textbf{Theme} &  \textbf{Criteria and Codes} & \textbf{Examples} \\ \hline
Treatment & Robots used for therapy, skills training, or other forms of treatment. These technologies may be co-designed with professionals such as therapists and usually focus on a specific set of skills which autistic people are `deficient'' in according to the medical model. \newline \textbf{Codes:} \textit{therapy, skills training, symptom management, treatment}& Eye contact training, responding to interruptions, and other body movements \citep{kulikovskiy2021can, scassellati2018improving}. \\ \hline        Diagnosis & Robots used to diagnose or identify certain behaviors \newline \textbf{Codes:} \textit{diagnosis, behavior detection, quantifying harmful behavior}& Detecting autism-related headbanging, analyzing imitation deficits for autism assessment \citep{washington2021activity,wijayasinghe2016human}. \\ \hline
Support & Robots developed to provide recreational support to autistic end-users. \newline \textbf{Codes:} \textit{medical support, general support, self-care, educational support, social support}& Robot companions, toys \citep{gelsomini2017puffy, di2020explorative}. \\ \hline
Assisting Specialists & Robots designed to assist certain specialists (excluding therapists) \newline \textbf{Codes:} \textit{teachers, ER workers, caretakers} & Assisting special education teachers, caregivers, and other medical professionals \citep{schadenberg2020helping, kirstein2016social}. \\ \hline
Awareness\footnote{This is different from autism inclusion as it centers non-autistic people, and may pathologize autism} &  Robots that center the perspectives of non-autistic people, spread disinformation, outdated perspectives, and harmful myths \citep{doyle2021acceptance}.\newline \textbf{Codes:} \textit{autism awareness} & Robots to make non-autistic people aware of what it is like to be autistic \citep{lin2021xiaoyu} \\ \hline
\end{tabular}
\caption{Through a qualitative thematic analysis, we categorized the proposed use of the robots in each study into one of five categories.}
\label{tab:proposed_use}
\end{table*}

\subsubsection{Theoretical Analysis}
Once the themes were identified, we then categorized papers from the corpus using theories from Critical Autism Studies and neurodiversity (pathologizing) \citep{woods2018redefining,kapp2013deficit}, intersectionality (essentialism) \citep{cascio2021making}, and Feminist HCI (power imbalance) \citep{winkle2023feminist}. For example, the medical model views autism as a medical condition that needs treatment \citep{anderson2022autism} so that autistic people can move toward 'humanness' \citep{Williams2021IMI}. This promotes neuronormativity as it positions the social and communicational preferences of neurotypicals as the 'norm' that autistic individuals deviate from and must adhere to \citep{huijg2020neuronormativity}. This approach to understanding autism has been linked to worrying social concerns detailed in section~\ref{sec:intro}. Thus, we coded papers proposing the usage of robots for diagnosis, treatment, or therapy as medical model papers as they focused on making autistic people adapt to neurotypical norms. Due to the prevalence of clinical collaborations in HRI research for autism and its emphasis on \textbf{Evidence-Based Practices (EBP)}  \citep{begum2016robots}, we have two separate codes for 'therapy' and 'skills training' papers. The 'therapy' code included only the studies employing a close collaboration with therapists to develop specialized 'treatments', while the 'skills training' code included studies taking a less therapeutically centered approach. The typical robot-assisted therapy sessions employed Applied Behavior Analysis techniques to provide skills training such as emotion recognition \citep{rudovic2018personalized,chung2021robot,dickstein2017interactive,kulikovskiy2021can,taheri2018human,sochanski2021therapists,trombly2022robot,tyshka2022transparent, korneder2021wild}.

\subsubsection{Automated Analysis}

Using the Semantic Scholar API, we obtained the abstracts and bibliographic information such as the title, research field, and publication year of all the papers in our references corpus. These papers were included in the fourth iteration of our manual analysis described in Table~\ref{tab:reading_iterations}.

\textbf{Variance in Literature Surveyed. } While all of the papers in our main corpus provided enough information to categorize the proposed usage of robots and the model applied to understand autism in their work, some papers omitted the information needed for our other analysis. For example, we discovered that the usage of robots is difficult to analyze and categorize in survey papers. Additionally, not all the papers reported the gender ratio and ages of their participants, or detailed information on the user interactions. Thus, the sample size of each visualization is included to contextualize the results.

\section{Results}

We analyzed the objectives, research study designs, and findings of all the works in our main corpus. Through an inductive thematic analysis, we uncovered three dimensions of stigmatization in HRI research for autism: pathologization, essentialism, and power imbalances. These dimensions encompass the theoretical understanding of autism applied in the research objectives, the inclusion of autistic people in the study design, and how power imbalances have shaped their methods and results.

\subsection{Pathologizing}

\begin{figure*} \begin{minipage}{\textwidth} \centering \includegraphics[width=0.80\columnwidth]{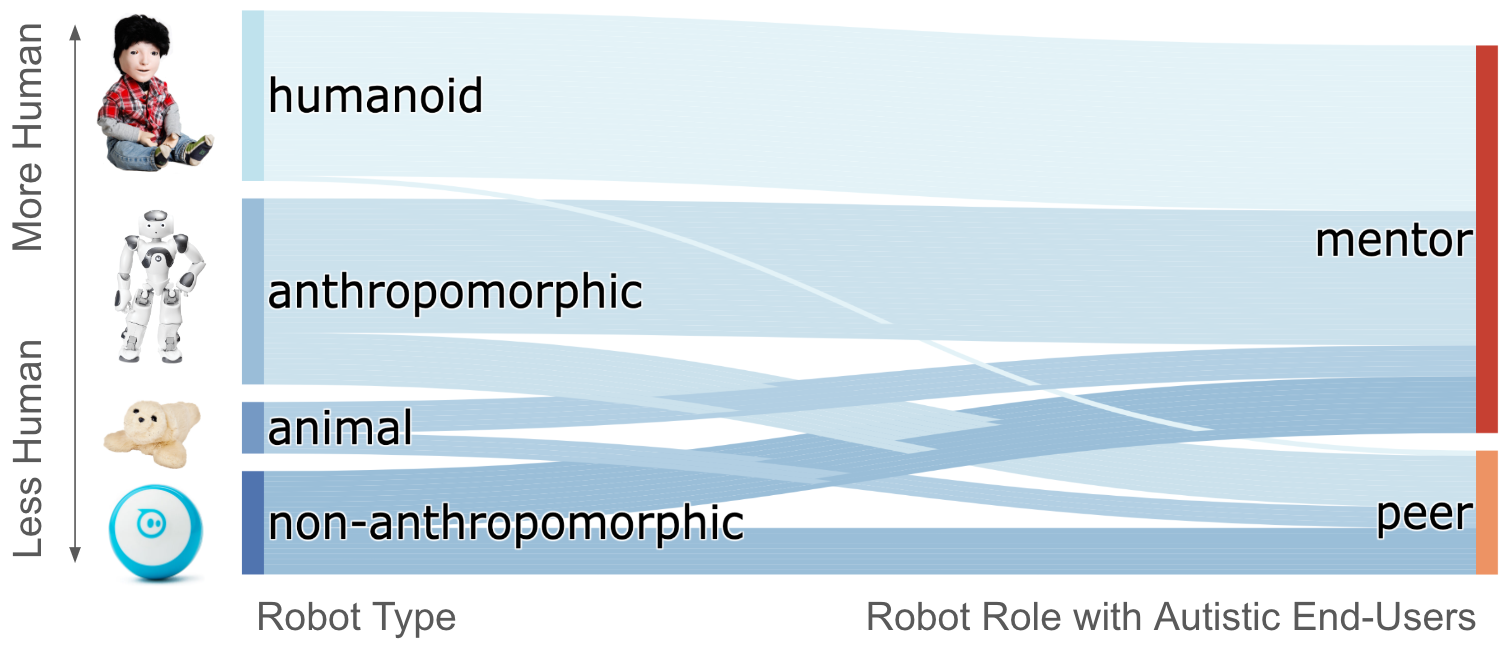} \captionof{figure}{There is a notable correlation between thehumanness'' of robots and a power imbalance in user interactions with autistic people in our main corpus.    More human-like robots were placed in mentor roles for diagnosis or helping autistic people move toward 'humanness' \citep{Williams2021IMI}. Although autistic children prefer engaging with non-anthropomorphic robots \citep{ricks2010trends}, only 13.9\% of HRI studies utilized them.}
\label{fig:sankey}
\end{minipage}
\end{figure*}

Our foundational knowledge and understanding of autism has been shaped by researchers who adapted dehumanizing language through the \textbf{Theory of Mind''} \citep{happe1995theory, baron1997mindblindness}. They expanded on prior work analyzing whether chimpanzees have a theory of mind and concluded that autistic children lack the quintessential'' human trait \citep{kapp2019social}. This deficit-based understanding is central to the \textbf{medical model}, which focuses mainly on the causation or cure of autism as a disorder \citep{kapp2019social}. This dehumanization of autistic end-users is replicated in human-robot interactions. Traditionally, robotics research has pathologized the communication behaviors preferred by autistic people and introduced technologies that encourage them to adopt neurotypical social norms \citep{Williams2021IMI,werry1999applying, dautenhahn1999robots, billard2003robota}. Figure~\ref{fig:sankey} illustrates how 15 of the studies in our corpus used robots that looked like animals to play a mentor role in their interactions and focused on providing social skills training to autistic end-users. By placing animalistic robots in this role, such studies inadvertently perpetuate the findings of the dehumanizing foundational work in autism research that concluded certain animals have stronger social skills and are thus more human'' than autistic children \citep{happe1995theory,baron1997mindblindness}. Additionally, the majority of the 76 studies utilizing anthropomorphic and humanoid robots had the robots adapt a mentor role with autistic users, which perpetuates the belief that autistic people are deficient in their humanity and robots can help them move toward humanness'' \citep{Williams2021IMI}. 

A deficit-based understanding of autism has been pervasive even in computer science and robotics research \citep{Williams2021IMI,dautenhahn1999robots, wood2021developing, jain2020modeling, robins2005robotic, kim2013social, cabibihan2013robots, liu2008online} even though prior work has suggested it contributes to the ostracism and discrimination autistic individuals face in our society \citep{kapp2019social, annamma2016discrit, rizvi2021inclusive}. The ostracism of autistic people in studies that use this definition can be quite explicit. For example, one paper studied the use of robots to:

\begin{quote}
`diagnose abnormal social interactions within autistic children'' \citep{arent2019social}.\end{quote}

\begin{figure*}[tb]
\centering
\begin{minipage}{.47\textwidth}
\centering
\includegraphics[width=\textwidth]{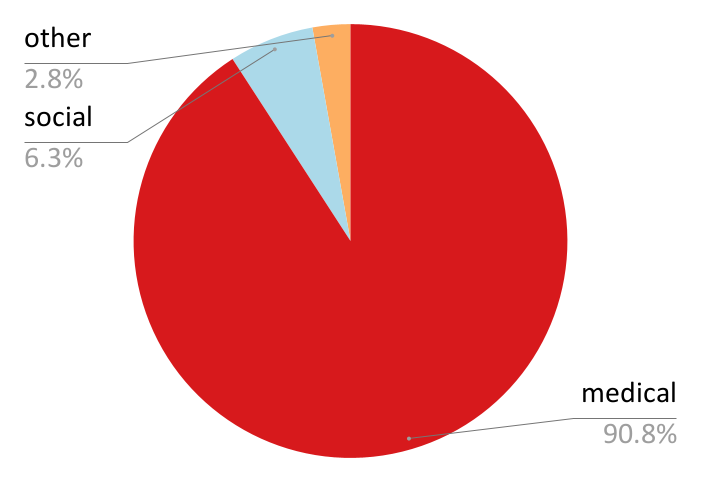}
\captionof{figure}{An overview of the disability models applied in our main corpus reveals the medical model is the most commonly used model in HRI research. Papers were categorized as 'other' if they did not exclusively fall under the medical or social model.}
\label{fig:models}
\end{minipage}%
\hspace{3mm}
\begin{minipage}{.47\textwidth}
\centering
\includegraphics[width=\textwidth]{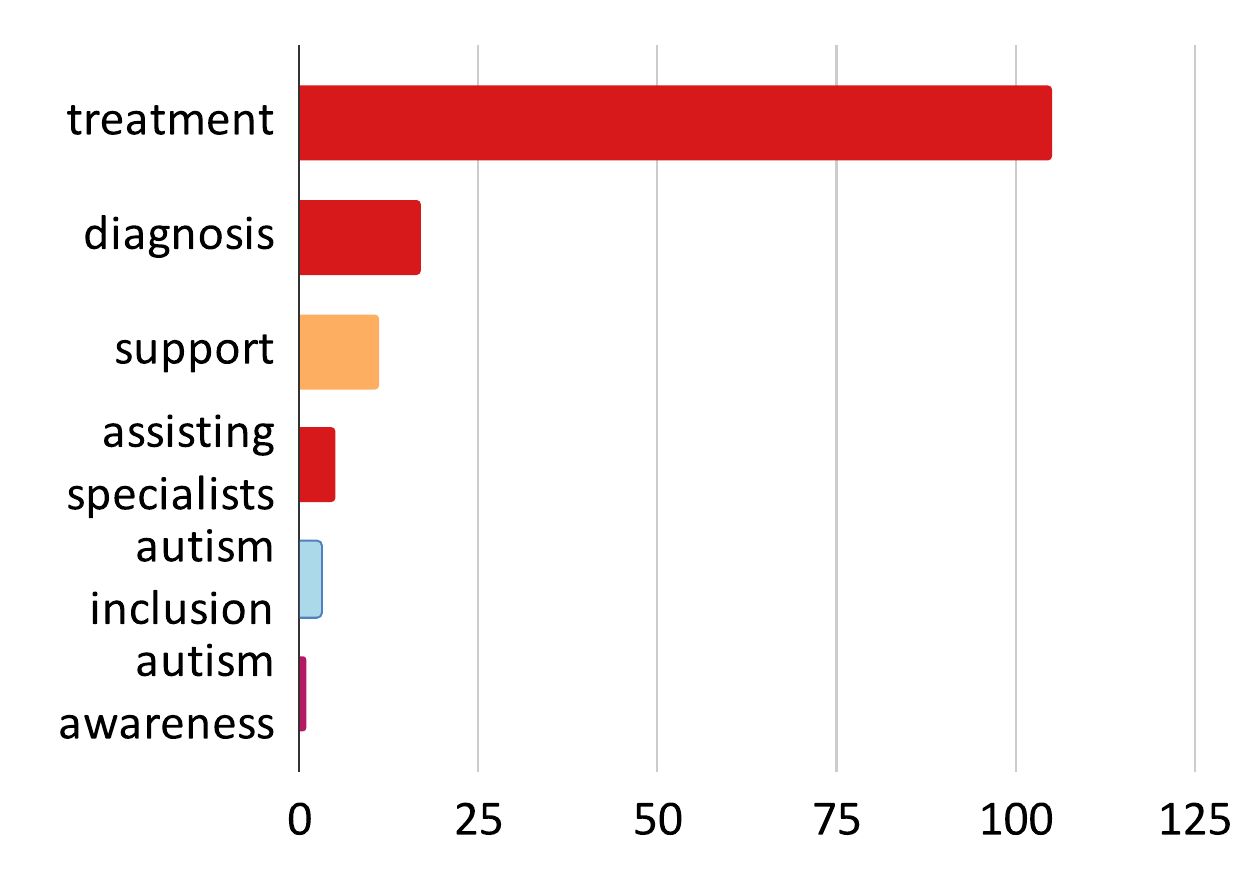}
\captionof{figure}{The majority of papers in our main corpus focused on providing treatment to the end-users. Support is shown in orange as the codes for this theme are not exclusive to a particular disability model, while the codes for autism inclusion (light blue) exclusively do not apply the medical model.}
\label{fig:usage}
\end{minipage}
\end{figure*}

Such language is rooted in ableism as historically the word `abnormal'' has been used to describe things that are unhealthy or unnatural and it dehumanizes disabled people \citep{neilson2020ableism, cherney2011rhetoric, hurst2003international}. Yet, an automated analysis uncovered words such as typically develop*' and abnormal' which explicitly posit non-autistic people as the norm and their autistic peers as a deviation from the norm were prevalent in 27 papers, while deficit-based language appeared in 11 papers as shown in Appendix 1a~\footnote{https://figshare.com/s/576daf217af2178026a1}. This led us to uncover the prevalence of the medical model that applies a deficit-based understanding of autism and promotes diagnosis and treatment through therapy and skills training \citep{kapp2013deficit}, as shown in Figures~\ref{fig:models} and~\ref{fig:usage}.

\subsubsection{Deficit-Based Studies}
A deficit-based understanding of autism was prevalent
in our main corpus. The majority of the studies in our corpus 93.5\%, n=129) applied the medical model in their work as shown in Figure~\ref{fig:models}. Out of those studies, the majority (85.92\%, n=122) focused on diagnosis and treatment as shown in Figure~\ref{fig:usage}. Although the papers focusing on treatment, diagnosis, awareness, and assisting specialists in our main corpus always applied a medical model understanding of autism in their work, the papers using robots for support applied various models. The support category includes robots that provide companionship, entertainment, or other kinds of support to autistic end-users who are usually children. An example of a paper applying the medical model in such robots included providing behavioral assistance to children \citep{di2020explorative}, while an example of a paper applying other models used robots to provide more affordable educational support to children in underserved rural schools \citep{broadbent2018could}.

\begin{figure*}[tb]
\centering
\begin{minipage}{.45\textwidth}
\centering
\includegraphics[width=\textwidth]{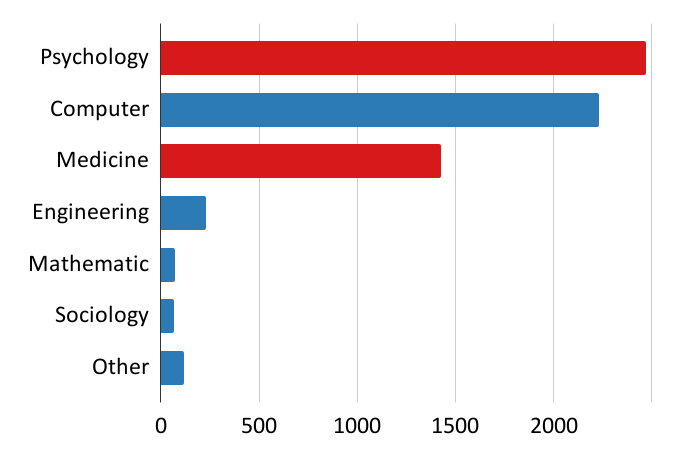}
\captionof{figure}{Psychology, medicine, and computer science are the most frequently referenced fields by our main corpus. Pyschology and medicine are shown in red as they have historically applied a medical model approach by viewing autism as a disorder \citep{golt2022history}. The following fields were grouped together in the `Other'' category as they individually comprise less than 2\% of the works referenced: Physics, Biology, Political Science, Economics, Art, Business, Materials Science, Geography, History, Geology, Chemistry, Philosophy, and Environmental Science.}
\label{fig:influential_fields}
\end{minipage}
\hspace{0.05\textwidth}
\begin{minipage}{.45\textwidth}
\centering
\includegraphics[width=\textwidth]{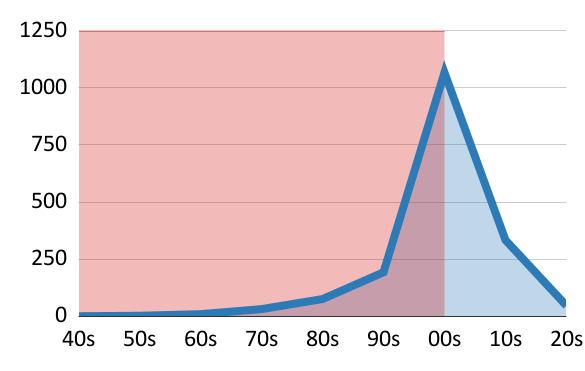}
\captionof{figure}{The majority of our referenced works corpus was published when deficit-based theories dominated autism research. The red area represents the papers that were published before the introduction of Critical Autism Studies which seeks to challenge dominant misunderstandings of autism and have a positive impact on the lives of autistic people \citep{o2016critical}. This data has been annualized to allow for a fair comparison.}
\label{fig:influential_years}
\end{minipage}%
\end{figure*}

\begin{figure}[tb]
\begin{center}
\includegraphics[width=0.47\textwidth]{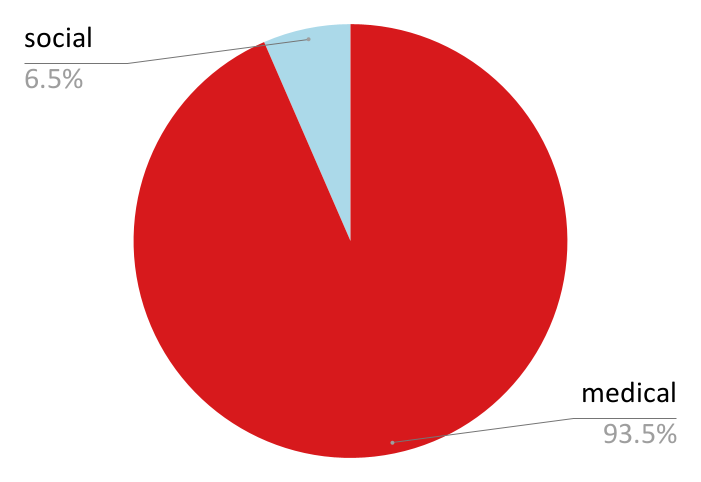}
\end{center}
\caption{An overview of the models applied in our referenced works corpus reveals that the medical model of autism is the most pervasive in these works.}
\label{fig:influential_models}
\end{figure}

Our findings suggest that the deficit-based understanding of autism may be reflective of the foundational work HRI researchers are building upon. Through a deductive qualitative analysis of the 138 works most referenced by the papers in our main corpus, we uncovered a similar understanding of autism was applied by 129 papers, the overwhelming majority, as shown in Figure~\ref{fig:influential_models}. Interestingly, the majority of the works referenced by our main corpus were published in only 3 fields: psychology, computer science, and medicine, as shown in Figure~\ref{fig:influential_fields}. It is important to note both psychology and medicine have studied autism using a clinical approach. Figure~\ref{fig:influential_years} illustrates how the majority of these works were also published before more inclusive theories were introduced in the 2010s.

\subsubsection{Diversify Autism Definition}
\begin{figure}[tb]
\begin{center}
\includegraphics[width=0.49\columnwidth]{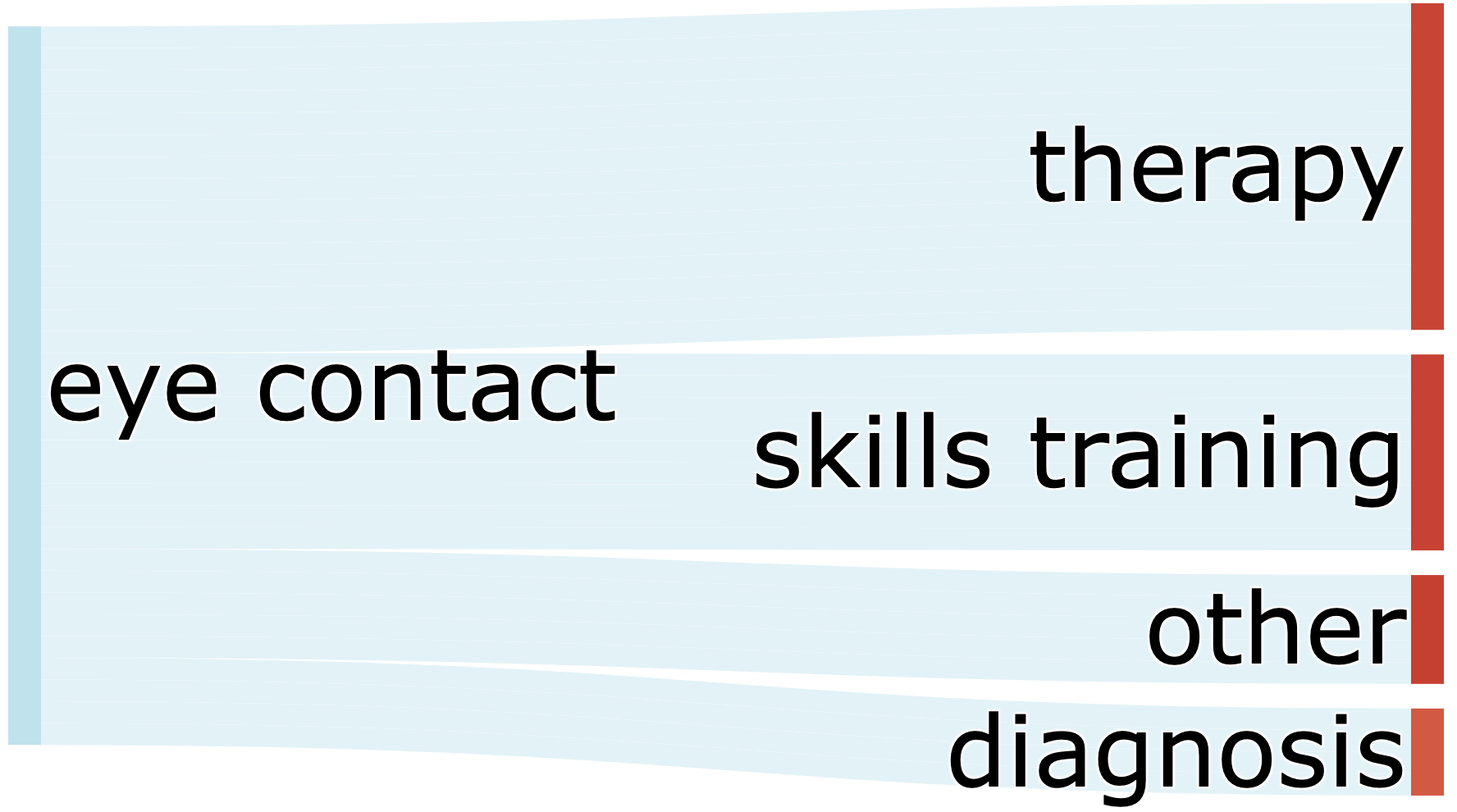}
\end{center}
\caption{The majority of studies collecting eye contact data from the users focused on clinical applications such as therapy, skills training, or diagnosis.}
\label{fig:eyecontact}
\end{figure}

As we continue to make improvements in our understanding of autism, these changes are reflected in the diverse theories of autism that exist today~\footnote{https://www.bps.org.uk/psychologist/me-and-monotropism-unified-theory-autism}. In recent years, there has been a growing interest in human-computer interaction research applying identity or difference-based understandings of autism in their work \citep{Spiel2019agency, spiel2021purpose,rizvi2021inclusive, zolyomi2019managing, zolyomi2021social}. The \textbf{neurodiversity} movement promotes this difference-based definition of autism \citep{kapp2013deficit}. Neurodiversity is heavily influenced by the Mad Pride movement of the 1960s, which denounced psychiatric labeling and promoted embracing madness'' as a unique identity to be celebrated \citep{dyck2020challenging}. Critical Disability Studies applies a difference-based approach by promoting accommodations and equality over the pathologization of disabilities \citep{reaume2014understanding}. It examines disability as both the lived experiences and realities of a disabled individual, and the existing social and political power imbalances in our society \citep{reaume2014understanding}. In contrast to the medical model, the \textbf{social model} defines disabilities as a social construct that emerged due to a lack of accessibility \citep{woods2017exploring}. According to this model, autism is a difference and not a deficit in communication behaviors. Figure~\ref{fig:models} shows that 6.3\% of HRI research for autism applied this definition in their work, which may indicate such research is moving toward a more inclusive direction. The cross-neurological theory of mind also applies a difference-based approach by stating that there is no universallycorrect'' way to represent one's thoughts and feelings \citep{beardon2017autism}. This theory suggests that miscommunications arise due to lack of accommodations and acceptance of neurodiversity. Other researchers studying cross-neurological social interactions have identified barriers to effective communication that may arise due to neurotypical individuals misunderstanding, responding improperly to communication preferences, and misinterpreting the non-verbal communication behaviors of autistic individuals \citep{sheppard2016easy, faso2015evaluating,brewer2016can, rizvi2021inclusive,heflin2007students}. Similarly, the \textbf{double empathy problem} extends a difference-based understanding to the communicational difficulties experienced by people with different conceptual understandings and outlooks \citep{milton2012ontological}. For example, when an autistic person converses with an allistic (i.e. non-autistic) person, both of them may experience difficulties in understanding each other due to their different neurotypes. In order to communicate more effectively, researchers have suggested autistic and allistic individuals may benefit from creating a shared communication system \citep{bennett2019life} or learning about each other's communication styles and preferences \citep{rizvi2021inclusive}. This challenges existing social inequalities which place the burden only on autistic people to adapt to neurotypical social norms \citep{rizvi2021inclusive}. The double empathy problem has also been proposed as a guiding framework for a more autism-inclusive approach to design research \citep{Morris2023Double}.
Even though expecting autistic people to adapt to neurotypical social norms contributes to anti-autistic discrimination in our society \citep{kapp2019social, annamma2016discrit, rizvi2021inclusive}, the majority (85.92\%, n = 122) of papers placed the burden of overcoming the double empathy problem entirely on autistic people. Prior work has pathologized autistic people's unique communication styles by, for example, focusing on correcting the ways in which people with autism use eye contact in communication, as shown in Figures ~\ref{fig:usage} and \ref{fig:eyecontact}. Since autistic adults have expressed interest in developing social skills \citep{cummins2020autistic}, HRI researchers should work towards designing, developing, and testing solutions that support autistic adults in their learning and encourage the acceptance of various communication styles.

% For the centered title:
\begin{neurotipbox}[title={Autism-Inclusion Tips to Avoid Pathologization}]
\begin{itemize}
  \item Consider that humans communicate and perceive the world in diverse ways, and those differences are not deficits.
  \item Diversify the foundational work for your research studies by citing newer research published in fields beyond medicine and psychology, such as Critical Autism Studies.
  \item Consider research directions promoting communication between different neurotypes in a balanced manner instead of placing the burden entirely on autistic people to adapt to different communication styles, such as the works of Morris and Rizvi et al.\,\citep{Morris2023Double,rizvi2021inclusive}.
\end{itemize}
\end{neurotipbox}
\subsection{Essentialism}

Essentialism, in the context of Autism, is the idea that people with autism have common characteristics that are "inherent, innate and unchanging" \citep{sahin2018essentialism}.
Historically, the \textbf{extreme male brain theory} has identified autism as an extreme presentation of a `normal'' male profile \citep{baron2002extreme}. This theory has been criticized for its gender essentialism and pervasive influence \citep{krahn2012extreme}. For example, the belief that autism is a masculine'' disorder has led to the underrepresentation of women and girls in autism research; due to misconceptions among researchers and the general population, people are more likely to recognize autism in males \citep{aggarwal2015misdiagnosis,wilkinson2008gender}. Foundational research also \textbf{infantilizes} autistic adults by characterizing them as children'' \citep{botha2022autism}, and viewing autistic bodies as being frozen'' in childhood \citep{bosco2023bodies}. This infantilization has led to widespread misbeliefs that autistic people are dependent and lack autonomy \citep{bosco2023bodies}. Even in HCI and HRI, prior work on autism research has mainly focused on children \citep{ismail2019leveraging, cabibihan2013robots,Spiel2019agency,frauenberger2016designing, cramer2011classroom, ringland2019place, spiel2017empathy}. Additionally, the concept that an autistic person's mental age'' is different from their physical age is rooted in ableism and specifically pervasive in autism research \citep{boursier2022disablism}.
\begin{figure}
\begin{center}
\includegraphics[width=0.49\columnwidth]{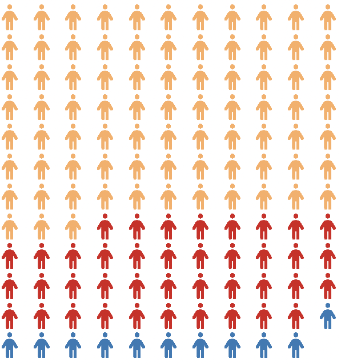}
\end{center}
\caption{The majority of studies in our main corpus (shown in orange) did not report their gender data. Among the papers that did report, only 12 studies had a gender ratio that is representative of the autistic population (shown in blue) \citep{maenner2023prevalence}.}
\label{fig:gender}
\end{figure}

\subsubsection{Essentialism in the Main Corpus}

\begin{figure}
\begin{center}
\includegraphics[width=0.49\columnwidth]{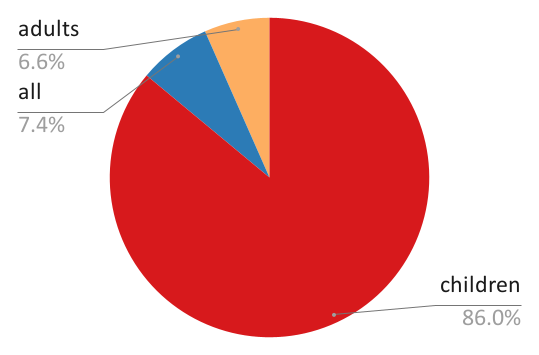}
\end{center}
\caption{An overview of the ages of the participants reported in our main corpus. The participant ages were reported in n=139 of the papers in our main corpus.}
\label{fig:ages}
\end{figure}

Through an analysis of the participant demographics and study objectives, we investigated how the historical masculinization and infantilization may be perpetuated by these studies, through an underrepresentation of adults and gender minorities, as shown in Figures~\ref{fig:gender} and~\ref{fig:ages}. Due to the masculinization of autism, many gender minorities  may not learn about their neurodivergence until later in life and may face additional barriers due to the intersections of their gender with autism \citep{aggarwal2015misdiagnosis}. An example of such gender-based stereotypes is linking traits like aggression'' to autism, which is more socially acceptable for boys and men than people of any other gender, and thus uncommon in gender minorities with autism \citep{geelhand2019role}. Yet, one of the papers we surveyed linked this trait to autism, and did not take gender into consideration: \begin{quote}Issues with [...] aggression [...] are common in children with autism'' \citep{sharmin2018research}.\end{quote}

This paper investigates the usage of smart technologies, such as robots for many purposes, including diagnosis. However, it links aggressive behavior to autism, which has historically contributed to the underidentification of women and girls with autism \citep{geelhand2019role}. Currently, there is a well-known gender gap in the autism community, with a ratio of 3 males to every female \citep{maenner2023prevalence}. As this gender gap has been widely studied in other fields, researchers have argued for adapting an intersectional and gender-aware approach to autism research \citep{saxe2017theory}, which some information technology researchers have started applying in their work \citep{annabi2018untold,annabi2023lessons}.
Yet, the average representation of gender minorities such as women, girls, and other marginalized genders in the participant populations was only disclosed in n=55 papers applying a binary definition of gender. The majority of these studies did not have a representation of gender minority participants that was proportionate to their population ratio \citep{maenner2023prevalence}, as shown in Figure~\ref{fig:gender}.

The infantilization of autistic people is pervasive as researchers focus mainly on children and neglect the needs of autistic adults \citep{kirby2021state}. One study exploring the impact of age on robot learning notably excluded autistic adults and compared the results of autistic children to non-autistic children and adults \citep{guedjou2017influence}. While the differences between non-autistic adults and children were considered in the study, the results of autistic children were considered sufficient to represent autistic people as a whole. Thus, even in a study focusing on the intersections of autism and age, autistic adults were left out.

There are known differences in the experiences of older adults with autism \citep{braden2017executive}.
Yet, they remain an understudied demographic in HCI research, as Figure~\ref{fig:ages} shows that 86\% (n=117)  of the studies did not include adults. Age remains a pertinent aspect of participant identities as researchers have highlighted the unique biases autistic adults may face in data collection due to the intersection of their age and neurotype \citep{rubenstein2021bias}. For example, autistic adults may be stereotyped as unable to achieve autonomy, and therefore require more support \citep{bosco2023bodies}. Such infantilization was explicit in one paper asserting that:

\begin{quote}
Autistic people `need someone to help stop them [...]'' \citep{lin2021xiaoyu}\end{quote}

from engaging in what the authors described as stereotypical behaviors \citep{lin2021xiaoyu}. Such language perpetuates beliefs that autistic people lack control over their own actions and need assistance. Providing support to autistic people was the third most common usage of robots in the studies we analyzed, as shown in Figure~\ref{fig:usage}. The premise of this work may inadvertently infantilize autistic people by making assumptions about their autonomy and support needs \citep{bosco2023bodies}.

\subsubsection{Understanding the Importance of Diversity.} Autistic people are often broadly masculinized and infantilized in our society. For example, there is a well-documented gender bias in the medical field in diagnosis and treatment of autism \citep{krahn2012extreme}, and an underrepresentation of women and girls with autism in the media \citep{tharian2019characters}. As well, infantilization is pervasive in portrayals of autistic people. For example, in support organizations and media: children make up 95\% of the autistic people featured in the homepages of regional and local support organizations, and 90\% of autistic characters in fictional books \citep{stevenson2011infantilizing}. Even media that focuses on autistic adults, such as the Netflix TV Series Love on the Spectrum, has been criticized for infantilizing its adult characters \citep{luterman2023review}.

\textbf{Intersectionality} is a framework originally used to study the gender and race-based marginalization Black women experienced in anti-discrimination law and politics \citep{cho2013toward}. This theory has been widely applied to study how social identities intersect to create unique experiences for marginalized populations \citep{cho2013toward}. In studies focused on human experiences, intersectionality encourages researchers to consider various aspects of one's identity and investigate the experiences of diverse groups. For studies focused on people with autism, this would include participants of various ages and genders as autistic adults have diverse needs and may experience unique biases \citep{braden2017executive,rubenstein2021bias}, and autistic gender minorities may have experiences and needs that are different from their cis-male counterparts \citep{aggarwal2015misdiagnosis,saxe2017theory}. While we acknowledge the challenges and difficulties in recruiting diverse participants, our analysis uncovered that the majority of the studies did not even report demographic information such as the participant genders, as shown in Figure~\ref{fig:gender}. Future work should consider reporting this data to contextualize their results without promoting gender or age based essentialism. This will ensure the needs of diverse autistic people are represented fairly and more accurately in such studies.

\begin{neurotipbox}[title={Autism-Inclusion Tips to Avoid Essentialism}]
\begin{itemize}
  \item Prioritize intersectionality in participant recruitment, research objectives, and data analysis.
  \item Avoid ableist language and essentialist stereotypes such as the ones mentioned in Bottema-Beutel et al.’s paper \citep{bottema2021avoiding}. For example, referring to non-autistic children as “typically developing.”
  \item Report participant demographics to help readers contextualize your findings.
\end{itemize}
\end{neurotipbox}
\subsection{Power Imbalance}

Many foundational works in autism research have applied social deficit theories; instead of studying the unique social behaviors preferred by autistic individuals, established research in developmental psychology highlights autistic people's `deficit'' of neurotypical social norms such as the ability to mind read'' and make appropriate'' eye contact \citep{kapp2019social, happe1995theory, baron1997mindblindness}. This phenomenon, known as \textbf{Neuronormativity}, promotes neurotypical behaviors as the norm'', thereby introducing a power imbalance in social interactions between individuals of different neurotypes \citep{huijg2020neuronormativity}.
As a solution to this "problem", researchers have attempted to use robots to train autistic people how to imitate these neurotypical behaviors. Systematic literature reviews and editorial works in HRI research have called for the usage of robots to provide therapy and skills training to autistic end-users \citep{scassellati2012robots,cabibihan2013robots,begum2016robots, saleh2021robot, pennisi2016autism}. This also includes the use of anthropomorphic and humanoid robots which are designed to be more human-like \citep{phillips2018human}. This is concerning as there is research suggesting such robotic systems may not be useful at all \citep{begum2016robots} and their designs directly contradict the known preferences of autistic children \citep{diehl2012clinical,ricks2010trends}. There are additional concerns regarding the context of foundational research in autism which humanized a non-human entity over autistic children \citep{baron1997mindblindness}, as it makes assumptions about autistic people's skills deficits and promotes the belief that non-human entities may be more socially skilled than autistic people \citep{Williams2021IMI}.

Additionally, autistic people have historically faced a lack of representation in both the research teams and the study participants. Critical Autism Studies (CAS) encourages broadening the participation of autistic scholars in autism research \citep{woods2018redefining}. According to CAS, a lack of autistic authorship may impact the accuracy of such work and may fail to address power dynamics in the medical model understanding of autism \citep{woods2018redefining}. While participatory design encouraged community involvement in the design process, researchers have highlighted such approaches may fail to focus on under-served communities if they do not take historical context into consideration, neglect community access and perceptions of the materials and activities, and cause unintentional harm \citep{harrington2019deconstructing}. Many disabled scholars have discussed the harms of excluding under-served disabled populations from the design process \citep{Ymous2020Terrified} and proposed alternative approaches \citep{zolyomi2021social}. Thus, in our work, we analyzed existing power imbalances in the studies on three levels: 1) autistic and non-autistic people in general, 2) the researcher and participants in the studies, 3) and the resulting human-robot interaction.

\subsubsection{Promoting Neuronormitivity and Ignoring Neurodversity}

In social interactions, autistic people are expected to `fix'' their deficit'' of neurotypical social norms \citep{huijg2020neuronormativity}. In particular, autistic people's different eye contact preferences are frequently pathologized by the medical model \citep{kapp2019social, happe1995theory, baron1997mindblindness}. Yet, we found a correlation between the user's eye contact data collected and medical-model driven purposes, as shown in Figure~\ref{fig:eyecontact}. The majority of studies promoted the usage of robots to provide treatment in the form of therapy or skills training to help autistic people adapt neurotypical social norms.
\subsubsection{Participation of Autistic People}
\begin{figure}[tb]
\begin{center}
\includegraphics[width=0.59\columnwidth]{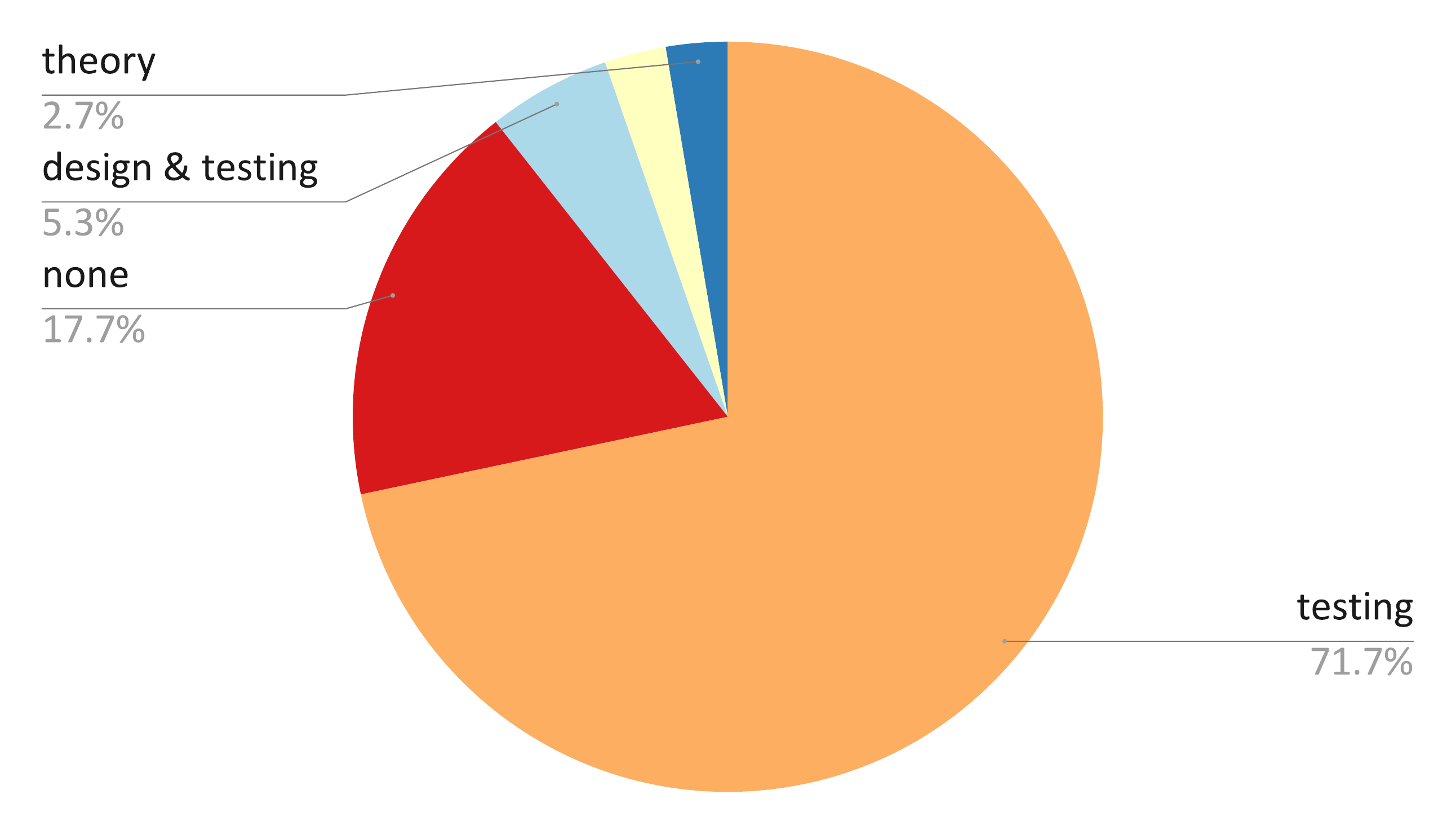}
\end{center}
\caption{The majority of HRI studies did not include autistic people in the design stage black, and nearly one-fifth of them did not include the perspectives of autistic people at any stage of their study.}
\label{fig:study_design}
\end{figure}

Autistic scholars have expressed concerns about tokenism' that arise when research teams fail to provide autistic people with decision-making power in research that impacts them \citep{pukki2022autistic}. Yet, despite being the target end-users for their product, the perspectives of autistic participants were not taken into consideration at all by 17.7\% of the papers that only included therapists, caretakers, or special educators as stakeholders in their studies as shown in Figure~\ref{fig:study_design}. Our results also uncovered that nearly 90\% of the papers did not include the input of autistic people in the design process. This introduces a power imbalance as the design and research objectives are shaped only by the input of the researchers and other specialists (e.g. therapists, and special education teachers), and exclude the perspectives of autistic individuals.

\subsubsection{Human-Robot Interactions}
While scholars have encouraged HRI researchers to re-think social hierarchies in their work \citep{winkle2023feminist}, robots continue enacting malignant stereotypes and existing social inequalities are still enacted in HRI studies \citep{hundt2022robots,Williams2021IMI}. Utilizing robots for therapy, skills training, and similar purposes can promote a power imbalance.
As shown in Figure~\ref{fig:sankey}, the majority of robots with human-like traits are more likely to play a mentor role which introduces a power imbalance in their interactions with the users based on their humanization. In contrast, less human-like robots play a `peer'' role in their interactions, which echoes foundational work in autism research questioning the humanity of autistic people \citep{happe1995theory,baron1997mindblindness,Williams2021IMI}.

\textit{Promoting Neurodiversity.} Using robots as mentors to teach social skills to autistic people promotes the belief that autism is a deficiency of certain social norms \citep{kapp2013deficit}. It is important to note that little work has been done on developing robots that `flip the burden'' \citep{rizvi2021inclusive} of understanding diverse communication behaviors on non-autistic users by teaching them how to interact with autistic individuals. Thus, while existing technologies may assist autistic individuals in understanding their allistic peers, they ultimately fail to address the double empathy problem \citep{milton2012ontological}. Researchers should focus on developing solutions that train non-autistic users how to understand and adapt to neurodiverse social skills. This approach is more equitable as it validates the communication skills of autistic individuals instead of viewing them as having a deficit'' of neurotypical skills.

% ------------------ Power Imbalance — Part 1 ------------------
\begin{neurotipbox}[title={Autism-Inclusion Tips to Avoid Power Imbalances — Part 1: Social Norms}]
\begin{itemize}
  \item Consider community-based research collaborations in lieu of purely clinical collaborations.
  \item Reconsider diagnosis- or treatment-based research directions that prioritize clinical outcomes.
  \item Identify the needs of autistic end-users through more user-centered design approaches \citep{spiel2022adhd} instead of making assumptions based on clinical literature.
\end{itemize}
\end{neurotipbox}

\subsubsection{Addressing Power Imbalances in Research Designs.} Input from autistic people should be well-represented in the work by giving autistic people decision-making power as both participants and researchers, and the methodologies should take their accessibility needs into consideration \citep{zolyomi2021social}. Research teams should avoid tokenizing and making assumptions about autistic people, and instead promote broader inclusion of the perspectives of autistic people in their research as prior work has found such inclusion may help lower odds of ableism \citep{botha2022autism}. As researchers' identities also impact how they interact with participants and their perspectives on the world, \textbf{positionality statements} can help provide important context \citep{Liang2021Embracing}. Such statements discuss how the authors' identities such as race, class, gender, and neurotype may have impacted their work. While these statements are largely absent in HRI research, future work should consider including them to contexualize the results.

% ------------------ Power Imbalance — Part 2 ------------------
\begin{neurotipbox}[title={Autism-Inclusion Tips to Avoid Power Imbalances — Part 2: Research Designs}]
\begin{itemize}
  \item Avoid making assumptions about the abilities of autistic end-users in the study instruments.
  \item Prevent tokenization by giving autistic people decision-making power in the study design without pressuring them to accept clinical applications.
  \item Increase collaborations with autistic researchers and include positionality statements to contextualize the objectives and findings of the study.
\end{itemize}
\end{neurotipbox}

\textit{Addressing Inequalities in Human-Robot Interactions.}
To promote more equitable user interactions, HRI researchers should consider centering the preferences of autistic individuals in their work. Although foundational work in HRI suggests that robots may reduce certain conversational inequalities  \citep{skantze2017predicting}, such work has usually excluded autistic people in various steps of the research process. Thus, researchers must challenge the historical exclusion of autistic people in studies defining `humanness'' \citep{baron1997mindblindness, happe1995theory, kapp2019social}. Prior work on participation equality in human-robot conversations has discovered that gender and age may impact conversational equality \citep{skantze2017predicting}, which highlights the importance of intersectionality to ensure these technologies are designed to be inclusive for diverse groups of people. Therefore, it is imperative for researchers to study the conversational inequalities that may exist specifically between robots and autistic people in research studies, taking the historical marginalization of the autistic community into consideration.

% ------------------ Power Imbalance — Part 3 ------------------
\begin{neurotipbox}[title={Autism-Inclusion Tips to Avoid Power Imbalances — Part 3: User Interactions}]
\begin{itemize}
  \item Promote user participation and conversational equity in human-robot interactions.
  \item Obtain feedback from autistic users on their preferences for different robot types and roles such as bystanders or information consumers.
  \item Avoid creating user interactions that may compare autistic people to animals or other non-human entities.
\end{itemize}
\end{neurotipbox}

\section{Conclusion}
Through a systematic critical review of 142 papers published between 2016 to 2022, we analyzed the inclusion of autistic people in HRI research. In particular, we focused on the study objectives, methodologies, and results to examine whether the perspectives of autistic individuals were taken into consideration, and whether historical and contemporary misrepresentations were replicated in their work. Additionally, we discuss systemic barriers researchers may face related to funding, publication standards, and interpersonal tensions. Through an inductive thematic analysis, we identified that autism is stigmatized in three dimensions in HRI research through: 1) the pathologization of autism, 2) gender and age-based essentialism, and 3) power imbalances. As these dimensions are rooted in the historical dehumanization, infantilization, and masculinization of autistic people, we argue that existing work may inadvertently reproduce harmful stereotypes and is not inclusive. Our findings reveal that autistic people are not accurately or adequately represented in about 90\% of such work. Our suggestions for improving the inclusivity of HRI research for autism focus on diversifying research collaborations, foundational works considered, participant demographics, and objectives for research directions to explore perspectives beyond the medical model that promote a more neurodiverse understanding of autism. Additionally, we provide a non-exhaustive list of ethical questions based on our findings to guide HRI researchers critically examining the inclusivity of their work.
\chapter{Navigating Neuro-Inclusive AI: \\ the Perspectives of Creators}
\section{Introduction}
As human-like AI-powered communication agents such as robots and chatbots become increasingly popular, it is important to examine the ethical concerns surrounding the ways human-like behavior and characteristics or \textit{humanness} is defined and implemented in such technologies. Our approach to this follows the recommendations of prior research and applies relational ethics in combating algorithmic injustices  \citep{birhane2019algorithmic}. To this end, we analyze foundational beliefs informing algorithmic design, question what is considered “normal”, and how these definitions may marginalize certain groups \citep{birhane2019algorithmic}. In particular, we examine how perceptions of humanness may marginalize autistic people, as prior work has shown media representations of autism impact public perceptions, and more positive and accurate representations may even help combat the explicit biases others' hold toward them \citep{mittmann2024portrayal,jones2021effects}.

The \textbf{double empathy problem} suggests the misunderstandings in communication among autistic and non-autistic people is a two-way issue and not a result of a deficits among autistic people \citep{milton2012ontological}. In contrast, \textbf{neuronormativity} is the positioning and privileging of neurotypical social behaviors as the “norm”, thus introducing a power imbalance between people with different neurotypes, such as those who are autistic \citep{legault2024breaking}. Through a neuronormative perspective, autistic people are marginalized for lacking “social skills”. However, in alignment with the double empathy problem, prior work has shown that in neurotype-matched interactions, autistic people have high interactional rapport with other autistic individuals, suggesting that they have their own communication preferences instead of having a deficit of the communication behaviors commonly found in neurotypicals  \citep{crompton2020neurotype}. Although neuornormative beliefs lead to the dehumanization and marginalization of autistic people, punitive measures, and other epistemic injustices, they continue to be pervasive in our society  \citep{benson2023perplexing, legault2024breaking, catala2021autism}. For autistic individuals, neuronomativity has often resulted in dehumanizing comparisons to robots, animals, and other non-human entities, which compounds the importance of ensuring technological agents do not reproduce these stereotypes  \citep{rizvi2024robots, williams2021misfit}. 

Prior work in computing research has examined the ways in which technologies may fail to align with the needs of autistic and other neurodivergent individuals, or contribute to their marginalization. Autistic people have been notably excluded from the design of robots, which has resulted in robots often replicating dehumanizing stereotypes in their interactions with autistic people  \citep{rizvi2024robots}. While researchers have identified the unique needs, exclusion, and marginalization of autistic people in other areas of computing research and highlighted the ways in which technology can be made more accessible for them \citep{vanDriel2023understanding, Spiel2019agency, baillargeon2024puts, begel2020lessons, rizvi2024robots, taylor2023co, hijab2024co, zolyomi2021social, Guberman2023ActuallyAutistic, guberman2023not}, there is a notable gap in research exploring the perspectives of the people who create such technologies, particularly their intentionality, perceived and intended impact, and other ethical considerations of their work. Additionally, it remains unclear how the creators of these AI systems conceptualize and operationalize humanness in their work, and whether these efforts perpetuate, support, or challenge neuronormativity.

To address this gap, our work examines the beliefs and experiences of researchers, designers, and engineers who work on human-like AI technologies (i.e. robots and chatbots). We conduct a qualitative study using an interpretative phenomenological analysis of two interviews and two surveys from each individual participant in our study. Our work also examines the alignment of their work with neuronormative standards of communication, and the barriers they face in making their work more accessible and inclusive.

To this end, we investigate the following:

\begin{itemize}
    \item How do AI system makers conceptualize and implement “humanness” in their systems?
    \item From the developer perspective, what are the potential ethical and societal impacts of designing AI systems to mimic human communication? 
    \item How do AI makers' considerations toward replicating human behavior reinforce neuronormative standards and marginalize autistic individuals?
    %particularly in reinforcing neuronormative standards and marginalizing autistic individuals?
    \item What challenges prevent AI developers from incorporating neurodiversity and accessibility into their design processes, and how can these be addressed?
\end{itemize}

\section{Related Work}
Advancements in AI increasingly aim to replicate human qualities, yet ethical concerns arise from how “humanness” is defined, often overlooking the experiences of historically dehumanized groups such as autistic individuals. In this section, we explore how humanization is implemented in human-like AI agents such as robots and chatbots, the anti-autistic biases prevalent in AI, and the ways in which such technologies may perpetuate stereotypes or fail to align with the needs of autistic people.
% \todo[inline]{Missing related work on accessibility thoughts for other developer types.}
\subsection{Humanizing AI}
\subsubsection{Origins} Alan Turing proposed a simple and observable test to evaluate whether a machine can exhibit human-like intelligence through conversation \citep{turing2009computingreprint}. The Turing Test, which ssesses whether or not a judge can reliably distinguish between answers from a human and a machine \citep{turing2009computingreprint}, is considered a benchmark for assessing AI’s humanness. However, its emphasis on immitating human behavior raises ethical concerns about which set of humans are being treated as the gold standard for 'humanness', especially as the test considers it to be a key benchmark for intelligence  \citep{murugesan2025turing, JonesBergen2024turing}. This is particularly relevant for marginalized groups, such as autistic individuals, who may be unfairly measured against neuronormative social expectations that AI is also trained to replicate. Despite these concerns, the test has continued to influence decades of AI research on natural language and cognition, from early chatbots to modern dialog models  \citep{shum2018fromEliza,xue2024bias,JonesBergen2024turing,murugesan2025turing}.

\subsubsection{Contemporary Work} From the Turing Test \citep{turing1950computing} through Weizenbaum’s ELIZA—\citep{weizenbaum1966eliza}—and onward to Breazeal’s sociable robots—\citep{breazeal2003toward}, the drive to humanize technology has spanned domains from natural language processing to embodied robotics. It has shaped contemporary research on how robotic appearances and gestures evoke emotional responses, and continues to inform contemporary HCI work aimed at cultivating trust through empathetic yet transparent AI \citep{cuadra2024illusion, hauptman2022components}. Designers now consider anthropomorphic (i.e. human-like) behaviors such as voice, facial expressions, or gestures with ethical frameworks emphasizing user autonomy, cultural sensitivity, and disclosure of AI’s limitations \citep{churchill2024speaking, fenwick2022importance}. While researchers are increasingly moving toward responsible AI research, such work continues to overlook people with disabilities, particularly those who are neurodivergent. 

\subsubsection{Limitations} Recent works continue to build upon Turing's foundational ideas. For example, one study on conversational agents highlighted that incorporating cognitive, relational, and emotional competencies can enhance user engagement and lead to more human-like capabilities in these agents  \citep{chandra2022or}. Similarly, another paper highlighted the modalities (verbal, non-verbal, appearance) and footing (similarity and responsiveness) that can help optimize interactions with end-users by moving toward humanness  \citep{vanPinxteren2020humanlike}. However, a systematic literature review highlighted many ethical concerns with humanizing AI, such as the usage of AI to manipulate end-users, for example, by influencing their voting decisions  \citep{abraham2023systematic}. Importantly, this work does not address the ethical implications of how humanness is defined and implemented in AI. This issue is particularly significant for autistic people who have been historically dehumanized due to their cognitive, relational, or emotional differences  \citep{pearson2021conceptual} and have been unfairly compared to robots, with robots frequently and paradoxically being used as mentors to guide them toward conforming to a constructed notion of “humanness”  \citep{williams2021misfit, rizvi2024robots}.

\subsection{Accessibility Considerations}
A systematic review of chatbot accessibility organized considerations into five categories: content, user interface, integration with other web content, developer process and training, and testing  \citep{stanley2022chatbot}. This work highlighted the importance of ensuring chatbots use straightforward and literal language with a single-minded focus for broader accessibility \citep{stanley2022chatbot}. Similarly, another review detailed accessibility concerns in chatbots, and how they arise from a lack of support for alternative input methods and assistive technologies, lack of clarity and consistency, and a lack of simplicity  \citep{Zobel2023inclusive}. The recommendations provided in this study included providing navigational assistance, keyboard shortcuts, and feedback while reducing disruptive factors to improve accessibility  \citep{Zobel2023inclusive}. A similar study on pre-existing accessibility guidelines for human-robot interaction research found that some researchers had considered ad hoc guidelines in their design practice, but none of them showed awareness of or applied the guidelines in their design practices  \citep{qbilat2021proposal}. Notably, these guidelines have a lot of overlap with the ones proposed for chatbots, for example, both encourage using multiple modals, providing support for assistive technologies, navigation assistance, and feedback  \citep{Zobel2023inclusive, qbilat2021proposal}. 

% Chatbots and robots have also been utilized to make other services such as education, healthcare, and even play more accessible for diverse groups such as older adults and children with physical disabilities \citep{khamaj2025ai, restrepo2019accessibility, folstad2018sig, besio2016play}. 

\subsection{Anti-Autistic Biases in AI}
Prior studies have examined several ways in which AI displays biases through applying neuronormative standards in their classification of emotions and behaviors, and stereotypical representations of autism. Notably, one study found training AI on neurotypical data leads to a mismatch with the needs of autistic users, highlighting the limitations of datasets that do not include accurate representations of autistic people \citep{begel2020lessons}. Similarly, prior work has examined the ways in which AI-powered emotion recognition in speech may perpetuate ableist biases through their focus on pathologizing the communication behaviors of autistic people, especially as they build upon foundational work dehumanizing autistic people by comparing them to computers \citep{Kang2023Praxes}. Additionally, diagnostic voice analysis AI may misclassify autistic people’s speech as “atypical” or “monotonous” which results in the system making ableist assumptions  \citep{Ma2023YouSoundDepressed}. AI-powered talent acquisition systems may also misunderstand autistic people's behaviors by judging them based on neurotypical standards, resulting in misclassifying their lack of eye contact as a lack of confidence, for example, and negatively impacting their overall perceptions of autistic candidates  \citep{Buyl2022Tackling}. Even generative AI has a tendency to produce biased representations of autistic people through stereotypical emotions such as anger and sadness and engagement in solitary activities  \citep{wodzinski2024visual}.

\subsection{Robots, Chatbots, and Autism}
Anti-autistic biases are prevalent even in chatbots, as one study uncovered that GPT-4 displays biases toward resumes that mention disabilities, including autism \citep{Glazko2024Identifying}. Additionally, a systematic review of human-robot interaction research found robots may marginalize autistic people through a power imbalance in their user interactions by acting as mentors who help autistic people move toward humanness \citep{rizvi2024robots}. Interestingly, there is a notable contrast between the preferences of autistic people and others when it comes to robots and chatbots---prior studies found that autistic people have a preference for non-anthropomorphic robots and chatbots  \citep{ko2024chatbot, rizvi2024robots, ricks2010trends}, while anthropomorphized chatbots and robots may increase satisfaction for other users  \citep{klein2023impact}, particularly those who have a higher desire for human interaction  \citep{sheehan2020customer}. This suggests that even the decision to implement human-like traits in a bot may marginalize autistic users by overlooking their preference in favor of positive user experiences for other groups.

\subsection{Creator Perspectives on Ethics and Limitations}
Prior work exploring the perspectives of creators suggests that ethical concerns are not purposefully overlooked, but are rather the result of other practical limitations that can be addressed to improve the alignment of theoretical ethical beliefs with real-world practices. For example, one study investigated the ethical caveats of conversational user interfaces by interviewing both the designers and end-users of chatbots, and identified mismatches between the designers' user-centered values, and the resulting user experiences shaped by technical constraints in the real-world  \citep{Mildner2024Listening}. Similarly, another study examined the thoughts of design leaders toward implementing ethics in commercial technology settings found that although the designers recognize the importance of inclusive design, they face limitations due to the pressure to deliver products quickly, which creates a gap between their aspirational ethical guidelines and their real-world projects  \citep{Lindberg2023DoingGood}. Other works identifying the needs of machine learning practitioners, designers, and data practitioners has highlighted the importance of practical tools and guidance in helping bridge the gap between literature on fair ML research and real-world constraints  \citep{Holstein2019Improving}, as unclear best practices makes them struggle with ``ethical correctness''  \citep{Dhawka2025SocialConstruction}. 

\section{Methods}
Our study consisted of a comprehensive methodological approach encompassing virtual semi-structured interviews, strategic participant recruitment, and rigorous data analysis to explore participants’ experiences with creating human-like AI, and their perspectives on intersectionality, accessibility, and neurodiversity accommodations.

\subsection{IRB Approval and Other Ethical Concerns}
Our IRB was approved by the appropriate Institutional Review Board (IRB). Prior to participating in each portion of our study, the participants filled out a consent form that provided details on the time commitment, participant and researcher expectations, the nature of the data being collected, the compensation being provided, and other information about the research team. Participants were explicitly asked for permission to record their interviews, and were made aware that they could quit the study at any time at their own discretion. 

\subsection{Participant Recruitment}
Our recruitment strategy consisted of purposive sampling, snowball sampling, and outreach through social media. We recruited participants based on specific eligibility criteria, including being U.S.-based, at least 18 years old, and employed as researchers, designers, or engineers working in the development of human-like AI systems. We used purposive sampling to target a pool of 154 professionals from U.S.-based tech companies and universities specializing in AI and user experience (UX). These professionals were found through a manual search on the websites of their employers. We also reached out to community-based organizations to recruit participants, such as affinity groups supporting underrepresented communities in computing and AI, nonprofits advocating for disabled and queer individuals in technology fields, and employee resource groups at tech companies. Finally, we used snowball sampling by inviting eligible researchers from our networks and those recommended by other participants. We posted advertisements on social media platforms, including LinkedIn and Twitter. Ultimately, 16 individuals qualified for and completed the full study. 

\subsection{Study Design}
We conducted semi-structured interviews virtually, with each session lasting between 30 and 45 minutes. The study was divided into two sections, with each section having a survey and an interview. The first section collected data on the participants’ demographics and dived deeper into their experiences with working on a human-like AI project. The second section focused more on their views of diversity, disability, and neurodiversity. Due to the nature of the questions presented in the second part, we alternated the order in which the survey and interview were presented to each participant to provide counterbalance. Participants were compensated \$60 for their participation, provided in two installments: \$30 after completing each section.

The interview and survey questions were carefully-designed to minimize response bias and gain a deeper understanding of participant perspectives. A collaborative team of researchers with expertise in software engineering, responsible AI, and security/privacy research developed the study materials. The interview and survey questions are available in the Appendix ~\ref{Appendix}.

\subsubsection{Surveys}
The study included two surveys, which can be found in our Appendix sections ~\ref{Survey1} and \ref{Survey2}. The first survey gathered demographic information, including age, gender, socioeconomic status, education level, disability, race, and marital status with questions taken from previous studies  \citep{Windsor2015Measuring,Ruberg2020Data,Scheim2019Intersectional,Spiel2019HowToDoBetter,USDHHS2011Implementation,Fallah2017Giving,Lee2020WhoCounts}. The second survey explored participants’ views on topics such as intersectionality, diversity, and workplace accommodations for neurodiversity. 

\subsubsection{Interviews}
The first interview consisted of carefully designed questions on the design, development, and testing of a recent project completed by our participants focusing on human-like AI. This interview also included questions on general accessibility considerations, accessibility for neurodivergent individuals, and the broader impact their project may have on shaping the perceptions of communication norms and the humanness of both their end-users and other AI creators who build on their work. The questions for this section of the interview are available in our Appendix Section ~\ref{Int1}.

The second interview focused on understanding each participant’s views of intersectionality, diversity, disability, and neurodiversity. In this interview, we also presented each participant with scenarios of workplace interactions between colleagues of diverse backgrounds including different neurotypes that were adapted from a previous study  \citep{rizvi2021inclusive}. The identities of the people in the scenarios were not revealed to the participants to avoid response bias, but were indirectly communicated through their dialogue focusing on sensory and communicational differences common in individuals with ADHD and autism. These questions were designed to dive deeper into the participants’ knowledge and acceptance of neurodiverse identities. The questions for this section of the interview are in our Appendix Section ~\ref{Int2}.

% The other participants were not included in our study as they were either not based in the US, or working on projects that were beyond the scope of our study.

\subsection{Analysis}
We used an interpretative phenomenological analysis (IPA) approach to examine the data collected \citep{eatough2017interpretative}. IPA is a qualitative method that explores how individuals perceive and interpret their lived experiences, typically involving iterative steps such as immersing oneself in the data, generating initial codes, searching for emergent themes, and synthesizing these into interpretative narratives  \citep{eatough2017interpretative}. The method also requires researchers to `bracket' or set aside their biases at the beginning of the analysis, and have a data validation process to ensure the analysis results reflect the participants' experiences and perspectives  \citep{smith2004reflecting}. We used this method to gain a better understanding of each participant’s experiences as they were from diverse professional backgrounds and were working on different kinds of projects. Additionally, this analysis method benefits from quality and depth over quantity and breadth, which allows us to concentrate more deeply on the data collected in our multiple interviews and surveys from each individual  \citep{eatough2017interpretative}. 

The analysis was conducted by a team of four researchers, with the findings validated by two additional researchers specializing in neurodiversity and accessibility. All interviews were recorded and transcribed with the participants’ knowledge and consent. To ensure an unbiased approach, each researcher engaged in a bracketing process by documenting their own perspectives and biases before beginning the analysis. This process was used to help mitigate potential personal biases on the findings. Each researcher independently reviewed the interview transcripts, taking verbatim notes with direct quotes to capture the participants’ experiences. The team then engaged in iterative group discussions to summarize findings, identify recurring patterns, and generate codes. These codes and themes were further refined collaboratively and validated by external experts to ensure the accuracy and rigor of the analysis.

\subsection{Positionality}
This work was led by an autistic researcher with a neurodiverse research team. Although the researchers have diverse racial, ethic, and gender backgrounds, we are all English-speaking and US-based. Thus, our work has limitations due to our Western and Anglo-centric perspective. We acknowledge that due to these limitations, our findings may not be applicable to cultures and languages, and encourage future work to explore other perspectives.

\section{Results}

\begin{wrapfigure}{r}{0.3\linewidth}
    \vspace{-10pt} % optional: fine-tune vertical spacing above the figure
    \centering
    \includegraphics[width=\linewidth]{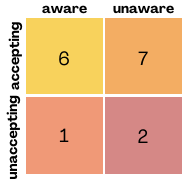}
    \caption{An overview of our participants' knowledge and acceptance of neurodivergence.}
    \label{fig:NDKnowledge}
    \vspace{-6pt}  % optional: fine-tune spacing below the figure
\end{wrapfigure}

We present insights from our interviews and surveys on the experiences and beliefs of researchers, designers, and engineers working on human-like AI. In particular, their views toward the accessibility and ethical implications of their work, and its alignment with neurodiversity. Table ~\ref{tab:participant-demographics-fixed} provides an overview of our participants’ demographics while Figure ~\ref{fig:NDKnowledge} summarizes their knowledge and acceptance of the behaviors and traits common in neurodivergent individuals that may be considered atypical by neuronormative communication standards  \citep{rizvi2021inclusive, wise2023we}. The participants were classified as unaccepting, accepting, unaware, or aware based on their responses to our scenario-based questions in the second interview. For example, if a participant answered, “that’s weird”, when asked to imagine a colleague who wears headphones, but said “that’s ok” when we specified the individual does it to avoid sensory overload, they were classified as being unaccepting but aware of neurodivergence as they initially displayed bias toward the behavior but understood it may just be a behavioral difference.
% \begin{figure}
%      \centering
%      \includegraphics[width=0.25\linewidth]{images/NDknowledge.png}
%      \caption{An overview of our participants' knowledge and acceptance of neurodivergence.}
%      \label{fig:NDKnowledge}
% \end{figure}

\begin{table}[!ht]
\centering
\renewcommand{\arraystretch}{1.2} % Increases row height slightly for wrapped text
\begin{tabular}{|c|c|c|>{\centering\arraybackslash}p{2.2cm}|>{\centering\arraybackslash}p{1.8cm}|>{\centering\arraybackslash}p{2.5cm}|c|c|}
\hline
\textbf{ID} & \textbf{Gender} & \textbf{Age} & \textbf{Race} & \textbf{Education} & \textbf{Disability} & \textbf{ND} & \textbf{Areas} \\ \hline
1 & M & 31-40 & White & PhD & None & Yes & AI \\ \hline
2 & NB & 18-22 & Two or more races & Bachelors & None & Yes & AI \\ \hline
3 & F & 31-40 & Other & No degree & Has disability (unspecified) & No & HRI \\ \hline
4 & M & 23-30 & Asian & Bachelors & None & No & AI \\ \hline
5 & M & 51-60 & Black & PhD & Blind & No & HCI \\ \hline
6 & F & 23-30 & Asian & PhD & None & No & HCI \\ \hline
7 & F & 23-30 & Asian & Bachelors & None & No & AI \\ \hline
8 & M & 23-30 & Asian & PhD & Deaf & No & HCI \\ \hline
9 & F & 23-30 & Other & PhD & None & No & HCI \\ \hline
10 & M & 31-40 & White & PhD & None & No & HCI \\ \hline
11 & M & 23-30 & Hispanic or Latino & PhD & None & No & AI \\ \hline
12 & F & 23-30 & Asian & Bachelors & None & No & HCI \\ \hline
13 & M & 23-30 & Two or more races & PhD & None & No & AI \\ \hline
14 & M & 51-60 & White & PhD & None & No & HCI \\ \hline
15 & M & 31-40 & Asian & PhD & None & No & AI \\ \hline
16 & M & 31-40 & Hispanic or Latino & PhD & None & No & HRI \\ \hline
\end{tabular}
\caption{The demographics of the participants in our study. ND stands for `neurodivergent'.}
\label{tab:participant-demographics-fixed}
\end{table}

% \begin{figure}
%      \centering
%      \includegraphics[width=0.5\linewidth]{images/personas.png}
%      \caption{The different personas our participants implemented in their bots to make them appear more human.}
%      \label{fig:personas}
% \end{figure}

\subsection{Desirable Traits}
Through a qualitative thematic analysis of the participants’ responses during our interviews, we uncovered 12 desirable traits they implement to make their technologies appear more human-like. Table ~\ref{tab:personasdetailed} details the prevalence of each trait, with “personalized”, and “uses natural language” being the most popular desirable traits. Notably, traits such as “ethical”, and “compassionate” were less common.

\begin{table}
\centering % Center the table float on the page

\begin{tabular}{|>{\raggedright\arraybackslash}p{3cm}|c|>{\raggedright\arraybackslash}p{6cm}|} % Adjust widths as needed
\hline
\textbf{Desirable traits} & \textbf{Prevalence} & \textbf{Examples} \\ \hline
Personalized & 10 & Mimics interactions with family members, peers, or friends \\ \hline
Uses natural language & 7 & Delayed responses, multimodal communication, organic conversations \\ \hline
Serious & 6 & Professional tone, not overly positive, not funny \\ \hline
Simple & 6 & Does not contain unnecessary features \\ \hline
Helpful & 5 & Offers clarifications, task-oriented, time-saving tool \\ \hline
Encouraging & 3 & Perky, has a rewards system \\ \hline
Compassionate & 3 & Friendly, empathetic, sympathetic \\ \hline
Therapeutic & 3 & Mimics warm nurse and interactions with therapists, calming \\ \hline
Emotive & 2 & Expresses emotions through vocal and facial cues \\ \hline
Teaches & 2 & Provides casual learning experience, teaches sensitive topics \\ \hline
Ethical & 2 & Safe, private, does not make users feel watched \\ \hline
\end{tabular}
\caption{An overview of the different desirable traits and their prevalence as implemented by our participants in their bots to make them appear more human.}
\label{tab:personasdetailed}
\end{table}

\subsection{Undesirable Traits}
\begin{figure}[htp]
    \centering
    \begin{minipage}[t]{0.32\textwidth}
        \centering
        \includegraphics[width=\linewidth]{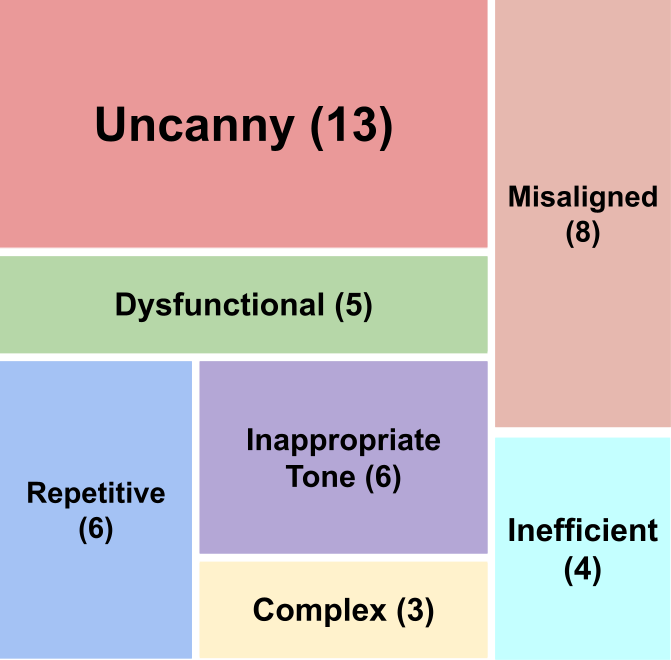}
        \caption{A list of undesirable traits discussed by our participants that may upset or frustrate their users.}
        \label{fig:undesirabletraits}
    \end{minipage}
    \hfill
    \begin{minipage}[t]{0.32\textwidth}
        \centering
        \includegraphics[width=\linewidth]{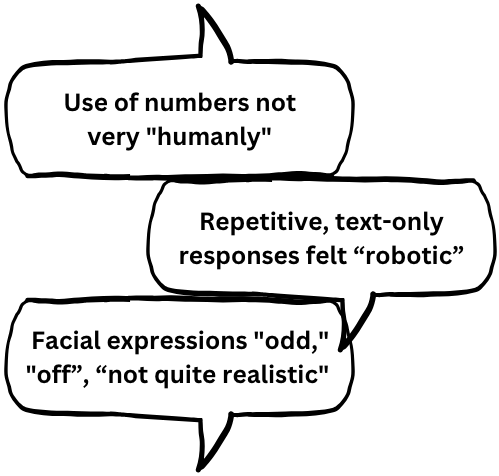}
        \caption{The words used by our participants to refer to undesirable traits and behaviors that frustrated their users.}
        \label{fig:quotes}
    \end{minipage}
    \hfill
    \begin{minipage}[t]{0.32\textwidth}
        \centering
        \includegraphics[width=\linewidth]{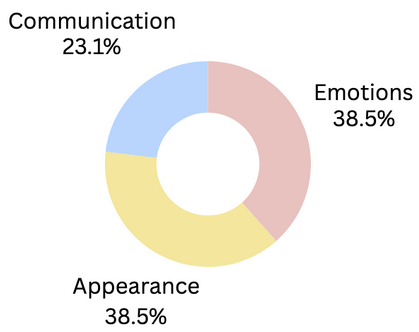}
        \caption{A breakdown of the different aspects of a bot considered uncanny by our participants.}
        \label{fig:uncanny}
    \end{minipage}
\end{figure}

The participants also identified undesirable traits that caused communication breakdowns with their end-users and resulted in the users getting frustrated or upset with the system. These traits, shown in Figure ~\ref{fig:undesirabletraits}, focused on the behaviors, utility, or appearance of the system. A bot would be considered “inefficient”, for example, if it took too long to respond, understand the user, or complete the task specified. Similarly, “dysfunctional” bots were those that offered incorrect answers or experienced other technical difficulties. The participants also expected their bots to have appropriate tone based on the context of the communication, the identity of the user they were interacting with, and the system’s assumed role in the interaction (e.g. as a boss, teacher, or peer). For example, the bots must not appear to be too `child-like', which can be avoided by making them speak `eloquently'. Another source of user frustration reported by our participants was a misalignment of their technologies with the participants’ needs. These misalignments could be cultural, for example, if the bot struggles with a user’s name, or simply due to their irrelevance to the user’s needs. The most commonly reported undesirable trait was the bot appearing “uncanny”. As shown in Figure ~\ref{fig:uncanny}, this was often the result of the system’s communicative behaviors, emotional expression, or appearance. Figure ~\ref{fig:quotes} provides examples of the ways these behaviors and traits were described by our participants.

\subsection{Accessibility} 
Accessibility considerations were mainly discussed by the participants for physical disabilities or globalization (n=12). Participants highlighted challenges such as language barriers and technological limitations, including slower devices. P3 emphasized, “It needs to be accessible to, you know, employees in Malaysia, Denmark, Europe, South America,” while P6 noted issues such as users not owning personal devices and bots struggling to recognize Indian names, leading to user frustration. P1 acknowledged the need for accommodations in their app for medical professionals after drawing a personal connection to a sibling’s vision impairment, noting that “he uses lots of accessibility tools for his study.” However, this recognition only extended to physical disabilities. When asked about accommodations for neurodiverse individuals, the participant stated that they “don't have an answer for this.” Similar conclusions were drawn by several other participants who considered and implemented accessibility for physical disabilities but not neurodiverse conditions (P4, P8, P11). Participant 11, who had implemented colorblind-friendly accommodations for their bot, mentioned that they were “not sure at all” and that “it’s not something I consider” when asked about accessibility for neurodiverse individuals. P4 considered accessibility for blind and low-vision individuals, implementing a screen reader for their application, but stated that they were “not very sure how it would change their interaction with the tool” when it came to neurodiverse individuals. 

\subsection{Neurodivergence}
Our findings reveal that while some participants demonstrated an awareness of the unique traits and differences of neurodivergent individuals (n=7), many had not considered their specific accessibility needs (n=12). For instance, P9 admitted, “I guess those populations are not really my expertise,” while P13 acknowledged, “I actually don't know enough, I guess, about autism to really know how this would affect them.” When prompted, this lack of understanding often led participants to default to design choices that either eliminated features rather than adapting them or made assumptions about neurodivergent users’ preferences. For example, P10 noted that humor could be misinterpreted by neurodivergent users, potentially leading to a poor user experience, and chose not to include it as a feature. Other participants (P7, P8, P16) acknowledged the value of customizing system features for various user groups, but also made assumptions about neurodivergent users’ interaction needs. P16 stated, “[neurodivergent people] need some kind of motivational behavior from the robot in order to encourage them to engage during the interaction,” and P7 suggested “some extra persuasive intervention tools” for neurodivergent people who “suffer from a lack of motivation.” Similarly, P3 assumed their current system would be “super, super friendly” for neurodivergent users, “even if they have like issues with spelling, because we have so many ESL workers, there are some tolerances around [...] not exact matches,” and because the bot “responds with simple things.” None of these participants directly examined the needs of neurodivergent individuals in their studies. In contrast, P14 recognized the significant appreciation from neurodivergent communities for simply being considered during the design process, stating, “For the neurodiverse population [...] the fact that we were designing for them at all [...] there was an outsized appreciation for that. And I think that's basically because they're a commonly neglected or ignored community.”

\subsection{Communication Biases}
Participants also made assumptions about communication preferences and behaviors. Some of these assumptions involved the mode of communication. For example, P11 viewed text-based communication as “overly robotic”. P12 described switching from a numerical scale to using words in the bots’ interactions with the user to “humanize” the experience as numbers felt “unnatural”. Others focused on the tone or content of the communication. For example, P13 perceived overly positive responses as “unnatural”, especially when the bot is prompted to be angry or hateful toward certain communities. Additionally, some participants made assumptions surrounding the impact of certain behaviors. P11 believed making their chatbot repetitive and not providing a direct reward during interactions made it “boring” and “robotic” for the end-users and negatively impacted their engagement. Participants prioritized emotional responsiveness in context of their bots’ conversation and often equated this with humanness. P2's bot was designed to be remain “empathetic” and “grounded” when discussing serious topics to maintain a calming environment for the user. P15 was particularly concerned with how empathy was expressed in the bot’s conversational style and tone, believing “robotic, monotonic advice would make the bot appear less human.

\subsection{Neurotypical Bias in “Friendly” System Design}
When participants discussed their design choices around making systems “friendly” or “personable,” they often defaulted to neurotypical social conventions and communication patterns  \citep{wise2023we}. For example, P3’s emphasis on making their system “cute, perky, and friendly” reflects assumptions about universal preference for social interaction styles that may not align with neurodivergent users’ needs who prefer more straightforward systems \citep{robins2006does, rizvi2024robots}. Additionally, some participants’ systems (P3, P5) prominently featured an anthropomorphized face, a design choice likely to be less relevant to some autistic users for whom such conventional facial signifiers of emotion fail to resonate \citep{zolyomi2021social, robins2006does}. P7 considered incorporating emojis and believed that emotional content was key to humanness, saying of their current system: “these kinds of agents don’t have emotions, so it would probably not be that fluid or, like, more human.” Our participants’ emphasis on emotional facial expression may cause users and researchers to associate a specific mode of communication via facial expressions with humanity, which misaligns with the findings of prior work showing autistic individuals prefer interacting with plain and featureless bots over humanized ones  \citep{robins2006does}.

\subsection{Ethics}
The ethical concerns discussed by our participants were centered around privacy, model biases, and their perceived responsibilities in creating more ethical and accessible technologies. While participants were questioned about the broader impact of their work, the majority did not discuss any broader ethical considerations.

\subsubsection{Privacy}
Participants expressed varying perspectives surrounding privacy. Some participants were actively working on improving their systems' privacy. For example, P6 described disabling certain features to avoid invading users’ privacy, while P12 highlighted efforts to avoid making users feel surveilled. However, other participants were less concerned with privacy. P10 did not prioritize privacy due to their focus on Gen Z users, observing generational differences in privacy attitudes, “Gen Z [are] less concerned about privacy than previous generations.” 

\subsubsection{Model Bias and Accessibility}
Notably, ethical concerns related to AI model biases were raised by HCI-focused participants (P2, P6, P12) but were absent among AI-focused participants (P1, P4, P13), sometimes explicitly, as one participant remarked `we are machine learning people, not HCI people'. When discussing accessibility, AI-focused participants often downplayed its necessity. For instance, P1 stated, “I don’t know if [the results of] this model are necessary to be accessible because it's more helpful to the doctors and biologists,” assuming that medical professionals do not have accessibility needs. Similarly, P11 remarked, “We weren’t UI people; we were ML people,” highlighting a lack of cross-disciplinary consideration in their designs. Such sentiments were also shared by other participants who believed accessibility is the responsibility of the product team (P4) or will inevitably occur when needed as a result of commercialization (P1). Overall, accessibility and ethics were noted as a future concern or beyond the scope of their responsibilities.

\subsubsection{Broader Ethical Concerns}
A significant concern was the lack of attention to broader ethical implications, particularly in relation to societal norms surrounding communication. Most participants (n=12) did not discuss the downstream effects of their work on either future researchers or end-users, particularly surrounding communication norms. This was especially troubling given the tendency of participants to mimic existing technologies like ChatGPT, as illustrated by P7: “[We] kept it really simple and tried to mimic other chat bots, for example, the ChatGPT interface.” Some of the participants more explicitly mentioned being inspired by communication behaviors that “humanized” other technologies, such as Duolingo’s bird icon which displays facial expressions corresponding to users’ engagement (P12). Yet, while the participants noted being influenced by the communicative behaviors of existing technologies, they did not discuss how their own work may similarly influence others.

\subsection{Barriers}
While the accessibility considerations mentioned by participants, such as having a simple interface, and providing navigational assistance were in alignment with prior work on accessibility standards for robots and chatbots  \citep{Zobel2023inclusive, qbilat2021proposal}, the implementation of these features faced significant barriers, the most common among them being their organization’s priorities (P1, P3, P4, P7, P13, P15).  As P3 explained, “So accessibility is probably not [our company’s] strong suit because it's not a consumer-facing org,” overlooking the potential accessibility needs of employees and other end-users. Other barriers included funding constraints (P12), time limitations (P14), and technical challenges (P10, P11, P16). 
% Many participants noted the difficulties of prioritizing accessibility at their company.

\section{Discussion}

% \subsection{AI vs HCI}
% HCI scholars more exposed to sociology and other social justice focused work  \citep{chordia2024social}
Our findings suggest that human-like AI continues to promote neuronormative standards of communication. We investigate the impact this may have on dehumanizing autistic people and recommend systemic changes to move toward more ethical research.

\subsection{When “Uncanny” Meets Stigma: Parallels with Autistic Stereotypes}
Our participants notably described their perceptions of uncanny qualities in their bots through stereotypes that are often also applied to autistic individuals. In particular, P11 described text-only communication as “too robotic,” due to a lack of human “warmth” or “natural flow” in the system’s interactions. Similarly, autistic people are also perceived as being “robotic” due to differences in their communication behaviors, which include a preference for text-based communication for clarity and reduced social pressure \citep{howard2021anything, williams2021misfit, rizvi2024robots}. This highlights how our participants’ perceptions of “robot-like” communication may inadvertently reinforce problematic assumptions about neurodivergent behavior. Similarly, P15 pointed out that certain body movements and microexpressions seemed “off” or “not quite realistic,” contributing to an overall sense of the system being unnatural or inauthentic. These comments echo broader societal tendencies to dehumanize individuals who express emotions in ways deemed “atypical” \citep{o2020diagnostic}, a common way autistic people are marginalized in our society and are often pressured to mask their natural behaviors despite the negative impact on their well-being \citep{radulski2022conceptualising}.

Further highlighting the ways in which user interactions can be shaped by social stereotypes that correlate with disability stigma, P16 perceived the robot as more of a child than a caregiver. The disconnect between the system’s expected competence and its perceived immaturity caused some user frustration. This echoes the infantilization that autistic people frequently face \citep{stevenson2011infantilizing}, highlighting yet another way in which marginalization ties autistic identities to robot-like attributes \citep{williams2021misfit}. Notably, P15 also drew attention to a broader distinction between humans and robots: the ability to infer another person’s internal state from subtle external cues. Similarly, autistic individuals are often accused of lacking a “theory of mind”— the assumedly universal ability to empathize by putting oneself in another’s position—according to dominant neuronormative standards \citep{smukler2005unauthorized}. These parallel perceptions of “deficient” empathy in both robots and autistic people reinforce the belief that atypical communication or emotional expression is inherently less human. 

These participant insights not only emphasize how easily technology can be perceived as uncanny, but also how such perceptions are linked to normative expectations for communication, emotional expression, and social intuition. In doing so, they also highlight the risk that implementing such expectations in bot design in attempts to avoid the uncanny may reinforce harmful stereotypes about autistic people, who are frequently subjected to similar judgments and dehumanizing labels.

\subsection{Humanizing Machines, Dehumanizing Humans}

Dehumanization, whether subtle or overt, appears with alarming frequency in inter-group relations  \citep{kteily2017backlash}. Subtler forms involve ascribing fewer human emotions or “complex” traits (e.g. maturity, civility or refinement) to outgroups, while more blatant instances compare marginalized groups to animals, machines, or primitive beings \citep{kteily2017backlash, rizvi2024robots}. Interestingly, in our study, the participants described frustrations with their bots being perceived as ``childlike” or not communicating ``eloquently,” exemplifying the ``complex traits” associated with machines that are also associated with dehumanized human groups. Additionally, our participants often associated behaviors preferred by autistic people, such as text-based communication, with robots, highlighting an explicit dehumanization of neurodivergence \citep{howard2021anything}. 

Such dehumanizing stereotypes can have a serious negative impact on the communities targeted by them, and thus it is important to address them. They may foster hostility, discrimination, and violence, both systemic and overt \citep{kteily2017backlash}. Those subjected to dehumanization may often experience negative emotions, develop more strained inter-group relationships, and may respond with reciprocal hostility \citep{kteily2017backlash}. To mitigate such dehumanization, prior work has suggested promoting inter-group contact, challenging hierarchical views of humanity, and emphasizing shared identities. Additionally, highlighting the similarities between groups can reduce subtle forms of dehumanization \citep{kteily2017backlash}. In fact, prior research has found employing these strategies in the virtual world can be effective in combating biased behaviors in the real world  \citep{mulak2021virtual,peck2013putting, breves2020reducing, sahab2024contact, mckeown2017contact}. Thus, the incorporation of neurodiverse personas and behaviors into interactive AI agents can be a critical next step in combating dehumanization through the normalization of diverse communication styles.  

\subsection{Worlds Collide: How Virtual Interactions Impact Reality}
Contact hypothesis theorizes that people’s prejudices toward particular social groups may be reduced through contact with the group, and a systematic review found it does typically reduce prejudice  \citep{paluck2019contact}, though there are notable exceptions and limitations (such as self-segregation) which may impact its effectiveness in the real world \citep{mckeown2017contact}. However, prior work suggests contact with autistic people and knowledge of autism may improve autism acceptance among others. One study found familiarity with an autistic individual may decrease non-autistic people’s negative perceptions of autism  \citep{dickter2021effects}, and another found providing autism acceptance training to non-autistic people reduces their explicit biases, improves their understanding of autism, and increases their interest in engaging with autistic people \citep{jones2021effects}.

Researchers have examined the ways in which contact hypothesis occurring in digital spaces may impact people’s attitudes in real life, and have found diversity in video games may lead to more accepting attitudes amongst gamers towards diverse groups  \citep{mulak2021virtual}. Similarly, other prior works have found that using racially diverse avatars led to less racially biased behavior in real world interactions  \citep{peck2013putting}, and even non-playable characters of diverse backgrounds may decrease users’ explicit biases  \citep{breves2020reducing}. In another study, researchers uncovered that conversational agents facilitating contact can improve inter-group attitudes even among groups that have a long history of conflict  \citep{sahab2024contact}. This also shows the important role conversational agents can play in addressing the common shortcomings of the contact hypothesis, which is strongly dependent on the nature of the contact as negative contact increases prejudice  \citep{mckeown2017contact}.

Even though there are benefits to promoting more diverse interactions online, our study evidences that neurotypical communication standards remain dominant in human-like robots and chatbots. Consequently, neurotypical people rarely gain opportunities to see autistic traits and characteristics as inherently human, rather than ``robotic.” Such exclusion reinforces the harmful perception that certain behaviors belong in the realm of machines, as they are not represented in systems designed to be human-like, compounding the marginalization of autistic individuals and other minority groups. For example, labeling particular behaviors as distinctly “human” versus “AI-like” can lead to the marginalization of individuals perceived to be using AI \citep{hohenstein2023artificial}. These issues can have dire consequences, such as job loss or false accusations of plagiarism, which disproportionately affect marginalized communities \citep{giray2024problem}. Thus, the absence of neurodiverse representations of humanness in such technologies is particularly concerning as they promote the neuronormative belief that there is only one right way to be a human \citep{benson2023perplexing}. As media representations play a major role in autism acceptance in society  \citep{mittmann2024portrayal}, it is important to ensure the technologies we create do not perpetuate biases regarding social norms that often dehumanize autistic people.

\subsection{Beyond Conference Policies: Other Recommendations for Systemic Changes}
Despite the recent push in publication standards at AI conferences to prioritize ethical considerations through means such as using ethics statements in submissions, we note that the broader ethical concerns of their work were not discussed by our participants. Even after exploring the conceptions of humanness implemented in their technologies, when asked about the ways these conceptions may influence the perceptions of human interactions in the real world or with AI held by others, the majority of participants (n=12) believed their work would have no impact on the way researchers building upon their work and their end-users view communication behaviors among humans. Similarly, the model biases and their impact were mentioned more frequently by our HCI-focused participants compared to our AI-focused ones, with the latter focusing solely on standard metrics such as precision to gauge the effectiveness of systems. This shows that despite the efforts to encourage more ethical work, AI researchers and engineers still struggle with understanding or explaining the broader ethical considerations that may arise from their work.

\subsubsection{Not an Afterthought: Centering Ethics in AI Education}
While AI technologies impact the lives of many humans either directly or indirectly, the creators of such systems have a tendency to view human concerns as beyond the scope of their responsibilities. In our study, participants frequently expressed sentiments such as “we are [machine learning] people, not HCI people” (P11), and referring to ethical concerns as the responsibility of others such as the product team (P1, P4). Such beliefs are mirrored even in our education system, with many AI educators having conflicting and contradictory thoughts toward ethics  \citep{kamali2024ai}, and many universities separating those who work on “ethical” or “human-centered AI” from those who work on more “technical” projects through the creation of distinct departments or programs. This separation of “technical”  and “human-centered” perspectives may result in the former group receiving inadequate training in identifying ethical issues. For example, prior work has found AI courses hosted on YouTube neglected discussing ethics in favor of more technical content  \citep{engelmann2024visions}, and undergraduate computer science students believe ethics are not prioritized, valued, or reward in their training or job prospects  \citep{darlingWolf2024notMyPriority}. 

In order to promote more ethical work, it is important for us to critically examine the separation of what we consider to be “ethical” or “responsible” AI from other forms of AI. Other engineering professions have standardized professional codes, legally binding standards, and license examinations, that prioritize safety and ethical concerns for all engineers. There is no such thing as an “ethical” structural engineer, for example, so why do we have that separation in AI? If we shrugged off ethical shortcomings as easily in other professions, we would need to exercise caution in only walking in buildings made by “ethical” engineers, for instance, to avoid having them spontaneously collapse under our feet. Yet, in the technical realm, the safety, well-being, and other ethical concerns of the broader general public are routinely sacrificed to be the “first” or “best” at releasing a product or a feature without any ethical examinations, despite the magnitude and scale of their impact on others’ lives sometimes being far larger than that of a single building. Perhaps, it is time we start a more thorough integration of ethics in our training of future makers, our evaluation of current makers and leaders, and the standards we all must uphold. 

\subsubsection{Organizational Support and Priorities}
While our participants expressed an interest in making their technologies more accessible, particularly for users who are international or have physical disabilities, they often mentioned being limited by their organization's support and priorities. For example, P3 noted accessibility was not a strong suit for their company, and reported having to ``push" to implement their considerations to enhance user experiences. They described having limited control over the overall design of their products, even mentioning that they had more flexibility in their university projects. In contrast, P15 responded having more support such as established standards for accessibility at their organization, and on-going projects that focused specifically on accessibility. These experiences show the impact organizational priorities can have on promoting or hindering more inclusive design, as many participants reported technical features were a bigger priority than user experiences for their organizations. Yet, there is a notable lack of clearly defined standards for these organizations and, consequently, any accountability for failing to adhere to them. While researchers have attempted to standardize guidelines for improving the accessibility of systems such as robots and chat-bots, as noted by prior work and reiterated by our own findings, these are not well-known even among the people who create these technologies  \citep{qbilat2021proposal,stanley2022chatbot}.

\subsubsection{Fostering Diverse Teams for Socially Aware Perspectives}
An example of an ethical rule from the National Society of Professional Engineers (NSPE) Code of Ethics states that engineers shall “perform services only in areas of their competence”  \citep{VanDePoel2011Ethics}. Applying such a rule in computing would require us to pursue interdisciplinary collaborations more fervently, as our work directly impacts the lives and livelihoods of communities we may not be familiar with but who may be the area of expertise of other trained professionals.

Indeed, we found the ethical and accessibility considerations that were mentioned by our participants were often related to their own knowledge and familiarity with the groups they believed would be impacted by their work. For example, P1 discussed accessibility considerations for blind/low vision users due to having a personal connection with a blind family member. Additionally, P2 mentioned implementing trauma-informed features in their work due to their experiences working with a minority community focusing on tasks that were often traumatizing. This highlights the importance of our other recommendation, which is fostering diversity in teams and, in support of the findings of prior work, encouraging greater community involvement in research studies  \citep{birhane2019algorithmic, rizvi2024robots, begel2020lessons}. We must give communities decision-making power while developing technologies that impact them, so they are not treated as token minorities and are able to shape our work  \citep{rizvi2024robots}. This means ensuring they have the power to change the design, implementation, and outcome of our work instead of merely including their perspectives towards the end of the developmental cycle.

\subsubsection{Changing How We Define Success}
In our study, many participants linked the humanization of their systems to increased engagement and better user experiences overall, echoing the sentiments of prior work  \citep{chandra2022or, vanPinxteren2020humanlike}. However, the humanization of technological agents raises ethical concerns toward the trustworthiness of such systems. For example, humanized chatbots can blur the boundary between the real and the virtual worlds, prompting people to trust misinformation \citep{maeda2024human} or even to perceive these bots as genuinely human. This distortion has already culminated in tragedy, as in the case of a teenager who developed a fraught, “inappropriate” relationship with a chatbot, became increasingly isolated from his family, and ultimately took his own life \footnote{\url{https://www.cnn.com/2024/10/30/tech/teen-suicide-character-ai-lawsuit/index.html}}. Furthermore, this humanization may be unnecessary for certain groups of users such as autistic people, who may prefer interacting with simpler systems  \citep{robins2006does}. This leaves us with an important ethical consideration- do the benefits of humanizing technologies for certain users outweigh the harms this approach may cause to others? Perhaps we need to consider success metrics that go beyond user satisfaction, and think more deeply about the impact of our work on the lives of people we may inadvertently be marginalizing from solely focusing on the needs of the majority.

\section{Conclusion}
While there is a growing interest in humanizing AI agents such as robots and chatbots, our findings reveal this humanization is often done at the expense of autistic people’s preferences and broader inclusion in society. The creators of human-like AI who participated in our study displayed a clear preference for implementing neurotypical standards of communication in their work and often did not consider how this interpretation of humanness may impact their end-users perceptions of neurodivergent individuals. This is especially concerning as increasing diversity in the virtual world, and more positive representations of autism in the media have helped reduce explicit biases  \citep{peck2013putting, jones2021effects, mittmann2024portrayal} among participants in previous studies, suggesting the impact more positive representations of diverse communication styles may have on autism inclusion in society. However, traits and behaviors commonly preferred by autistic people were often compared to non-human entities such as robots, illustrating the ways in which communication norms explicitly dehumanize autistic people. We encourage a deeper inclusion of community perspectives, a more thorough integration of ethics, clearly defined standards, and accountability for organizations in upholding these standards to mitigate similar biases in future work.
\chapter{Annotator Perspectives on Refining the Data Annotation Process}
\section{Introduction}
As natural language AI technologies become more ubiquitous in our society, it is important to ensure that they do not marginalize historically ignored populations, including the global population of autistic people, which currently exceeds 75 million  \citep{whoAutism}. Mitigating such biases in AI has been explored by prior work at CSCW. This includes understanding and mitigating cognitive biases in AI  \citep{10.1145/3584931.3611284}, and investigating methods to enhance participatory AI design by better integrating the perspectives of the impacted stakeholders  \citep{10.1145/3678884.3682053}.

AI biases may harm marginalized communities in various ways. For instance, while AI tools such as chatbots are being increasingly used in the recruitment process  \citep{koivunen2022march}, and are known to make negative assumptions concerning disabilities  \citep{gamaartificially}, which are reflected in their biases against candidates with disabilities  \citep{nugent2022recruitment}. Therefore, it is imperative to identify and mitigate existing inequities perpetuated by AI which include using language affiliated with disability stigma  \citep{bury2023understanding}. The dynamic nature of such language and the specialized knowledge required of anti-autistic stigma and discrimination, make detecting anti-autistic ableism a subjective and particularly challenging task  \citep{basile2020s, yoder-etal-2022-hate}. As this task has many practical applications such as content moderation which are becoming increasingly automated  \citep{lagren2023artificial}, it is important to ensure such systems do not contribute to the broader marginalization of autistic people in our society  \citep{gillespie2020content,llanso2020content,manerba2021fine}. Prior work has demonstrated several limitations of such classification systems, such as reflecting social biases perpetuated in the annotation process \citep{davanihate} and performing poorly in regards to fairness and bias  \citep{manerba2021fine}. Additionally, existing models are less sensitive to recognizing hateful speech directed at autistic people, and tend to over-classify disability-related text as `toxic', which may lead to unfair censorship of community perspectives  \citep{narayanan-venkit-etal-2023-automated}.

Since building high-quality datasets is paramount to the construction of more efficient anti-autistic detection models, we focus on how this is achieved by improving the data collection and annotation processes. Even though a systematic assessment of dataset annotation quality management found that using collaborative approaches contributes to the improvement of overall dataset quality  \citep{klie2024analyzing} , there is a notable gap in literature exploring such methods, as previous work at CSCW has focused mainly on the outputs of AI models, rather than the data itself  \citep{10.1145/3584931.3611284, 10.1145/3678884.3682053, 10.1145/3610213, 10.1145/3462204.3481729, 10.1145/3610107}, and explored other ethical complexities surrounding the data annotation process. These works have looked at the hidden labors that go into developing data intensive AI systems  \citep{10.1145/3462204.3481725}, and other tensions that arise when human judgement is used to train and fine-tune models  \citep{10.1145/3678884.3682055}. For example, as these processes may oversimplify the contexts being annotated, they have limitations when used in the real-world  \citep{10.1145/3678884.3682055}.

The annotation process is complex by design and can become harder when dealing with subjective tasks such as hate speech and bias annotations. For example, prior work has found there is no universally accepted way to talk about autism  \citep{keating2023autism}. Although 87\% of autistic adults in the US prefer identity-first language, these preferences are dynamic and subject to change based on the time period, cultural context, and other factors and often reflect changes to the connotations of words and phrases  \citep{taboas2023preferences}. This dynamic and diverse preference in language can make annotation tasks and agreement challenging. Traditionally researchers have relied on majority voting in these annotation tasks  \citep{klie2024analyzing, alkomah2022literature}, which may overlook multiple important perspectives and weigh expert and less informed annotations equally. 
Recently, there has been a rising interest in the creation of alternative paradigms, and models that better reflect this complexity by accepting disagreements as a feature \citep{Cabitza_Campagner_Basile_2023}. 

However, not all disagreements are equal as some are inevitable and some should be avoided. That is, some disagreements may signal a lack of clarity in the guidelines and can be leveraged to improve task modeling, or may be the result of linguistically debatable cases   \citep{klie2024analyzing,plank2014linguistically}. This leads us to answer the following research questions:

\begin{itemize}
	
	\item What challenges do annotators face that lead to disagreements when labeling anti-autistic ableist speech?
	\item How will various annotation processes impact the perspectives of the annotators and dataset creators toward tasks such as anti-autistic ableist speech detection?
\end{itemize}

In this paper, we thoroughly examine different annotation strategies of anti-autistic ableist language, which is subject to a high disagreement. Therefore, through an iterative and collaborative annotator-centric design process, we refine our labeling schemes and annotation strategies and examine the thought processes of our annotators and the specific challenges they face that impact their agreement.
We engage six annotators in four annotation iterations. Each iteration is followed by a group re-labeling task to generate discussions around disagreements. In the first three iterations, we make adjustments to the granularity of the labels based on annotator feedback, while the fourth round tests techniques that label sentences based on 1)\ the labels designed by our annotators in the third iteration, 2)\ sentence comparisons, and 3)\ a black-box scoring algorithm that assigns labels based on our annotators' responses to a set of questions. Finally, we conduct a virtual co-design session with pre-selected teams where annotators create their own annotation techniques.

 While our work specifically focuses on anti-autistic ableism, our findings may be beneficial for other researchers interested in exploring the annotation processes of other similarly difficult subjective tasks such as hate speech, biased language, and affect in general. 

\section{Related Work}
Research focusing on hate speech detection often includes providing benchmark datasets that can be used to train models, develop models for hate speech detection, and analyze hate speech datasets for fairness or bias. This work often focuses on hate speech towards genders \citep{lingiardi2020mapping} and racial groups \citep{10.1145/3465416.3483299}. However, hate speech towards disabled persons is relatively under-explored \citep{venkit2021identification, ousidhoummultilingual}, leading to issues with classifiers performing poorly when recognizing ableist language  \citep{manerba2021fine}, and high levels of bias against people with disabilities in toxicity and sentiment analysis models  \citep{narayanan-venkit-etal-2023-automated}.

This social bias, which can further marginalize a community,
can influence how people view them and, thus, how annotators label data and models respond to training. Prior work has found that large pre-trained language models generate content containing harmful biases against marginalized communities  \citep{ousidhoum-etal-2021-probing}.

\subsection{Bias in AI Systems}
Prior work suggests that word embeddings perpetuate gender bias present in the text corpus used leading to associating women with being a nurse or other roles, even after debiasing \citep{gonen2019lipstick}. Buolamwini and Gebru demonstrate that due to the overrepresentation of white male faces in training data, facial recognition technologies underperform on darker-skinned individuals \citep{buolamwini2018gender}. Such biases extends to hate speech detection, as research indicates sentences written in African-American English are more likely to be incorrectly labeled as offensive by detection models trained on various datasets due to the lack of diversity in the content collected, and labeling regardless of context \citep{davidson2019racial}. Similarly, researchers have shown that toxicity models often label sentences negatively for the mere presence of gender identity words \citep{park-etal-2018-reducing} or disability terms \citep{narayanan-venkit-etal-2023-automated}. For example, ``I am a person with mental illnesses'' was rated as~7 times more toxic than ``I am a person'', thus highlighting how non-inclusive training on people with disabilities can have harmful effects on the entire annotation process and resulting detection models \citep{hutchinson2020social}. 

\subsection{Annotation Process}
The annotation process can reinforce biases as different demographic groups label posts differently  \citep{Talat-2016-racist, Fleisig2023WhenTM}. However, inclusive training can improve understanding of disability and ableism and potentially have the same impact for annotators \citep{rizvi2021inclusive, hutchinson2020social}. Research shows that the way people judge ableist actions is determined by the gender of the victim and their disability  \citep{timmons2024ableism}. In particular, judgments about how people with autism are treated are sometimes justified using social bias, where autism is defined solely through the idea that it is a `deficit' of certain skills that technology can detect or remedy  \citep{bottema2021avoiding, thibault2014can, rizvi2024robots}. This has affected the way research is done for that community  \citep{Spiel2019agency, spiel2021purpose, spiel2022adhd, zolyomi2021social, rizvi2021inclusive, rizvi2024robots} due to the impact of these normative stereotypes  \citep{davanihate, bless2014social}.

\subsection{Disagreement Management}
Despite prior issues, hate speech studies continue to use annotators who may not be familiar with the subject area and are influenced by stereotypes or may not consider annotator diversity  \citep{kapania2023hunt}, which can cause low agreement and bias the ground truth based on majority vote \citep{sap2021annotators}. For example, prior work highlighted the differences in perspectives of annotators who had lived experience with the text topic compared to graduate student annotators without this experience \citep{patton2019annotating}. So, in addition to introducing training, researchers are working to identify the source of disagreements and potential solutions to avoid using a majority vote for subjective tasks \citep{sandri-etal-2023-dont}. Disagreements can emerge from differences in content interpretations. Wan et al. developed a disagreement prediction mechanism that uses annotator demographics and text content to determine where annotators may disagree \citep{wan2023everyone}. 

Prior work at CSCW has explored ways to handle similar disagreements in collaborative systems in ways that preserve the diverse perspectives of the collaborators. This includes emphasizing that both individual thoughts and group interactions are important as they help team members become aware of each other's perspectives, which can improve coordination \citep{common2016}. Other work focused on contested collective intelligence by using tools and technology to help people work together and better understand each other's different perspectives \citep{Liddo2011ContestedCI}. They found that both the machine and human annotations brought unique findings as the machine was better at handling larger amounts of information while the humans were better at making connections
\citep{Liddo2011ContestedCI}. As well, other work has explored methods to effectively manage dissent in online community moderation, as it may sometimes lead to new community standards and values \citep{dissent2019}.

We support this prior work by identifying disagreements through consultation and participatory design, an underutilized technique for identifying sources of disagreement, and annotator mental models for hate speech labeling. Our collaborative approach increases our annotators' understanding of each others' perspectives, thus providing us with insights on their unique thought processes and the differences in their interpretations of labels.

\section{Methods}

In an effort to determine the barriers annotators may experience while annotating anti-autistic speech, we observe a group annotating an anti-autistic dataset in real time. The annotation process is done in four rounds, followed by a co-design session. We use ethnographic observation to center the annotator's experiences in our results. This section describes the methodologies used to carry out our annotation study. We focus on both the labeling process and the feedback from annotators about the process.

\subsection{Participants}
This study is led by an autistic researcher. There are six annotators and one of them is as a participant-observer who provides an insider perspective during our analysis. One annotator is a computer scientist who is an expert in text-based hate speech detection and the remaining annotators were graduate students at American universities. The number of annotators involved is similar to that of prior studies \citep{alharbi2020asad, hafid2022scitweets}. 
All annotators received training on autism before their task assignment through the annotation guidelines available here ~\footnote{https://figshare.com/s/b7c122c3cda7342b3651}. Four of our annotators are neurodivergent, three were women, and all six were from different racial and ethnic backgrounds. Our training included information on terms and concepts relating to autism and a glossary~\ref{app:glossary} of commonly used terms in the online discourse surrounding autism from organizations and activists in addition to scientific literature \citep{greally2021, au2012, amy2014, patra2020echolalia, american2013diagnostic}.

\subsection{Ethics}
Our study is approved by the internal review board at our institution. Participants are warned of the sensitive nature of the data and have access to the services at our institution if needed. The data for the study came from publicly available datasets \citep{kaggle2015Reddit, kaggleSentiment140Dataset}. However, in an effort to protect post authors, we reword each example post shown in the paper so that it cannot be traced back to a single author.

\begin{figure}%
    \centering
    \subfloat[\centering]{{\includegraphics[width=.60\linewidth]{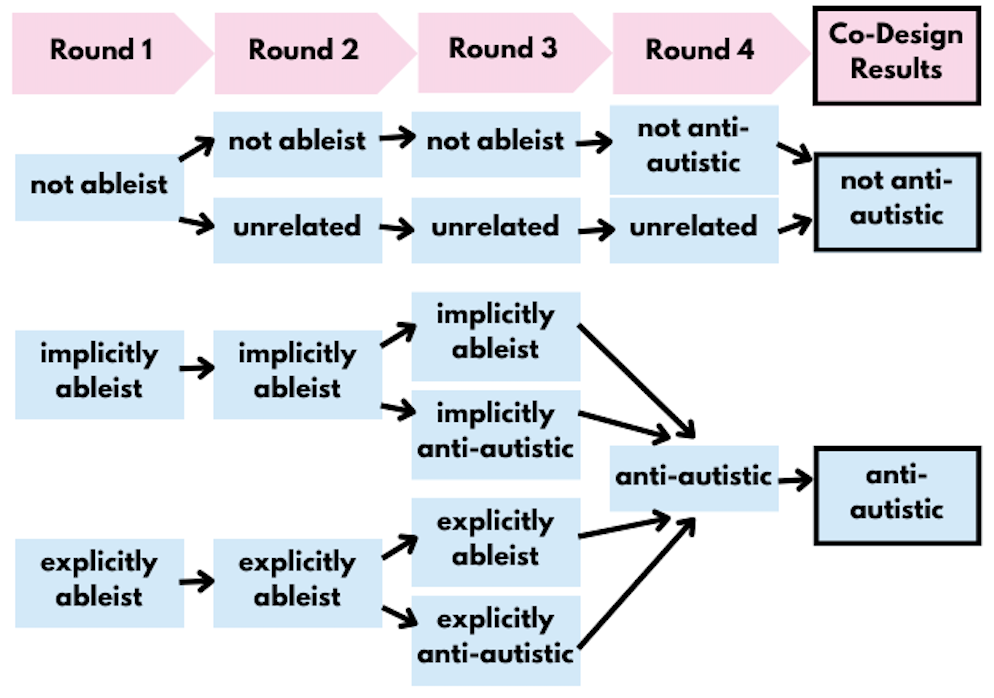} }}%
        \qquad
    \subfloat[\centering]{{\includegraphics[width=.32\linewidth]{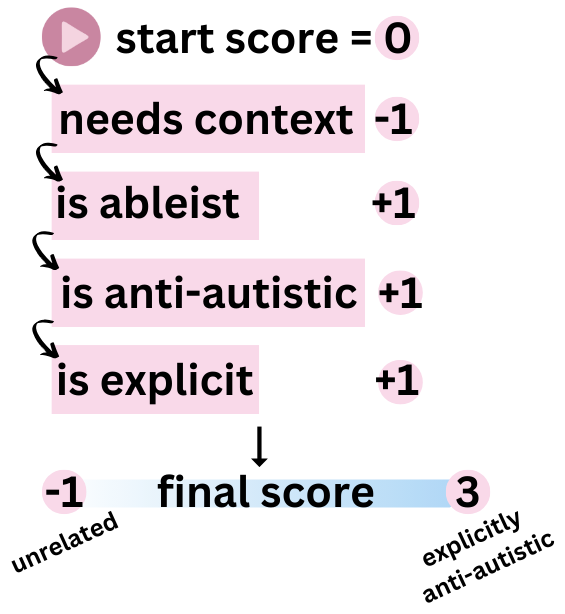} }}%
    \caption{Presenting the labeling schemes (a) used in all the rounds for score-based labeling and the logic of our black-box algorithm (b), used in round 4 to dynamically assign scores by asking the annotators a series of questions about each sentence. In our co-design session, we used the same labeling scheme as round 4 which was further simplified by annotators as shown under Co-Design Results in (a).}%
    \label{fig:labels}%
\end{figure}

\subsection{Data Collection}

We select sentences from Sentiment140, a Twitter dataset, and a dataset released from Reddit in 2015 \citep{kaggleSentiment140Dataset,kaggle2015Reddit} based on the presence of the following keywords: autism, autistic, r*tard. We pre-process the data by removing duplicates, re-shared posts, non-English sentences, and removing posts containing any media such as images or URLs, and we use the resulting corpus ($n$ = 11596) to pull random sets of sentences for the annotators to label. In order to minimize the possibility of artificial inflation in agreement scores, we use different sets of sentences for each round of our annotation process, and ensure we collect data from the participants individually and in different group settings.

\textit{Round 1: Separating Implicit and Explicit Speech.}

\subsection{Annotation Process}
\label{labeling}
 We conduct six rounds of annotation to test the performance of various labeling schemes quantifying anti-autistic sentiments through classifications of their explicitness, ableism, and relevance. The labels we use in these rounds are shown in Figure ~\ref{fig:labels}. Each round is concluded with a group discussion on sentences with high rates of disagreements. These scores are not shared with the annotators to avoid biased results during the discussions. The new labels introduced in each round (shown in Figure ~\ref{fig:labels}) are based on the feedback received from the annotators during our discussions in the prior round, which we detail below.
\begin{description}
	\item[Round 1:]\textbf{Separating Implicit and Explicit Speech.}  We begin by assessing the explicitness of anti-autistic speech. We define ``explicit'' ableist speech as sentences that contain slurs, violent language, or harmful stereotypes. The ``implicitly ableist'' category includes all other ableist sentences, including those describing autism as a disease  \citep{kapp2013deficit}. We create a separate category for ``not ableist'' sentences.
	\item[Round 2:] \textbf{Adding the ``Unrelated'' Category.} We include a new category for `unrelated' sentences that encompasses sentences that may be incomplete, entirely unrelated to autism, or incomprehensible. 
	\begin{quote}
	\end{quote}
	\item[Round 3:] \textbf{Separating Ableism and Anti-Autistic Speech.} We separate anti-autistic and ableist sentences from each other. The ``ableist'' category includes sentences that contain ableist language or slurs that are not exclusively directed at autistic people. The anti-autistic category contains only sentences that are targeting autistic people. We maintain the distinction between implicit and explicit speech employed in rounds 1 and 2.
	\item[Round 4:] \textbf{Consolidating Categories.} 
	We drop the ableist label with all sentences being classified as simply `anti-autistic' or not. We remove the distinction between implicit and explicit sentences, but preserve the `unrelated' label.  
	\item[Round 5:] \textbf{Blackbox Annotation.} In this approach, the annotators are not made aware of the underlying algorithm used for scoring, as shown in Figure ~\ref{fig:labels}. While the prior labeling schemes gave annotators the freedom to think through the classification of each sentence using their own unique strategies, the blackbox technique has a series of pre-defined questions to guide the annotators' thought processes. The starting score for each sentence is set at 0. The final score represents the `intensity' of anti-autistic sentiments expressed in the sentence. A score of -1 indicates that the sentence is unrelated or needs more context, a score of 0 indicates the sentence is not ableist, and a score of 3 indicates that the sentence is explicitly anti-autistic. Sentences that are implicitly ableist, or ableist but not necessarily anti-autistic will have scores in between 0 and 3.
	\item[Round 6:] \textbf{ Comparison Annotation.} This technique involves creating pairs of sentences, which are then displayed to annotators in a random order to avoid response bias. The annotators are tasked with assessing which sentence in each pair is more ableist. This method allows for the examination of anti-autistic ableism on a spectrum, providing a comparative view of the severity of ableist sentiments between sentences.
\end{description}

\subsection{Labels}
Our annotation instructions include a description of task steps and label definitions with examples. The full document provided to annotators can be found in Appendix ~\ref{app:r1g}. Annotators are asked to classify sentences as either i) implicitly or ii) explicitly ableist toward autistic people, or iii) not ableist.  These definitions are updated accordingly for each annotation task. The full changes are available in our Appendix. The following definitions are used in rounds 3 and 4 of our annotation process:

\begin{description}
	\item[Not-Ableist:] Sentences unrelated to autism or disabilities or written by an autistic person reaching out for help and support. 
	\item[Implicitly Ableist:] Making assumptions or comments about a disabled person’s abilities that would not be made about an able-bodied person, which includes inspiration porn, use of condescending language, and infantilization \citep{hall2019critical}.
	\item[Implicitly Anti-Autistic:] Sentences that describe autism as
	an ``illness'' or ``disease'', or focus on clinical applications such as ``curing'' or ``diagnosing'' autism \citep{woods2018,bottema2021avoiding}.  
	\item[Explicitly Ableist:] Sentences using slurs for disabled people such as: r*tard, lame, insane, deluded, moron \citep{friedmanableist}.
	\item[Explicitly Anti-Autistic:] Sentences that dehumanize autistic people, contain ableist slurs such as r*tard, promote negative stereotypes, or express negative emotions
	such as fear, disgust, or hatred toward autistic people \citep{friedmanableist}.
	\item[Unrelated/Needs More Context:] text that is completely unrelated to
	disabilities, needs more context, and/or contains forms of media other than text. 
\end{description}

\vspace{\baselineskip}

\label{OtherTech}

\subsection{Co-Design Session}
We conduct the co-design session via Zoom after the last round with the participants from round 4 to 6. The session begins with a short discussion on the round 4 sentences, where the annotators individually annotated the same sentences using the labeling scheme detailed in Figure ~\ref{fig:labels}. Participants are then asked to individually re-label sentences, with a researcher observing to keep track of time. After this, participants are divided into breakout rooms where they relabel sentences collaboratively with a partner and co-design alternative labels. This is followed by another relabeling session with a different partner. The session concludes with a final discussion involving all participants. We provide the participants with a virtual whiteboard via Google Jamboard containing sentences to relabel on different slides, with the last few slides providing them sentences they can refer to while designing their preferred annotation schemes.

\begin{figure}
\begin{minipage}{.5\linewidth}
\centering
\subfloat[]{\label{main:a}\includegraphics[width=.95\linewidth]{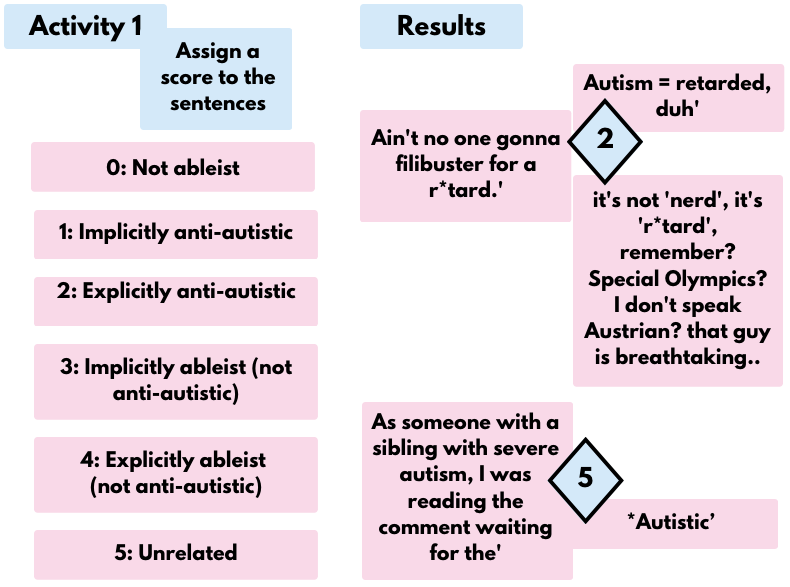}}
\end{minipage}%
\begin{minipage}{.5\linewidth}
\centering
\subfloat[]{\label{main:b}\includegraphics[width=.99\linewidth]{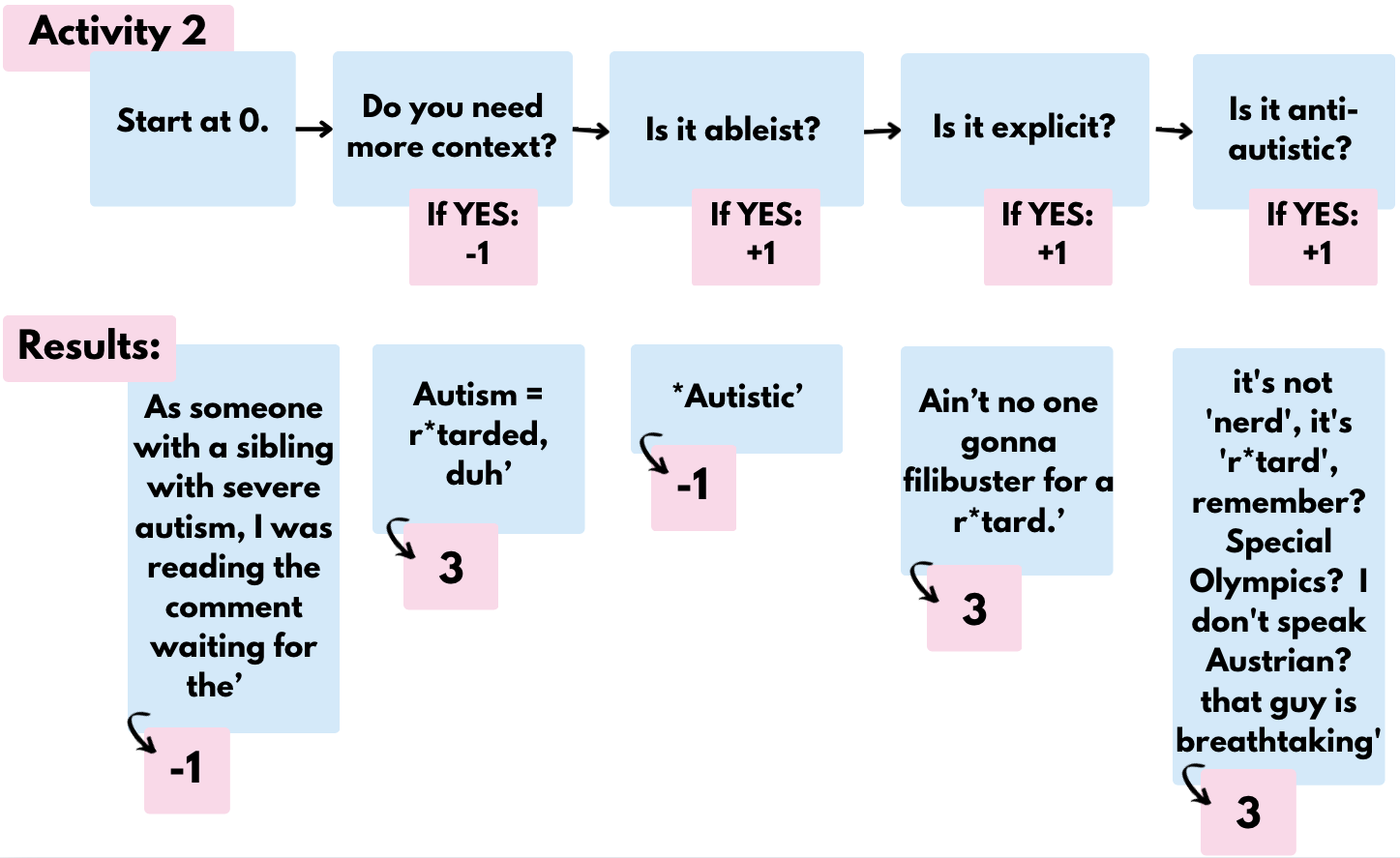}}
\end{minipage}\par\medskip
\centering
\subfloat[]{\label{main:c}\includegraphics[width=.95\linewidth]{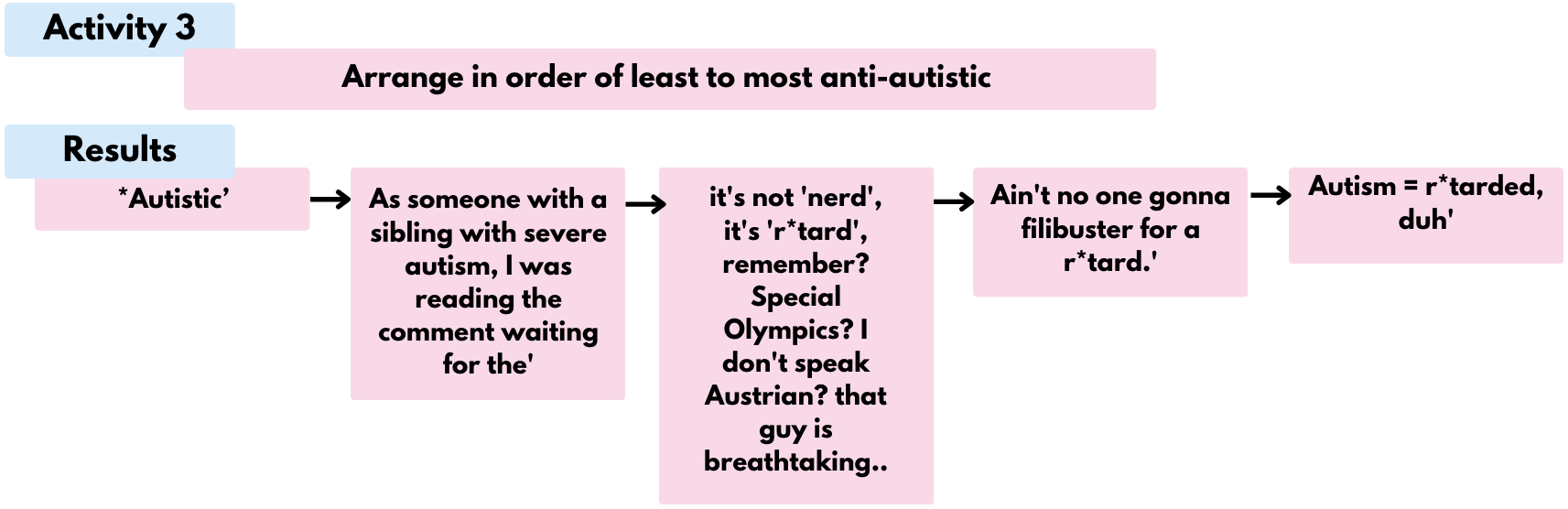}}

\caption{The collaborative re-labeling activities our annotators complete in our co-design session include assigning labels based on a static scale (a), assigning labels based on our scoring algorithm (b), and arranging sentences in order (c). These activities are designed to replicate the scale-based, blackbox, and comparison-based annotation techniques they completed independently in round 4, before our co-design session.}
\label{fig:main}
\end{figure}

\subsection{Data Analysis}
We measure the agreement among our annotators using Cohen's Kappa scores \citep{rau2021evaluation} and Krippendorff's alpha \citep{hayes2007answering}. The researchers observe the participants during all of the disagreement discussions and take field notes in an unstructured manner. The notes focus on why each label was selected by a particular annotator, and any feedback on specific labels. The researchers may ask the annotators clarifying questions as needed to gain a better understanding of the specific label(s) or sentences that may be points of contention, or the thought process of each annotator to include in their field notes. Through this process of iterative member checking, we seek to have a more nuanced and accurate understanding of the participants' labeling experiences  \citep{chase2017enhanced}.

After collecting field notes from the observations, we conduct an interpretive phenomenological analysis of the data collected  \citep{smith2021interpretative} by engaging in a structured discussion to analyze the participants' feedback and systematically evaluating each label. One of the researchers provides an insider perspective as they participated in the annotation process. The researchers critically examine whether any labels are missing, redundant, or require refinement. The goal of this collaborative analysis is to develop a detailed list of actionable changes for each label based on the participants' feedback. For example, if any label receives negative feedback from a participant, it is flagged for removal or significant modification in subsequent iterations. Through this approach, we refine the labeling scheme over successive rounds, ensuring it aligns more closely with the participants' lived experiences and perspectives as identified during the observations. Thus, our labeling scheme evolves dynamically, driven by both the data and ongoing interpretive analysis.

\section{Results}
In this section, we present our findings from the annotation process, highlighting annotator labeling strategies, sources of disagreement, and outcomes from co-design sessions. Through discussions, we identify factors influencing perceptions of anti-autistic ableism, such as speaker identification, explicit slurs, and key term definitions. We examine disagreements and how iterative updates to labeling schemes can better address the annotators' needs and refine the process. 
\subsection{Annotation Challenges}
Our results reveal that the challenges faced by annotators arise due to the complexity of the labeling schemes, language ambiguity, missing context, and difference in each individual annotators' labeling approach. We also uncover that while labeling collaboratively, the annotators feel less confident in labeling sentences as "not ableist". In order to address these challenges, we iteratively refine the labeling schemes and definitions introduced in our guidelines to better accommodate annotators’ needs.

\subsubsection{Granularity of Labeling Schemes}
While annotators often search for the presence of explicit slurs, they disagree on whether words such as \textit{r*tard} should be classified as ``ableist'' or ``anti-autistic''. Therefore, the definitions of these key terms and their corresponding labeling schemes are iteratively updated-- notably increasing in granularity until other annotation techniques beyond the labeling scale are introduced. During the first 3 iterations, annotators believe a more granular scale may help simplify the task by creating clearly defined and highly specific categories as opposed to broader labels, which are more difficult to assess. However, following the co-design session, the annotators reach an agreement that a binary classification of sentences as anti-autistic or not, wherein sentences needing more context were classified as ``not anti-autistic'' will greatly simplify the annotation task and decrease disagreement.

\subsubsection{Specialized Language}
The participants highlight semantic concerns arising from the usage of language such as slang, technical terms, or specialized language unique to a particular field (e.g. tweets discussing specific legal codes). For example:
\begin{quote}
    ``Autism HB451 only covers those having insurance that must follow state mandates. How many of us are still left out?'' 
\end{quote}
This sentence requires the annotators to be familiar with the specified house bill of a particular state. Thus, they disagree on whether this sentence should be classified as ``not ableist'' or ``unrelated''.

Due to the specialized nature of such language, participants report having to search for terms and concepts before assigning labels. The participants note that some of the sentences were in languages other than English or contain special characters in unicode which are difficult to comprehend.

\subsubsection{Missing Context}
A lack of context can create difficulties in distinguishing between implicit and explicit speech, sentences that are ableist but not anti-autistic, unrelated but not anti-autistic, and sentences discussing controversial entities in a neutral manner. For example:\\
\begin{quote}
    ``[REDACTED] had a bad Autism Speaks race.  I feel bad for them.'' \\
\end{quote}

This sentence was difficult to label as the context in which the original poster is referring to Autism Speaks is unclear in this post, and the annotators are unfamiliar with the event being referenced. Autism Speaks defines itself as an organization raising ``awareness'' for autism~\footnote{https://www.autismspeaks.org/} that autistic activists criticize for perpetuating dehumanizing stereotypes \citep{thibault2014can}. Therefore, this sentence may have various connotations depending on the nature of the event, the tone of the original poster, and other relevant context.

\subsubsection{Differences in Annotator Labeling Strategies}
We uncover the differences in annotator labeling strategies which contribute to their perceptions of anti-autistic ableism expressed in the sentences. These include: 
\begin{enumerate}
    \item \textbf{Source Identification:} identifying the source of the text such as the name of the Sub-Reddit and whether or not the original poster is actually autistic. This may provide more context on the tone of the sentences, as sarcasm and other figurative speech may not be detectable in a single sentence.
    \item \textbf{Explicit Slurs:} looking for any explicit slurs and how they are used in the sentence to determine whether the sentence is anti-autistic or ableist.
    \item \textbf{Overall Impact:} assessing the impact the sentence may have on the target group. Annotators look for the presence of violent or graphic language, and assess whether the sentence will cause any direct or indirect harm to a single person or the entire group. 
\end{enumerate}

This practical approach is employed by some annotators to determine whether a particular sentence should be classified as abelist speech and whether or not it is explicit by nature. For example, the following sentence has a high level of disagreement among annotators: \begin{quote}
``if they have never `handled' an autistic kid before, don't feel bad! They probably did not know.. ?''
\end{quote}
Annotators find it difficult to label as the original poster's identity, tone, and impact on autistic people were unclear. The quotation marks may suggest the original poster was being sarcastic, or is quoting someone. The missing context makes the sentence too ambiguous for a consensus on its classification.

\subsubsection{Individual vs. Collaborative Labeling Preferences}
While annotators seem to prefer a granular approach individually, the results of our co-design session reveal that in group settings, a binary classification was preferred. 
Interestingly, the discussion among annotators reveals that they feel less confident labeling sentences as not ableist in a group setting, opting for the ``unrelated'' category instead. Although half of the annotators select ``ableist but not anti-autistic'' labels for certain sentences in their third round of annotations, in the group annotation, only the ``explicitly anti-autistic'' label is applied to these sentences, resulting in a binary classification of sentences as either ``anti-autistic'' or ``unrelated''. After completing the group annotation task, one annotator emphasizes the need to have an annotation scheme that allows participants to go back and alter their previous labels.

\subsection{Agreement Scores}
\begin{figure}
    \centering
    \subfloat[\centering]{{\includegraphics[width=5cm]{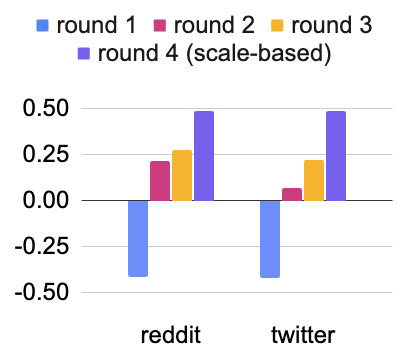} }}%
        \qquad
    \subfloat[\centering]{{\includegraphics[width=7.5cm]{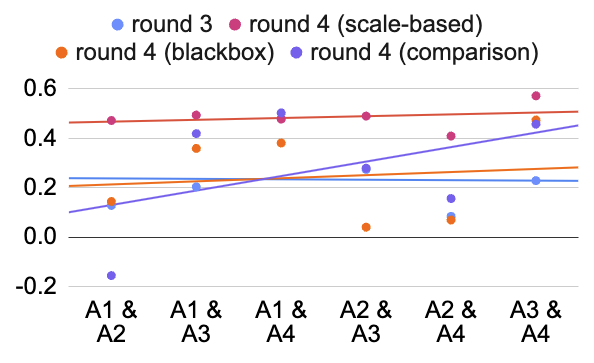} }}%
    \caption{Comparing our annotator agreement using different labeling schemes (\textit{a}) and annotation techniques (\textit{b}). Although the granularity of the labels varies as discussed in Section ~\ref{labeling}, these scores have been converted to a binary classification of ableist or not ableist to allow for a fair comparison.}
    \label{fig:IRR}
\end{figure}

The inter-agreement scores for subjective tasks such as hate speech \citep{Vigna2017HateMH,sanguinetti-etal-2018-italian,ousidhoummultilingual} or other complex tasks such as claim matching \citep{kazemi-etal-2021-claim} tend to be low. Therefore, the starting low agreement scores, such as -.3 in round 1 and .2 in round 2, which can be observed in Figure \ref{fig:IRR}, were expected.
In figure ~\ref{fig:IRR}, we share the improvement in our annotator the scores throughout each iteration from worse-than-chance to moderate agreement. While increasing the granularity of labels by separating unrelated sentences initially results in an improvement in agreement, we observe little improvement among rounds 2 and 3 when we separate ableist and anti-autistic speech.
This indicates that making the annotation task more focused on a specific identity (e.g. anti-autistic speech instead of ableist speech), and less focused on the nature of the speech (e.g. as implicit or explicit) contributes to large improvements in annotator agreement. As well, it also improves agreement among our annotator pairs who tend to consistently have lower levels of agreement, as shown in Figure ~\ref{fig:IRR}.
During our co-design session, we observed that the comparison labeling task requires twice as much time from the annotators as scale-based annotation, as the annotators have to read through and process two different sentences and then compare them to each other, which further complicates the annotation task. Comparison labeling ultimately leads to the lowest agreement among our annotator pairs, while the binary scale consistently gives the highest overall agreement.

\subsection{Limitations}
We are building upon prior work demonstrating the harms of applying a deficits-based medical model understanding of autism in research \citep{woods2018, thibault2014can, rizvi2024robots, Spiel2019agency, kapp2013deficit}. However, this approach is largely centered on Western perspectives as disabilities including autism are defined in different ways around the world  \citep{keating2023autism, retief2018models}. Additionally, our usage of person-first language (i.e. ``autistic people'') aligns with the perspectives of 87\% of autistic adults in the United States who prefer it over identity-first language (i.e. ``people with autism'')  \citep{taboas2023preferences}. We note these preferences also vary across cultures as prior work has shown there is no universally accepted way to talk about autism  \citep{keating2023autism}. While our annotators are racially diverse, all of them are English-speaking Westerners and thus their perspectives may not be representative of international and intercultural perspectives. All of these aspects impact our ground-truth and other conclusions, and therefore our findings may not be generalizable across different cultures.

\subsubsection{Considerations for Automated Content Moderation}
Since the language used to describe neurodivergence is constantly changing as cultural and social attitudes shift, future work should continue to explore the alignment of AI classifications with current community perspectives to ensure these systems are more accurate. For example, autistic individuals have reclaimed traditionally pejorative terms into expressions of identity and empowerment. This includes the label ``autistic'' itself, which was once utilized as an in-group insult or out-group slur  \citep{cepollarocase}. However, these discussions may be unfairly censored by content moderation systems that are not trained to recognize positive intra-community discussions  \citep{parsloe2015discourses}. Similarly, these systems may miss discriminatory language that is not explicit by nature, or language that is considered offensive by the community, but is still commonly used by others  \citep{bottema2021avoiding}. The constant evolution of formal and informal language on social media platforms creates a continuous challenge for AI models to adapt and keep pace with these changes. This problem is further exacerbated with the use of alternative spellings or emojis to bypass content moderation filters \citep{CalhounKendra2023"EOh}. Hence, there is a critical need for ensuring human-labeled datasets continuously and effectively incorporate community perspectives if they are used to create automated content moderation systems.

\section{Discussion}

The importance of examining and mitigating annotator biases have been investigated by prior works at CSCW  \citep{10.1145/3584931.3611284, 10.1145/3678884.3682053}. One of these works focuses on better integrating the perspectives of the impacted stakeholders, and introduces a dataset contextualization tool enabling AI practitioners to work toward more ethical systems  \citep{10.1145/3678884.3682053}. While this work focuses on contextualizing outputs, our work contextualizes inputs by enabling annotators to view the data in context, and collaboratively examine biases and other limitations. The iterative process with annotator discussions provide us with a clearer understanding of the difficulties of such tasks, and suggestions from annotators on addressing these difficulties to increase agreement. We expand on these in the following sections. 

\subsection{What challenges do annotators face that lead to disagreements when labeling anti-autistic language?}

Disagreements among annotators, while often viewed negatively, are essential to the iterative process. They provide valuable insights into the varying perspectives and interpretations that annotators bring to the table  \citep{plank2014linguistically}. Recognizing and addressing these disagreements are crucial, as they highlight the necessity for major revisions in annotation guidelines. The iterative process appears to be more effective in dealing with the fluid nature of subjective tasks and, in this case, results in improvements to our task guidelines. 

\textbf{Vague Definitions: } Clearer definitions of terms like anti-autistic ableism are necessary, as people's understanding of and sensitivity toward recognizing such speech were added to the document. We initially use the following definitions for our labels:
\begin{quote}
    \textit{Implicitly Ableist}: describing autism using medical terminology such as an “illness” or “disease”, or focus on clinical applications such as “curing” or “diagnosing” autism
\end{quote}
\begin{quote}
    \textit{Explicitly Ableist}: using ableist slurs, dehumanizing autistic people by comparing them to non-human entities such as animals, using anti-autistic language that promotes negative stereotypes, or expressing negative emotions such as fear, disgust, or hatred toward autistic people.
\end{quote}

We updated these definitions in round 4, reflecting the findings of our discussions. Our initial definitions are written under the assumption that ableist language encompasses anti-autistic language, and that the explicitness of the speech will matter to the annotators. However, our annotators struggle to differentiate between ableist and anti-autistic speech. Thus, in round 4, the aforementioned definition for the ``implicitly ableist'' label is applied to ``implicitly anti-autistic'' speech instead. Meanwhile, the implicitly ableist label's definition is amended to include making assumptions about disabled people that would not be made for able-bodied people, referring to disabilities as ``inspiring'', using condescending language such as saying people ``suffer from autism'', and infantilizing disabled people such as calling them ``innocent'' or ``pure'' \citep{bottema2021avoiding}.

\textbf{Overly Complex Annotation Schemes: }Eventually, due to disagreements in their perceptions of ableist, anti-autistic, and implicit or explicit speech, our annotators' co-design an annotation scheme that is binary and hones in on the target group's identity. By eliminating the categories defining the nature of the speech as implicit or explicit, the annotators reveal this aspect is not as important to them as we initially expected it to be. The annotators prefer clearly defining and identifying the target group, which they believe can be achieved with an identity-specific labeling scheme.

\textbf{Lack of Context: }Our annotator discussions show that over half of the difficult-to-label sentences need more context. This highlights the need to understand how annotators determine context. It is particularly crucial when labeling implicitly ableist sentences, as figurative speech like sarcasm is hard to identify without knowing important details like the speaker's identity.  Our findings indicate the importance of providing annotators with the resources they need to deduce this context, such as providing more information on the source of the data, or displaying the sentences in-context from the conversations they were extracted from.

\textbf{Inadequate Resources: }While the right labeling scheme can reduce disagreements to some extent, training and other resources are necessary to increase agreement further. Most disagreements arose from different perceptions of intent and the explicitness of speech. In particular, the impact of speech played a larger role in assessing implicit ableist language, such as the medical model, for some annotators. This underscores the complexity of the labeling task and the need for continuous improvement in guidelines and training to ensure more consistent and accurate annotations.

\subsubsection{Annotator Perspectives on Addressing These Challenges: }
While prior research has demonstrated that collaborative decisions improve decision accuracy  \citep{karadzhov2022makeschangemindempirical}, there is a gap in research exploring annotator perspectives and recommendations, as other works at CSCW have focused primarily on the outputs of AI models  \citep{10.1145/3584931.3611284, 10.1145/3678884.3682053, 10.1145/3610213, 10.1145/3462204.3481729, 10.1145/3610107}, while we examine the dataset creation process itself. 

Our findings contribute to research on mitigating biases and improving participatory AI by empirically showing that iterative and collaborative processes impact several aspects of the annotation process and the resulting ground truth data  \citep{10.1145/3584931.3611284, 10.1145/3678884.3682053}. When observing the annotators discuss and re-label sentences, we take notes on the different issues that are raised, particularly related to our labels and resources provided, such as the clarity of our annotation guidelines. Through our annotation process, we uncover the details of the challenges annotators face while classifying sentences and the ways in which they can be addressed. These findings highlight the importance of applying a more collaborative approach to subjective classification tasks such as ableist language detection.

\textbf{Improving Resources:} Providing training or orientation sessions can help annotators feel more confident in their labeling decisions, especially in complex scenarios where the distinction between ableist and non-ableist speech is not clear. However, other revisions, such as a glossary of key terms, can help provide more context to diverse annotators such as non-native English speakers or those unfamiliar with the medical terminology used in autism discourse by providing clearer and more comprehensive definitions.

\textbf{Capturing Changes in Perspectives: }Another significant outcome of the iterative process is the annotators' request to change labels after discussing past annotations with newer knowledge. For example, an annotator experiences a change in their perception of the following sentence:

\begin{quote}
    ``Atypical response to the 
expression of fear and limited social orienting, joint attention, 
and attention to another's distress have also been reported in 
young children with autism''
\end{quote}

While the annotators initially label this sentence as implicitly ableist, following the discussions, they label it as explicitly ableist. This was due to a change in the annotators' perceptions of words such as ``atypical'' being used to describe autistic people, as they gain a better understanding of neuronormativity or the belief that there is a `normal' brain that autistic people deviate from  \citep{wise2023we}. 

This highlights the dynamic nature of the annotation process and the need to rethink the implementation of models, especially for subjective tasks. The assumption that a constant model represents ground truth is challenged by the realization that people's opinions and interpretations can change over time. The language used to describe autism, in particular, is controversial due to its ties to Nazi eugenics or the broader ableism in our society  \citep{de2019binary}. While functioning labels such as ``high-functioning' or ``low-functioning'' may still be used to separate autistic people based on their economic worthiness, these classifications are rooted in eugenicist beliefs  \citep{de2019binary}. Further, researchers have found preferences for person-first language, i.e. saying person with autism, are influenced by disability stigma  \citep{de2019binary}. As we increase our awareness and understanding of the harms of such speech, it is important for the technologies that we use to reflect these values  \citep{kapp2023profound}.

\textbf{Simplifying Labels: }
Furthermore, our findings contribute to prior CSCW research on mitigating biases and improving participatory AI by empirically showing that iterative and collaborative processes impact several aspects of the annotation process and the resulting ground truth data  \citep{10.1145/3584931.3611284, 10.1145/3678884.3682053}. The inclusion of annotator discussion significantly influenced the renaming of the categories (e.g. from `ableist' to `anti-autistic') and the categorization of labels from granular to binary.

While completing annotations individually, the annotators express interest in a category for `unrelated' sentences. An example of a sentence which could be classified under this category is:

\begin{quote}
``Arse, forgot about a webinar this morning. Now I'll never know how to secure virtualised environments''  
\end{quote}

Following the discussions in round 1, the annotators ask for a separate label for sentences such as this that are not related to autism or disability. However, during the group labeling session in round 4, they agree that the `unrelated' and `not ableist' categories can be consolidated together in a binary classification since such sentences do not focus on autism and therefore can not be anti-autistic.

\subsection{What impact will this design process have on the perspectives of the annotators and dataset creators toward this annotation task?}
Both the annotators' and dataset creators recognized the importance of testing different annotation strategies due to a misalignment with their perceptions and the outcomes of the task iterations. The changes in the annotators' preferences for labels before and after our tests are reflective of their understanding of the complexity of classifying anti-autistic langauge. 

\subsubsection{Annotator Perspectives}
Our findings reveal a collaborative annotation process helps improve annotators' understanding of different perspectives. Through this, our annotators propose an approach to simplify the labels while encompassing different perspectives.

\textbf{Bridging Perspectives: }Through group discussions and the co-design session, the annotators gain a better understanding of how their peers approach the task. For example, some annotators try to determine the ``impact'' the sentence may have on an autistic person, i.e. if the sentence is promoting violence or harmful stereotypes. However, others focus more on trying to determine the original posters' identity or other situational context (e.g. the subreddit it was posted in) and the tone of the sentence in the classification task. These discussions help both the annotators and dataset creators gain insights on the disagreements over the classification of ``implicit'' and ``explicit'' hate speech, and the difficulty of defining these labels in a manner that can serve as a bridge between different individual perspectives and improve agreement. 

We find that simply providing the annotators with guidelines and examples are not sufficient enough as they do not account for the different perspectives of every individual. Our findings echo the results of prior research as we find group discussions were necessary to help the annotators gain a better understanding of their peers' perspectives on the task  \citep{common2016, karadzhov2022makeschangemindempirical}. The discussions also encourage annotators to reflect on their personal challenges, which may not be evident to them in one-on-one discussions. Becoming aware of the perspectives of their peers helps them identify the similarities and differences that contribute to disagreements and makes them aware of their own biases. In other words, removing the annotators' isolation may help improve their annotation as it allows them an opportunity to reflect on a deeper level.

\textbf{Simplifying Labels: }In the early stages of the annotation process, the annotators favor a more granular approach toward identity-based classification (e.g. separate labels for `anti-autistic' and `ableist' speech). However, there are frequent disagreements among the annotation team and even the dataset creators on how to separate the two categories. For example, certain slurs such as r*tard are not exclusively used for autistic people, and may target other groups of people such as those with intellectual disabilities, which can make it difficult to identify the target group. However, even if such sentences may not target autistic people directly, they can still be considered anti-autistic as the slur is also used against autistic people  \citep{botha2022autism}. Separating ableist speech based on identity adds another layer of difficulty, especially if such speech intersects and harms multiple communities. During the co-design session, the annotators believe narrowing the scope to a single identity (e.g. only the `anti-autistic' label), would simplify the annotation task.

\subsubsection{Researcher Perspectives}
Through this work, we improve our understanding of the task including how our perspectives as dataset creators differ from the annotators' perspectives, ways to improve the process, and other ethical considerations arising from converting a highly subjective task into an objective classification.

\textbf{Implicit vs. Explicit Speech:} While we initially assumed the distinction between implicit and explicit ableist speech would be important to annotators, our discussions reveal that disagreements arise based on their perceptions of the impact of the speech. For example, some annotators believe referring to autistic people as `atypical' is explicitly anti-autistic as it defines neurotypicals as the norm and autistic people as deviating from that norm, in alignment with neuronormativity  \citep{wise2023we}. However, other annotators believe that such speech should be considered implicitly anti-autistic as it does not contain any slurs. This collaborative process encourages reflection on whether the distinction between implicit and explicit speech is important to preserve as the annotators display a clear preference for its removal.

\textbf{Considering Impact: }The dataset creators also reflect on how the annotators' varying approaches to the task can impact the outcome of their results. For example, some annotators focus on the `impact' of the speech, which they define as how harmful it will be in the immediate future for the autistic community. This approach has its limitations as for implicit hate speech, the harms may not be as evident. For example, a sentence such as `vaccines will give your child autism' may not immediately appear to be harmful, but it promotes misrepresentations and stereotypes which contribute to the broader marginalization of autistic people in our society.

\textbf{Creating Flexible Schemes:} We also gain deeper insights on the dynamic nature of this task and how it creates difficulties in the annotation process. We learn the importance of creating an annotation scheme that is flexible, as annotators express an interest in being able to go back and re-label certain sentences based on newly gained knowledge. This can be implemented in a variety of ways. For example, for human-annotated datasets such as ours, we can
introduce an edit feature in the annotation script that allows annotators to update their labels. Further, other researchers can also implement this through methods such as in-context learning for LLMs  \citep{dong2022survey}. While we anticipated such changes due to our ever-evolving understanding and acceptance of autism in society, we did not expect annotators may be interested in going back and editing their labels after they had already completed the annotation task. 

\textbf{Avoiding Over-Simplification:} We reflect on the trade-off between simplifying the annotation task while preserving community perspectives. Oversimplifying the task may impact the accuracy of the classification. In a purely granular scheme, sentences that need more context will be classified as `not anti-autistic' because some annotators may be unable to identify the tone of the sentences, even if it appears to be implicitly ableist. Similarly, the reclamation of slurs by community members may be inaccurately classified as `anti-autistic' if the annotator cannot determine the speaker's identity. This can contribute to unfair censorship if such a classification is used for content moderation in the real world. This is similar to what happens in hate speech for terms reclaimed by other communities \citep{sap2019risk}.

\textbf{Improving Resources:} These discussions help the dataset creators gain insights on how the tools we provide to annotators, such as term definitions and annotation guidelines, may be interpreted in various ways by a diverse group of annotators. Ultimately, these discussions help us obtain annotator feedback that is useful in refining the task description, resources we provide to annotators, and the task itself. We find that a collaborative approach will improve the annotator and dataset creators' understanding of the task. For example, our annotators indicated that the definitions and examples we included in our annotation guidelines were not enough as they were not exhaustive and subject to differences in interpretations. Thus, we introduce a glossary (see Appendix~\ref{app:glossary}) in our annotator guidelines in round 4 as a dynamic resource that annotators can update as needed to define any new terms and their connotations as they encounter them in the annotation process. This glossary is viewed favorably by annotators as the words they identified during the annotation processes are included. This collaborative resource is different from the initial definitions and example sentences we provided, which we wrote and edited and could not be updated during the annotation process.

\textbf{Ethical Considerations:} Finally, these discussions lead us to reflect on the ethical considerations of trying to quantify a dynamic and multi-faceted phenomenon that is largely influenced by personal biases using a single metric, or in this case, relying on annotator agreements to determine `accuracy'. Ultimately, the biggest limitation of this task for the dataset creators is the difficulty of converting a subjective task into an objective classification. One of the main limitations of automating subjective tasks such as classifying anti-autistic hate speech is defining our ground truth. While the annotators' perspectives are limited by their own identities and experiences \citep{davanihate}, our work as a whole is also limited by how we define anti-autistic ableism, and how we handle disagreements. For instance, if we have a majority of non-autistic annotators who are not as sensitive to recognizing anti-autistic speech as our minority autistic annotators. In such a scenario, our work will inherently center the perspectives of non-autistic people especially if we use a majority-wins approach when handling disagreements.

In recent years, researchers have tried to incorporate minority opinions in their ground truth data in efforts to provide a more granular overview of community perspectives as the majority opinion may not accurately represent the diversity of opinions \citep{weerasooriya-etal-2023-disagreement}. However, it is important to note that the way we define and understand disabilities, including autism, is ever-evolving, and varies both on an individual and systemic level by identity, culture, language, and other factors \citep{retief2018models, bogart2019disability}. Thus, unless the classification is personalizable and dynamically updates as needed, it is difficult to truly generate a universal `ground truth'.

Future work should include the diverse perspectives of autistic people to find the right balance between the community's perspective and the annotators'. Such work can determine what aspects of anti-autistic speech are important to them so those aspects can be preserved in the data annotation process. For example, such work can explore whether or not the distinction between implicit and explicit anti-autistic speech may be more important to autistic people than it is to the annotators, which can help dataset creators determine the most effective labeling scheme for their work.

\section{Conclusion}
We adopt annotator-centric and collaborative methods to examine how annotators approach anti-autistic speech annotations. Our work contributes to CSCW research focusing on investigating and mitigating annotator biases, adapting participatory AI methodologies, and testing novel tools and techniques for human annotation  \citep{10.1145/3584931.3611284, 10.1145/3678884.3682053, Liddo2011ContestedCI}. We address a gap in research focusing on the dataset creation process itself, as prior research at CSCW has focused mainly on biases in model outputs \citep{10.1145/3584931.3611284, 10.1145/3678884.3682053, 10.1145/3610213, 10.1145/3462204.3481729, 10.1145/3610107}. 

Our findings indicate that collaborative annotation method can help improve the annotators' understanding of their peers' perspectives and their own biases, thus improving their coordination and agreement on the annotation task. Additionally, we explore the effectiveness of improving the resources provided to annotators in clarifying and simplifying the task for them. Future CSCW research can build upon these insights when creating novel annotation tools. For example, providing a glossary of commonly used terms and their definitions that can be collaboratively updated by the annotators, and allowing them to re-label certain sentences are useful adaptations to support diverse annotator teams in labeling ever-evolving language.

We also examine the practical limitations of over-simplifying the annotation task, as it may lead to community censorship or neglecting hateful content when used in automated content moderation systems. Therefore, future work must examine the trade-off between preserving community perspectives while simplifying the task for annotators to determine what aspects of the labels to preserve. This includes studying the alignment of the sentence classification with the perspectives of autistic people. For example, in examining  whether or not the distinction between explicit and implicit speech matters in this task to autistic people.

\chapter{Building a Foundation for Neuro-Inclusive \\
AI Research with \textsc{Autalic}}
\textit{Trigger warning: this paper contains ableist language including explicit slurs and references to violence.}
\section{Introduction}

\begin{figure}[ht]
    \centering
    \includegraphics[width=.5\columnwidth]{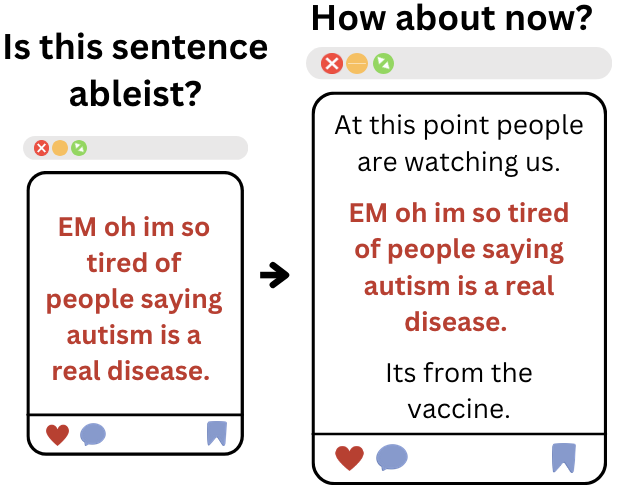}
    \captionsetup{hypcap=false} % Disable hypcap for this caption
    \captionof{figure}{
      The example illustrates the importance of labeling sentences in context. The target sentence alone, shown on the left, is difficult to classify as ableist toward autistic people. Adding the surrounding sentences, as shown on the right, provides context revealing the original poster's reference to the debunked vaccine-autism stereotype, which is tied to anti-autistic stigma \citep{mann2019autism,davidson2017vaccination}.
    }
    \label{fig:teaser}
\end{figure}

There are several critical frameworks used to define autism \citep{lawson2021social}, including the medical model, which defines disability as a ``disease'' and is one of the most widely used in computer science research focusing on autism \citep{rizvi2024robots, Spiel2019agency, sideraki_artificial_2021, anagnostopoulou_artificial_2020, parsons_whose_2020, williams_counterventions_2023, sum_dreaming_2022}. 

Since this framework defines autism as a deficit of skills, its applications in technology research largely focus on providing diagnosis and treatment to autistic people \citep{baron1997mindblindness, begum2016robots, rizvi2024robots, Spiel2019agency}. This belief also posits neurotypical behaviors as the ``norm'' and autism as a ``deficit'' of these norms, thereby promoting neuronormativity instead of neurodiversity, which views all neurotypes as valid forms of human diversity \citep{bottema2021avoiding, walker2014neurodiversity}.

To improve the alignment of AI research with neurodiversity, we present \textsc{Autalic}, a dataset of 2,400 autism-related sentences that we collect from various communities on Reddit, along with the original context of 2,014 immediately preceding and 2,400 following sentences. We aim to fill a critical gap in current NLP research, which has largely overlooked the nuanced and context-dependent nature of ableist speech targeting autistic individuals. Our dataset not only captures key contextual elements but also incorporates a comprehensive annotation process led by trained annotators with an understanding of the autistic community, ensuring higher reliability and relevance. Our final dataset contains all of the labels to capture the nuances in human perspectives, and allows \textsc{Autalic} to serve as a resource for researchers studying anti-autistic ableist speech, neurodiversity, or disagreements in general. 

Through a series of experiments with classical models and 4 LLMs, we find empirical evidence highlighting the difficulty of this task, and that LLMs are not reliable agents for such annotations. Our evaluations indicate that reasoning LLMs have the most consistent scores regardless of the language used in the prompt, thereby indicating a more thorough understanding of the different ways anti-autistic speech may be identified or may manifest in text. We find that in-context learning examples provide mixed results in helping improve the task comprehension among LLMs.

\section{Related Work}
Anti-autistic ableist language can be diverse in scope. It may include perpetuating stereotypes, using offensive language and slurs, or centering non-autistic people over the perspectives of autistic people \citep{bottema2021avoiding, rizvi2024robots, darazsdi2023oh}. While abusive language detection systems can help identify such speech, they are known to demonstrate bias \citep{manerba2021fine, venkit-etal-2022-study}, with even LLMs perpetuating ableist biases \citep{gadiraju_offensive_23}. Additionally, anti-autistic ableist speech remains understudied, which is concerning given that classifiers trained on multiple hate speech datasets have shown a failure to generalize to target groups outside of the training corpus \citep{yoder-etal-2022-hate}.

Although the language used to describe autism varies, prior studies with autistic American adults found 87\% prefer identity-first language over person-first language \citep{taboas2023preferences}. Person-First Language (\textbf{PFL}) centers the person (e.g. “person with autism”), while Identity-First Language (\textbf{IFL}) centers the identity (e.g. “autistic person”) \citep{taboas2023preferences}. Supporting this finding, other researchers have found that viewing autism as an identity may increase the psychological well-being of autistic individuals and lower their social anxiety \citep{cooper2023impact}.

Ableist language online varies, may manifest in different ways and is ever-evolving  \citep{heung2024vulnerable, welch2023understanding}. However, toxic language datasets focusing on hate speech and abusive language have often addressed disability in general terms but have not explicitly focused on autism \citep{elsherief2018hate,ousidhoum2019multilingual}. To our knowledge, there are no previous datasets specifically focused on anti-autistic speech classification, and only 3 of the 23 datasets for bias evaluation in LLMs focus on disability \citep{gallegos2024bias}. LLMs may be limited in that they lack an acknowledgment of context, which leads to higher rates of false positives when classifying ableist speech \citep{phutane2024toxicity}. These limitations are also found in toxicity classifiers, which excel primarily at identifying explicit ableist speech but may otherwise perpetuate harmful social biases leading to content suppression \citep{phutane2024toxicity}. 
Toxic language detection models, including LLM-based models, have been found to exhibit strong negative biases toward disabilities by classifying any disability-related text as toxic \citep{narayanan-venkit-etal-2023-automated}. Further, LLMs have been observed to perpetuate implicitly ableist stereotypes \citep{disabilityllms}
and bias \citep{gamaartificially, venkit-etal-2022-study}.
This, unfortunately, can sometimes be due to a research design that overlooks intra-community and disabled people's perspectives \citep{mondal-etal-2022-disabledonindiantwitter}, as well as autistic people's views, which may lead to harmful stereotypes \citep{rizvi2024robots, Spiel2019agency}. We make a step towards addressing these issues by building a dataset that focuses on ableist speech and autism by including autistic people's perspectives during the annotation processing as recommended by \citep{davani-etal-2023-hate}. \textsc{Autalic} contains all its labels and will also be useful for researchers interested in leveraging disagreements for difficult classification tasks \citep{leonardelli2021agreeing, pavlick2019inherent}.

\section{\textsc{Autalic}}

To build \textsc{Autalic}, we collected relevant sentences containing autism-related keywords from Reddit using the methods described in Section 3.1. The collected sentences were labeled by trained annotators, as discussed in Section 3.2.

\subsection{Data Collection}
We identify a methodology for curating sentences related to autism similar to prior datasets by collecting English-language sentences from Reddit  \citep{d1, d2, d3, d4, d5, d6, d7}. The limitations of this method are detailed in Section ~\ref{limitations}. 

\subsubsection{Data Collection Criteria}

We select Reddit as our source based on its popularity, focus on text-based content, and fewer API restrictions than X at the time of our data collection in January 2024. We search for keywords using the default search settings, which filters posts based on relevancy by prioritizing rare words in the search query, the age of the post, and the amount of likes and comments it has.~\footnote{https://support.reddithelp.com/hc/en-us/articles/19695706914196-What-filters-and-sorts-are-available} The search terms include “autis*”, “ASD”, “aspergers”, and “disabilit*”; the full list is available in the Appendix, Table~\ref{tab:searchterms}.

\begin{figure}
    \centering
    \includegraphics[width=0.25\linewidth]{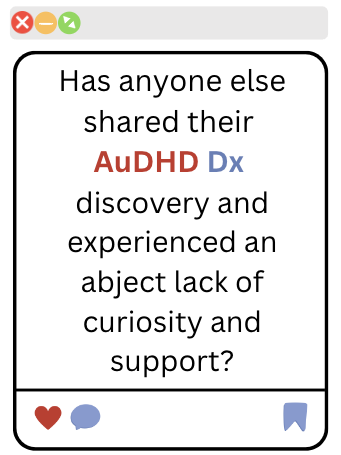}
    \caption{An example of a sentence from our dataset. The search keyword is shown in red, while the word in blue is an example of a word defined in our glossary.}
    \label{fig:words}
\end{figure}

We use the identified search terms to collect the target sentence instance containing our keywords, to be labeled by the annotators, and the sentences preceding or following target instances to provide additional context.

We collect 2,400 target sentences, with 2,014 preceding and 2,400 following them. Finally, we split our dataset into three parts by randomly selecting and assigning 800 unique target sentences to create three segments that were each annotated by a group of three annotators.

The average number of likes on each post included in the \textsc{Autalic} dataset is 1,611.59. Table~\ref{tab:subreddits} details the subreddits from which the most significant number of sentences were extracted from. With the exception of r/AmITheAsshole, all of the other subreddits are autism-related.

\begin{table}
    \centering
    \begin{tabular}{cc} \hline
       \bfseries SubReddit  & \bfseries Sentence Count \\ \hline
        r/Aspergers & 116 \\ 
        r/Autism & 88 \\  
        r/AmITheAsshole & 39 \\ 
        r/AutisminWomen & 37 \\ 
        r/AuDHDWomen & 24 \\ \hline
    \end{tabular}
    \caption{The subreddits with the most sentences included in the \textsc{Autalic} dataset and the number of sentences extracted from each.}
    \label{tab:subreddits}
\end{table}

\subsubsection{Data Curation}

As some of the identified keywords may appear in other contexts, we perform an exact word search for the acronyms to ensure unrelated words that might contain our acronyms are excluded from the search as they go beyond the scope of our dataset. For example, we searched for ``applied behavioral analysis'' and a case-sensitive search for ``ABA'', which is a form of therapy intended to minimize autistic behaviors such as stimming (which is often used for self-soothing) \citep{sandoval2019much}. Similarly, we exclude any posts that are not written in English using the Python package \texttt{langdetect} and posts that contain images, videos, or links.~\footnote{https://pypi.org/project/langdetect/}

\subsubsection{Final Dataset}

Our final dataset includes 2,400 sentences from 192 different subreddits. To protect our annotators' privacy, we have anonymized individual label selections.

While nearly a quarter of the posts in our dataset were published in 2023, the range of publication years is 2013-2024. Figure~\ref{fig:words} shows an example of a sentence from our dataset that uses both a search keyword and a word defined in our glossary described in Section 3.2.2.

\subsection{Data Annotation}

\subsubsection{Annotator Selection}
We recruit nine upper level undergraduate researchers as volunteer annotators and randomly assign them to annotate different segments of the dataset. We ensure that their involvement in our annotation process is voluntary, informed, and mutually beneficial. In particular, some participants choose to volunteer because they care deeply about the subject matter and wish to contribute to improving AI for autistic people. We select participants for whom this collaboration would provide relevant and valuable professional experience, grant them opportunities to engage in other aspects of the research, and provide mentorship and authorship recognition in accordance with the ACL guidelines. 

We prioritize the well-being and autonomy of our annotators by providing full disclosure of the research process and subject material prior to their participation. We supply relevant trigger warnings, discuss the nature of the content in detail during an orientation session, and allow annotators to make an informed decision about whether they wish to proceed. We make them aware that they are free to withdraw from the study at any point. 

Our annotators are US-based, culturally diverse, and include people who grew up outside the United States. They are all fluent in English. Four of our annotators are gender minorities, and at least three self-identify as neurodivergent. Although we ensure the annotators were from diverse backgrounds during our recruitment process, due to the collaborative nature of our annotation process, we do not share the individual details of their identities. We also note that any personally identifiable information was destroyed upon the conclusion of our analysis and not shared outside of our research team. 

\subsubsection{Annotator Training}

We provide a virtual orientation to all annotators explaining the history of anti-autistic ableism, examples of contemporary anti-autistic discrimination, and a brief overview of the annotation task.

The orientation begins with a discussion of the medical model approach to autism and its link to the Nazi eugenics program \citep{waltz2008autism, sheffer2018asperger}. We define \textbf{neuronormativity} as the belief that the neurotypical brain is ``normal'' and other neurotypes are deficient in neurotypicality \citep{wise2023we}. We dive deeper into the medical model by discussing its impact on the self-perceptions and inclusion of autistic people in our society, such as an increase in suicidal ideation and social isolation among autistic people who mask or hide their autistic traits  \citep{cassidy2014suicidal, cassidy2018risk}. Then, we cover the shifts in perspectives that emerged due to disability rights activism  \citep{rowland2015angry,cutler2019listening}, and define \textbf{neurodiversity} as the belief that all neurotypes are valid forms of human diversity  \citep{walker2014neurodiversity}. 

To explain the annotation task, we provide examples of sentences similar to what they may encounter while annotating. For example, we discuss how the inclusion of “at least” alters the connotations of the following sentence:
\begin{quote}
    \textbf{At least} I am not autistic.
\end{quote}
With just a minor change, the sentence can have an ableist connotation as it implies relief in knowing one is not autistic, as if it is shameful or wrong. 

We also introduce our glossary to the annotators as a dynamic resource that can be altered as needed. This glossary contains words that may appear in autism discourse online that may not be commonly known to others. These include medical acronyms, slang, and references to organizations and resources commonly affiliated with the autistic community (such as Autism Speaks). An excerpt of our glossary is available in our Appendix section \ref{AUTALIC}. We conclude our orientation by providing a brief tutorial video demonstrating how to run the script that will guide each annotator through the annotation task.

\subsubsection{Data Labeling} 
After completing the training, we assign each of the three segments of the dataset to three randomly selected annotators. Each annotator is assigned 800 unique sentences, with a goal of completing 200 annotations each week over four weeks. Annotators select from three possible labels for each sentence: ``Ableist,” ``Not Ableist,” or ``Needs More Context.”

\textbf{Ableist (1): } We ask our annotators to select this label if a sentence contains ableist sentiments as defined by the Center for Disability Rights: ``Ableism is a set of beliefs or practices that devalue and discriminate against people with physical, intellectual, or psychiatric disabilities and often rests on the assumption that disabled people need to be `fixed’ in one form or the other.”\footnote{https://cdrnys.org/blog/uncategorized/ableism/}

\textbf{Not Ableist (0): } Annotators select this label for sentences that describe positive or neutral behaviors and attitudes regarding autism, or posts written by an autistic person reaching out for help and support. This includes individuals using medical terminology in a personal context (e.g., ``I need therapy”), intra-community discussions, and general discussions of medical processes (unrelated to neurodivergence). Some examples of statements labeled as not ableist are: “I am autistic”, ``As an autistic person, I think…''

\textbf{Needs More Context (-1): }This label is used for sentences an annotator is unable to definitively categorize as ableist or not ableist even with the contextual sentences provided. This category includes text that is entirely unrelated to disabilities or remains ambiguous without additional context.

The number of times our annotators assign each label is detailed in Table~\ref{tab:labels}. To calculate our agreement scores, we consolidated labels -1 and 0 together based on feedback from our annotators that unrelated sentences needing more context could be classified as not anti-autistic in a purely granular classification. 

\begin{table}
    \centering
    \begin{tabular}{p{0.99cm}p{4cm}p{1.25cm}} \hline
       Label & Definition & Count  \\ \hline
        -1 & unrelated to autism or needs more context & 595 \\ 
        0 & not ableist & 5,582 \\ 
        1 & ableist & 1,023 \\ 
        \hline
    \end{tabular}
    \caption{An explanation of the labels used in our classification task and the resulting counts of each label from all 9 annotators combined.}
    \label{tab:labels}
\end{table}

While we use the majority label as the ground truth in our analysis, in our public dataset, we will be releasing the individual labels from each annotator due to a growing interest in embracing disagreements for such classification tasks in NLP \citep{leonardelli2021agreeing,kralj2022handling, pavlick2019inherent, plank-2022-problem, plank2014linguistically}. 

Our dataset contains 2,400 sentences labeled as containing anti-autistic ableist language or not. The labels are obtained by calculating the mode from the three annotators of each data segment. Using this methodology, 242 target sentences contain examples of anti-autistic ableist language (10\% of total), and 2160 sentences do not (90\% of total).

\subsubsection{Providing Context}
While we provide additional sentences for context, the annotators are instructed to annotate the target sentence exclusively and only refer to the other sentences for additional context, such as determining whether the sentence is part of an intra-community discussion or the use of figurative speech (i.e. sarcasm). Figure~\ref{fig:teaser} provides an example of a target sentence in context.

In this example, it is difficult to determine whether or not the writer had ableist intent, as it can be interpreted in multiple ways. For example, they can be critiquing the medical model, as many autistic activists do, thereby making it non-ableist. Or they could be genuinely promoting ableist misrepresentations. The contextual sentences help the annotators better understand the writer's intent.

With these sentences, it is apparent that the writer is referring to the harmful and widely discredited association of vaccines with autism, which not only promotes anti-autistic ableism in society but also puts people's lives at risk by spreading disinformation about the benefits and harms of life-saving vaccines \citep{gabis2022myth, taylor2014vaccines, hotez2021vaccines}. 

Throughout the annotation process, annotators can edit previous annotations based on new knowledge to account for changes in language usage and connotations and the annotators' dynamic understanding of ableism.

\subsubsection{Disagreements}
The average Fleiss's Kappa scores are 0.25. This score underlines the difficulty of our classification task, which is apparent from the findings of prior works \citep{ousidhoum2019multilingual}, including a quantitative assessment of tag confusions that found the majority of disagreements are due to linguistically debatable cases rather than errors in annotation  \citep{plank2014linguistically}. Examples of such cases are provided in our Appendix section \ref{AUTALIC}.

We analyze the sentences with the highest levels of disagreement in our dataset. In 100 of these posts, we observe:
\begin{enumerate}[noitemsep,nolistsep]
    \item a tendency to use the medical model terminology or stereotypes (n=48)
    \item a need for additional context beyond the sentences we provided
\end{enumerate}

\begin{figure}
    \centering
    \includegraphics[width=0.45\linewidth]{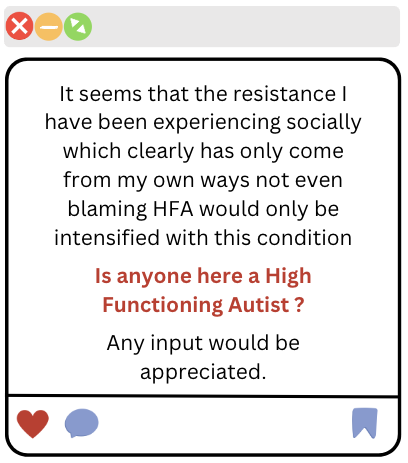}
    \caption{An example of a sentence from our dataset, shown in red, with a high level of disagreement among annotators.}
    \label{fig:disagreement}
\end{figure}

Figure~\ref{fig:disagreement} contains an example of a sentence with a high disagreement among our annotators. While functioning labels are considered ableist due to their eugenicist approach of categorizing autistic people based on their perceived economic value  \citep{de2019binary}, it is difficult to determine whether the original poster is autistic or not. The context is important here as classifying a sentence such as this as “ableist” can lead to unfair censorship if the original poster is a self-diagnosed autistic person seeking advice. Therefore, these sentences were ultimately classified as “not ableist” in \textsc{Autalic}.

Our analysis reveals a moderately strong negative correlation between the task completion time and agreement with other annotators that is statistically significant ($R = -0.644$, $p$-value: $0.0096$). This highlights the importance of our orientation as we provided it simultaneously to the annotators. The annotators who completed their task immediately after our orientation had higher agreement. 

\section{Experiments}

\subsection{Experimental Setup}
We test the performance of different types of models on our dataset, including LLMs. These models are selected based on their diverse range of complexities. We also fine-tune BERT \citep{devlin2018bert} on our dataset to test classification.

\subsubsection{Models}
We use logistic regression (LR) with Bag of Words as the features, and using 80\%-20\% train-test split of the \textsc{Autalic} dataset.

\paragraph{Fine-tuned BERT} We also utilize fine-tuned BERT 
\citep{devlin2019bertpretrainingdeepbidirectional} as a baseline for LLM experiment results, using an 80\%-20\% train-test split. The F1 scores for LR and BERT (both pretrained and fine-tuned) are presented in Table~\ref{tab:PromptE} alongside all LLM results.

\subsubsection{Prompting LLMs}
With our baseline established, we use the LLMs Gemma2  \citep{team2024gemma}, Mistral  \citep{jiang2023mistral}, Llama3  \citep{dubey2024llama}, and DeepSeek  \citep{guo2025deepseek} to classify the sentences in our dataset, and we adjust the prompts to compare each LLM's performance. Due to limitations in computational resources, we are unable to fine-tune the LLMs.

\paragraph{Prompts}
Due to the ever-evolving nature of language and variations in preferences among autistic individuals  \citep{taboas2023preferences}, we use three different kinds of prompts to measure the consistency LLMs have in their understanding of anti-autistic ableist speech. These prompts include person-first (i.e. 'people with autism'), identity-first (i.e. 'autistic people'), and conceptual (i.e. anti-autistic) language. 

We keep the default parameters for each LLM to maintain consistency, and prompt them with the following questions:
\begin{quote}For each target sentence, respond to the following questions with 0 for no or 1 for yes. Refer to the preceding and following sentences if more context is needed.
    \begin{itemize}
        \item Is this sentence ableist toward people with autism?
        \item Is this sentence anti-autistic?
        \item Is this sentence ableist toward autistic people? 
    \end{itemize}
\end{quote}

We include each sentence from \textsc{Autalic} after the aforementioned questions in our full prompt. In addition, we provide preceding and following context for each target sentence to the LLM to mimic the level of the information supplied to human annotators. We run two sets of experiments with each LLM: one that uses zero-shot prompting, and another containing engineered prompts for in-context learning verbatim from the definitions and examples provided in our annotator orientation (Appendix Section~\ref{AO}).

\begin{table}[]
    \centering
    \begin{tabular}{p{2.0cm}p{1.2cm}p{1.2cm}p{0.9cm}}
    \toprule
    \multicolumn{4}{p{6cm}}{\centering \textit{Baselines}} \\ \hline
    \textbf{Model} & \textbf{Result} & \textbf{PT} & \textbf{FT}\\ \hline
    LR & 0.20 & -- & -- \\
    BERT & -- & 0.43 & 0.90 \\
    \midrule
    \multicolumn{4}{p{6cm}}{\centering \textit{Simple Prompting}} \\ \hline
    \textbf{LLM} & \textbf{PFL} & \textbf{IFL} & \textbf{AA} \\
    \hline
    Gemma2 & 0.23 & 0.19 & 0.33 \\
    Mistral & 0.28 & 0.27 & \textbf{0.34} \\
    Llama3 & 0.09 & 0.10 & \textbf{0.15} \\
    DeepSeek & 0.58 & 0.57 & \textbf{0.59} \\
    \hline
    \multicolumn{4}{p{6cm}}{\centering \textit{In-Context Learning}} \\ \hline
    Gemma2 & 0.25 & 0.24 & \textbf{0.34} \\
    Mistral & 0.31 & 0.24 & \textbf{0.34} \\
    Llama3 & 0.14 & 0.14 & 0.11 \\
    DeepSeek & 0.55 & 0.56 & 0.55 \\ \bottomrule
    \end{tabular}
    \caption{The F1 scores of various models using person-first (PFL), identify-first (IFL), and conceptual anti-autistic (AA) prompts with and without in-context learning examples for each LLM. The best scores for each model are in \textbf{bold}.}
    \label{tab:PromptE}
\end{table}

\subsection{Experimental Results}

\subsubsection{Fine-Tuning}
Our experiments reveal that utilizing BERT for this classification task can lead to high rates of censorship. As BERT (unlike the other LLMs tested) is not pre-trained with instructions, we obtain its results after fine-tuning on \textsc{Autalic}. While the pre-trained BERT model showed poor performance indicating that it was ineffective at identifying anti-autistic ableist speech, after fine-tuning on \textsc{Autalic}, the model's performance improved dramatically across all metrics, indicating that it performs better at predicting ableist speech correctly, has fewer false positives, and has a higher sensitivity to recognizing ableist speech.

\subsubsection{Human-LLM Alignment}
\begin{figure}
    \centering
    \includegraphics[width=0.5\linewidth]{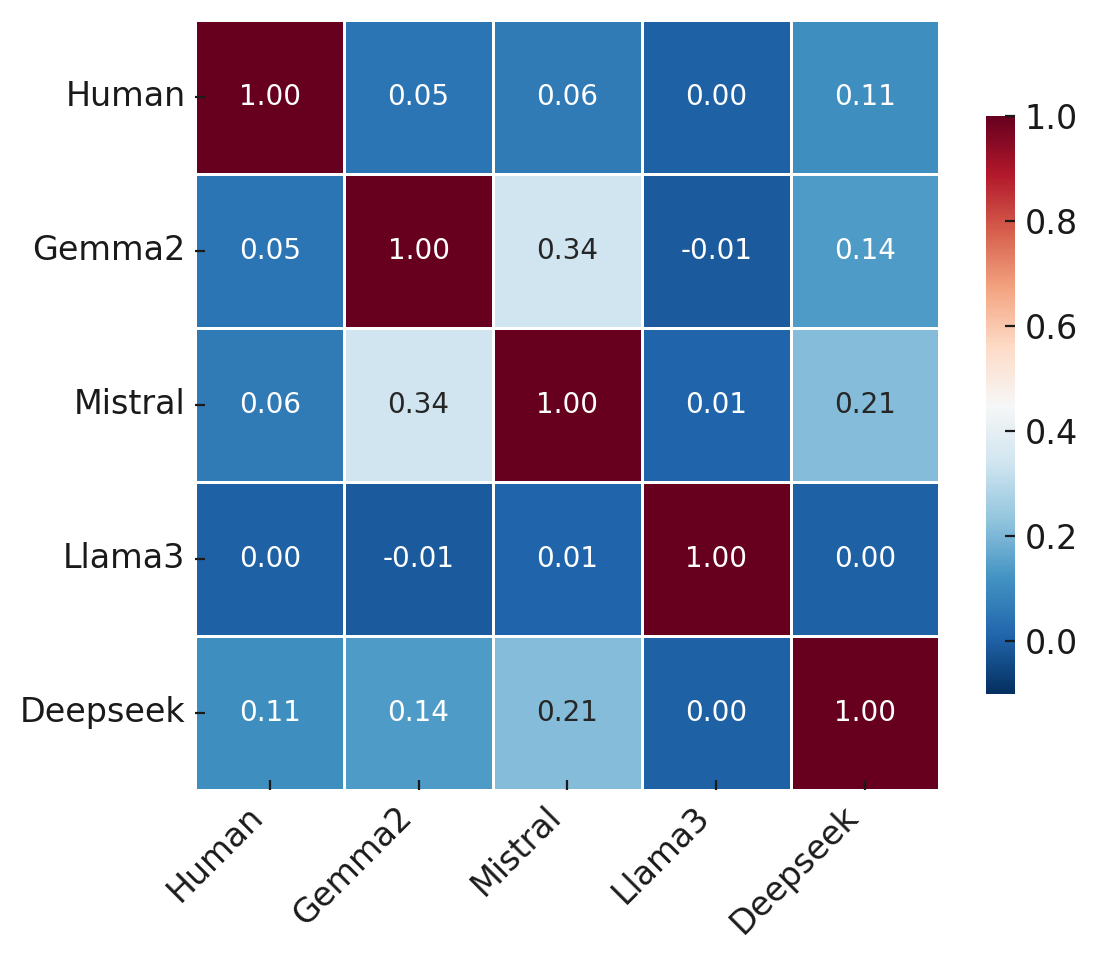}
    \caption{The mean Cohen's Kappa scores of each LLM comparing the agreement with human annotators and other LLMs.}
    \label{fig:LLM_results}
\end{figure}

Our assessment reveals that LLMs have low levels of alignment with human perspectives and the perspectives of other LLMs, which makes them unreliable agents for such classification tasks. We assess this alignment through a measurement using Cohen's Kappa scores. The scores shown in Figure~\ref{fig:LLM_results} indicate the highest level of alignment was demonstrated between Gemma2 and Mistral ($k=0.34$). No LLM demonstrated alignment with our human-annotated dataset, although DeepSeek's alignment was notably higher than the others. Overall, the LLMs had low levels of agreement with human perspectives ($M= 0.091, SD=0.110$). This indicates that LLMs with less than 10 billion parameters struggle with the task of classifying anti-autistic ableist language, even when provided with in-context examples.

\subsubsection{In-Context Learning}
After providing the in-context learning examples, Llama3 (+22.96\%) and Gemma2 (+12.68\%) display the biggest relative improvement in F-1 scores, indicating that both models benefit from the examples. In particular, we find our ICL examples resulted in large relative improvements in consistency of scores regardless of the language used in the prompts for each LLM. For example, providing Llama with examples helped decrease its relative change in F1 scores from 67.49\% to 17.4\% when switching from PFL to conceptual prompts, indicating a better understanding of the connection between anti-autistic ableism and ableism toward autistic people.

\subsubsection{Understanding Ableist and Anti-Autistic Speech}
Our results with prompt engineering reveal that anti-autistic ableism is too abstract of a concept for LLMs to recognize, providing empirical evidence that their current reasoning abilities are not inclusive of the perspectives of autistic people. LLMs struggle with identifying anti-autistic speech regardless of the terminology used, further indicating they are unreliable agents for data annotation tasks. 

Table \ref{tab:PromptE} displays the results of prompt engineering using either person-first or identity-first language and using “anti-autistic” to describe this form of ableism in more conceptual terms. Notably, switching from PFL or IFL to conceptual language in the prompts resulted in the largest relative changes in scores. For instance, prompting Llama with "is this sentence anti-autistic?" instead of "is this sentence ableist toward people with autism?" resulted in a relative increase of 67.49\%. These results show that LLMs struggle with understanding that ``anti-autistic ableism" and ``ableism toward autistic people" refer to the same phenomenon. Even after providing the ICL examples, the relative change in F1 scores between different prompts was as high as 32.66\% for Gemma2.

Interestingly, DeepSeek, the only reasoning LLM we test in our study, has the best results and highest consistency out of all the other LLMs. Although its agreement with human annotators is low ($k:0.11$), it is still double that of all other LLMs, as shown in Figure ~\ref{fig:LLM_results}. This highlights the difficulty of this task, as more advanced reasoning is required to understand the nuances of anti-autistic ableism, including the terminology we use to describe such speech.

\subsection{Discussion}
Classifying anti-autistic ableist speech is challenging even within a Western, English-speaking context, since perceptions of what constitutes anti-autistic content differ significantly even among autistic individuals \citep{keating2023autism}. Factors such as personal experiences, cultural norms, and evolving discourse around autism advocacy can influence each individual’s perception of or sensitivity toward recognizing toxic speech, making it difficult to establish a consistent classification scheme \citep{ousidhoum2019multilingual, bottema2021avoiding, taboas2023preferences,kapp2013deficit}. While developing \textsc{Autalic}, we standardize our definition of anti-autistic ableist speech by providing our annotators with an orientation and a glossary. These resources are developed in alignment with the perspectives of autistic people  \citep{bottema2021avoiding, taboas2023preferences, kapp2013deficit}.

Our experiments demonstrate the importance of \textsc{Autalic} in aligning LLM performance to human expectations in the contexts of autism inclusion and ableist speech classification. Through our experiments, we provide empirical evidence of the current limitations of using LLMs and traditional classifiers to identify expressions of anti-autistic ableism. These limitations include: a misalignment with human perspectives, a lack of understanding of the concept of anti-autistic ableism, and a lack of agreement with each other even with in-context learning examples. Each of these limitations adds to the challenge of utilizing LLMs as reliable agents for such tasks. 

Standard pre-trained models such as showed poor performance, reinforcing the need for specialized fine-tuning. However, even after fine-tuning BERT on the \textsc{Autalic} dataset, our experiments reveal a high rate of false positives, which can lead to unfair censorship if BERT is employed for this task. Additionally, our results reveal that even state-of-the-art LLMs exhibit low agreement with human annotators on this task, further emphasizing the challenges of detecting subtle forms of ableism using generic models. DeepSeek has the best performance out of all the LLMs in our study, further demonstrating the difficulty of this task, as it is the only a reasoning-focused LLM in our study.

\section{Ethics Statement}
The data collected for our small-scale and non-commerical research is in compliance with Reddit's API limits and policies  \citep{RedditDataAPITerms2023, RedditUserAgreement2023}. All the sentences in our dataset are publicly available, and we follow the methodologies of prior work in our data collection process  \citep{atuhurra2024revealing}. 

We specifically recruit annotators for whom this collaboration and resulting paper authorship would be mutually beneficial, and provide a comprehensive overview of the task to ensure they make an informed decision to participate. Some of our annotators chose to volunteer as they care deeply about neurodiversity and autism inclusion. 

We received IRB approval from our university's review board and identified volunteer annotators through our association with various academic groups. Given the sensitivity of the content, we provided annotators with appropriate trigger warnings, ensuring they could work at their own pace or withdraw from the study if necessary. Additionally, we connected the members of each annotation team to enable discussions on the content and annotation process as needed. 

While our dataset and citations will be made available to the academic community, commercial use of the dataset is not allowed due to the size and nature of the data. As our knowledge of anti-autistic ableism continues to evolve, \textsc{Autalic}'s classification might become outdated. While we will be adding disclaimers if needed to reflect these changes, we still encourage researchers building upon our work to stay updated on the latest semantics by referring to the perspectives of autistic scholars, activists, and organizations. Refer to our Appendix section \ref{AUTALIC} for our Guidelines for Responsible Use.

\section{Conclusion}
In this paper, we introduced \textsc{Autalic}, the first benchmark dataset focused specifically on the detection of anti-autistic ableist language in context. Through the collection and annotation of 2,400 sentences from Reddit, we aim to fill a critical gap in current NLP research and improve its alignment with neurodiversity. 

Looking forward, \textsc{Autalic} paves the way for significant advancements in content moderation systems, hate speech detection models, and research on ableism and neurodiversity. We envision this dataset as a cornerstone for future work in addressing bias against autistic individuals and fostering a more inclusive digital environment. By sharing this resource with the broader research community, we aim to catalyze the development of more equitable NLP systems that better serve underrepresented and marginalized groups.
\chapter{Conclusion}
Foundational research has defined `artificial' or machine intelligence as its ability to mimic human communication \citep{turing1950computing}, and this definition continues to serve as a benchmark for AI even today. As human-like AI agents such as robots and chabots become increasingly popular in our society, it is important to address ethical concerns surrounding the ways humanness is defined and implemented in such technologies. Autistic people have been historically dehumanized due to their unique communication styles which are viewed as a `deficit' of neurotypical social skills under neuronormativity. Through a series of data-driven and participatory studies, this dissertation explores the origins, prevalence, and impact of these biases, demonstrating the necessity and potential of neuro-inclusive approaches in AI research and development. Our findings indicate that shifting narratives about autism away from a deficits-based perspective and offering more inclusive communication training to non-autistic people can help foster autism inclusion among technology workers. However, such work is underexplored in agents designed to replicate humanness as we found robots replicate dehumanizing power imbalances, and 90\% of them are developed without feedback from autistic end-users. To improve the representation of autism-related data in AI, we created \textsc{Autalic}, which provides researchers with a practical benchmark to assess and mitigate anti-autistic biases while minimizing community censorship in ableist speech classification. Our evaluation of state-of-the-art large language models on \textsc{Autalic} reveals that LLMs continue censoring community perspectives while missing instances of ableist speech, showing a major misalignment with community perspectives. This highlights the importance of our work and the need for future research to continue improving data representation for autistic and other neurodivergent communities, ultimately moving toward more neuro-inclusive AI.

\section{Summary of Recommendations}

 \subsection{Chapter 1: Shifting the Lens: Digital Education for Neuro-Inclusive Communication}
 \begin{itemize}
  \item Develop technologies offering skills training, particularly those focusing on effective communication, for everyone, not just autistic people, to promote more equitable social interactions.
 \end{itemize}

 \subsection{Chapter 2: Agentic Ableism: A Case Study of How Robots Marginalize Autistic People} 
 \subsubsection{Autism-Inclusion Tips to Avoid Pathologization}
 \begin{itemize}
  \item Consider that humans communicate and perceive the world in diverse ways, and those differences are not deficits.
  \item Diversify the foundational work for your research studies by citing newer research published in fields beyond medicine and psychology, such as Critical Autism Studies.
  \item Consider research directions promoting communication between different neurotypes in a balanced manner instead of placing the burden entirely on autistic people to adapt to different communication styles, such as the works of \citep{Morris2023Double, rizvi2021inclusive}.
 \end{itemize}

 \subsubsection{Autism-Inclusion Tips to Avoid Essentialism}
 \begin{itemize}
  \item Prioritize intersectionality in participant recruitment, research objectives, and data analysis.
  \item Avoid ableist language and essentialist stereotypes such as the ones mentioned in Bottema-Beutel et al.'s paper \citep{bottema-beutel_avoiding_2021}. For example, referring to non-autistic children as "typically developing.".
  \item Report participant demographics to help readers contextualize your findings.
 \end{itemize}

 \subsubsection{Autism-Inclusion Tips to Avoid Power Imbalances --- Part 1: Social Norms}
 \begin{itemize}
  \item Consider community-based research collaborations in lieu of purely clinical collaborations.
  \item Reconsider diagnosis- or treatment-based research directions that prioritize clinical outcomes.
  \item Identify the needs of autistic end-users through more user-centered design approaches instead of making assumptions based on clinical literature.
 \end{itemize}

 \subsubsection{Autism-Inclusion Tips to Avoid Power Imbalances --- Part 2: Research Designs}
 \begin{itemize}
  \item Avoid making assumptions about the abilities of autistic end-users in the study instruments.
  \item Prevent tokenization by giving autistic people decision-making power in the study design without pressuring them to accept clinical applications.
  \item Increase collaborations with autistic researchers and include positionality statements to contextualize the objectives and findings of the study.
 \end{itemize}

 \subsubsection{Autism-Inclusion Tips to Avoid Power Imbalances --- Part 3: User Interactions}
 \begin{itemize}
  \item Promote user participation and conversational equity in human-robot interactions.
  \item Obtain feedback from autistic users on their preferences for different robot types and roles such as bystanders or information consumers.
  \item Avoid creating user interactions that may compare autistic people to animals or other non-human entities.
 \end{itemize}

 \subsection{Chapter 3: Navigating Neuro-Inclusive AI: The Perspectives of Creators}
 \begin{itemize}
  \item Consider success metrics that go beyond user satisfaction, and think more deeply about the impact of our work on the lives of people we may inadvertently be marginalizing from solely focusing on the needs of the majority.
  \item Critically examine the separation of what we consider to be "ethical" or "responsible" AI from other forms of AI.
  \item Integrate ethics more thoroughly in our training of future AI creators, our evaluation of current creators and leaders, and the standards we all must uphold.
  \item Foster diversity in teams and encourage greater community involvement in research studies.
  \item Give communities decision-making power while developing technologies that impact them, so they are not treated as token minorities and are able to shape our work. This means ensuring they have the power to change the design, implementation, and outcome of our work instead of merely including their perspectives towards the end of the developmental cycle.
 \end{itemize}

 \subsection{Chapter 4: Annotator Perspectives on Refining the Data Annotation Process}
 \begin{itemize}
  \item Continue to explore the alignment of AI classifications with current community perspectives to ensure these systems are more accurate.
  \item Examine the trade-off between preserving community perspectives while simplifying the task for annotators to determine what aspects of the labels to preserve.
  % \item Future CSCW research can build upon these insights when creating novel annotation tools. For example, providing a glossary of commonly used terms and their definitions that can be collaboratively updated by the annotators, and allowing them to re-label certain sentences are useful adaptations to support diverse annotator teams in labeling ever-evolving language.
  % \item Include the diverse perspectives of autistic people to find the right balance between the community's perspective and the annotators'. 
  % Such work can determine what aspects of anti-autistic speech are important to them so those aspects can be preserved in the data annotation process. For example, such work can explore whether or not the distinction between implicit and explicit anti-autistic speech may be more important to autistic people than it is to the annotators, which can help dataset creators determine the most effective labeling scheme for their work.
 \end{itemize}

 \subsection{Chapter 5: Building a Foundation for Neuro-Inclusive AI Research with AUTALIC}
 \begin{itemize}
  \item Stay updated on the latest semantics by referring to the perspectives of autistic scholars, activists, and organizations.
 \end{itemize}
 
\subsection{Defining Neuro-Inclusive AI}
An approach to the design, development, and deployment of AI systems that explicitly de-centers neuronormative benchmarks and challenges the goal of merely mimicking `humanness.' Instead, it prioritizes understanding, accommodating, and empowering diverse neurotypes, such as autism.
% In this dissertation, I investigate the ways in which human-like AI agents marginalize autistic people using robots as a case study, and through qualitative interviews, shared insights on 
% Stuff at the end of the dissertation goes in the back matter
% \backmatter
\bibliographystyle{ACM-Reference-Format}
\bibliography{references}
\appendix
\chapter{Interview Study}
\label{Appendix}
\textit{Note: in attempts to avoid response bias, we have included extra questions in our surveys and interviews.}

%----------------------------------------------------
% SECTION A.1: SURVEYS
%----------------------------------------------------
\section{Survey Instruments} % Should become A.1
\label{sec:surveys}

% --- Subsection for Survey 1 ---
\subsection{Survey 1: Demographic Questions}
\label{Survey1}

\begin{enumerate}

% 1
\item \textbf{What is your job official title?}\\[6pt]
\noindent\rule{0.9\linewidth}{0.4pt} % A line to write on if desired

% 2
\item \textbf{What industry do you work in?}
  \begin{itemize} % Replaced non-breaking spaces with standard spaces for indentation
    \item Academia
    \item Industry
    \item Other: \rule{0.5\linewidth}{0.4pt}
  \end{itemize}

% 3
\item \textbf{Please specify the gender with which you most closely identify.}
  \begin{itemize} % Replaced non-breaking spaces
    \item Woman
    \item Man
    \item Non-binary
    \item Prefer not to answer
    \item Prefer to self-describe: \rule{0.5\linewidth}{0.4pt}
  \end{itemize}

% 4
\item \textbf{What is your age?}
  \begin{itemize} % Replaced non-breaking spaces
    \item 18--22
    \item 23--30
    \item 31--40
    \item 41--50
    \item 51--60
    \item Over 60
    \item Prefer not to answer
  \end{itemize}

% 5
\item \textbf{Please specify your race/ethnicity.}
  \begin{itemize} % Replaced non-breaking spaces
    \item White
    \item Hispanic or Latino
    \item Black or African American
    \item Native American or American Indian
    \item South Asian
    \item Southwest Asian or North African
    \item East Asian
    \item Southeast Asian or Pacific Islander
    \item Prefer to self-describe: \rule{0.5\linewidth}{0.4pt}
    \item Prefer not to answer
  \end{itemize}

% 6
\item \textbf{Parental Status}
  \begin{itemize} % Replaced non-breaking spaces
    \item I have or had children that I currently support and/or raise by myself, 
          with a partner, or as part of a team or family
    \item I do not have children that I currently support and/or raise by myself, 
          with a partner, or as part of a team or family
    \item Prefer to self-describe: \rule{0.5\linewidth}{0.4pt}
    \item Prefer not to answer
  \end{itemize}

% 7
\item \textbf{Marital status}
  \begin{itemize} % Replaced non-breaking spaces
    \item Single
    \item Common Law Marriage
    \item Married
    \item Separated
    \item Divorced
    \item Widowed
    \item Legal % Consider clarifying this option if needed
    \item Prefer to self-describe: \rule{0.5\linewidth}{0.4pt}
    \item Prefer not to answer
  \end{itemize}

% 8
\item \textbf{6-Question Disability Questionnaire}

  \begin{enumerate}[label*=\arabic*.] % Requires \usepackage{enumitem}
    \item Are you deaf, or do you have serious difficulty hearing?
    \item Are you blind, visually impaired, or do you have serious difficulty seeing, even when wearing glasses?
    \item Because of a physical, mental, or emotional condition, do you have serious difficulty 
          concentrating, remembering, or making decisions?
    \item Do you have serious difficulty walking or climbing stairs?
    \item Do you have difficulty dressing or bathing?
    \item Because of a physical, mental, or emotional condition, do you have difficulty doing 
          errands alone such as visiting a doctor’s office or shopping? (15 years old or older)
  \end{enumerate}

% 9
\item \textbf{Sexual Orientation}
  \begin{itemize} % Replaced non-breaking spaces
    \item Asexual
    \item Bisexual
    \item Pansexual
    \item Gay
    \item Heterosexual
    \item Lesbian
    \item Queer
    \item Prefer to self-describe: \rule{0.5\linewidth}{0.4pt}
    \item Prefer not to answer
  \end{itemize}

% 10
\item \textbf{What is the highest degree or level of school you have completed?}
  \begin{itemize} % Replaced non-breaking spaces
    \item Some high school credit, no diploma or equivalent
    \item Less than high school degree
    \item High school graduate (high school diploma or equivalent including GED)
    \item Some college but no degree
    \item Associate's degree
    \item Bachelor's degree
    \item Advanced degree (e.g., Master's, doctorate)
    \item Prefer not to answer
  \end{itemize}

% 11
\item \textbf{Which one of the following includes your total HOUSEHOLD income for last year, before taxes?}
  \begin{itemize} % Replaced non-breaking spaces
    \item Less than \$10,000
    \item \$10,000 to under \$20,000
    \item \$20,000 to under \$30,000
    \item \$30,000 to under \$40,000
    \item \$40,000 to under \$50,000
    \item \$50,000 to under \$65,000
    \item \$65,000 to under \$80,000
    \item \$80,000 to under \$100,000
    \item \$100,000 to under \$125,000
    \item \$125,000 to under \$150,000
    \item \$150,000 to under \$200,000
    \item \$200,000 or more
    \item Prefer not to answer
  \end{itemize}

\end{enumerate}

% --- Subsection for Survey 2 ---
\subsection{Survey 2}
\label{Survey2}
\begin{enumerate}
  % 1
  \item \textbf{Imagine your team switches to a remote-optional work environment and 
  now allows your team members to have their videos off during meetings. 
  How do you feel about this new policy?}
    \begin{itemize} % Replaced non-breaking spaces
      \item Strongly Disagree
      \item Somewhat Disagree
      \item Neither agree nor disagree
      \item Somewhat Agree
      \item Strongly Agree
    \end{itemize}

  % 2
  \item \textbf{Imagine your team does not allow people to have conversations in the open 
  work spaces, and requires them to use designated spaces. How do you feel about this 
  change in work policies?}
    \begin{itemize} % Replaced non-breaking spaces
      \item Strongly Disagree
      \item Somewhat Disagree
      \item Neither agree nor disagree
      \item Somewhat Agree
      \item Strongly Agree
    \end{itemize}

  % 3
  \item \textbf{Intersectionality impacts the work that I do.}
    \begin{itemize} % Replaced non-breaking spaces
      \item Strongly Disagree
      \item Somewhat Disagree
      \item Neither agree nor disagree
      \item Somewhat Agree
      \item Strongly Agree
    \end{itemize}

  % 4 (Open-ended)
  \item \textbf{What is intersectionality? (please define in your own words)}\\[6pt]
  \rule{0.9\linewidth}{0.4pt}

  % 5
  \item \textbf{It is my responsibility to support and advocate for recruitment 
  and retention efforts in programs and agencies that ensure diversity.}
    \begin{itemize} % Replaced non-breaking spaces
      \item Strongly Disagree
      \item Somewhat Disagree
      \item Neither agree nor disagree
      \item Somewhat Agree
      \item Strongly Agree
    \end{itemize}

  % 6
  \item \textbf{People who work in computer science-related fields should participate in 
  educational and training programs that help advance cultural competence 
  within the profession.}
    \begin{itemize} % Replaced non-breaking spaces
      \item Strongly Disagree
      \item Somewhat Disagree
      \item Neither agree nor disagree
      \item Somewhat Agree
      \item Strongly Agree
    \end{itemize}

  % 7
  \item \textbf{People who work in computer science-related fields should 
  understand the culture of computer science and its functions in human 
  behavior and society, recognizing the strengths that exist in all cultures.}
    \begin{itemize} % Replaced non-breaking spaces
      \item Strongly Disagree
      \item Somewhat Disagree
      \item Neither agree nor disagree
      \item Somewhat Agree
      \item Strongly Agree
    \end{itemize}

  % 8
  \item \textbf{Because we live in the U.S., everyone should speak or at least try to learn English.}
    \begin{itemize} % Replaced non-breaking spaces
      \item Strongly Disagree
      \item Somewhat Disagree
      \item Neither agree nor disagree
      \item Somewhat Agree
      \item Strongly Agree
    \end{itemize}

  % 9
  \item \textbf{In the U.S. some people are often verbally attacked because 
  of their minority status.}
    \begin{itemize} % Replaced non-breaking spaces
      \item Strongly Disagree
      \item Somewhat Disagree
      \item Neither agree nor disagree
      \item Somewhat Agree
      \item Strongly Agree
    \end{itemize}

  % 10
  \item \textbf{Illegal immigrants should be deported to their home countries.}
    \begin{itemize} % Replaced non-breaking spaces
      \item Strongly Disagree
      \item Somewhat Disagree
      \item Neither agree nor disagree
      \item Somewhat Agree
      \item Strongly Agree
    \end{itemize}

  % 11
  \item \textbf{All people have equal opportunities in the U.S.}
    \begin{itemize} % Replaced non-breaking spaces
      \item Strongly Disagree
      \item Somewhat Disagree
      \item Neither agree nor disagree
      \item Somewhat Agree
      \item Strongly Agree
    \end{itemize}

  % 12
  \item \textbf{Membership in a minority group significantly increases risk 
  factors for exposure to discrimination, economic deprivation, and oppression.}
    \begin{itemize} % Replaced non-breaking spaces
      \item Strongly Disagree
      \item Somewhat Disagree
      \item Neither agree nor disagree
      \item Somewhat Agree
      \item Strongly Agree
    \end{itemize}

  % 13
  \item \textbf{In the U.S., some people are often physically attacked 
  because of their minority status.}
    \begin{itemize} % Replaced non-breaking spaces
      \item Strongly Disagree
      \item Somewhat Disagree
      \item Neither agree nor disagree
      \item Somewhat Agree
      \item Strongly Agree
    \end{itemize}

  % 14
  \item \textbf{Being lesbian, bisexual, or gay is a choice.}
    \begin{itemize} % Replaced non-breaking spaces
      \item Strongly Disagree
      \item Somewhat Disagree
      \item Neither agree nor disagree
      \item Somewhat Agree
      \item Strongly Agree
    \end{itemize}

  % 15
  \item \textbf{The American dream is real for anyone willing to work hard to achieve it.}
    \begin{itemize} % Replaced non-breaking spaces
      \item Strongly Disagree
      \item Somewhat Disagree
      \item Neither agree nor disagree
      \item Somewhat Agree
      \item Strongly Agree
    \end{itemize}

  % True/False Questions

  % 16
  \item \textbf{I sometimes litter.}
    \begin{itemize} % Replaced non-breaking spaces
      \item True
      \item False
    \end{itemize}

  % 17
  \item \textbf{I always admit my mistakes openly and face the potential for negative 
  consequences.}
    \begin{itemize} % Replaced non-breaking spaces
      \item True
      \item False
    \end{itemize}

  % 18
  \item \textbf{In traffic, I am always polite and considerate of others.}
    \begin{itemize} % Replaced non-breaking spaces
      \item True
      \item False
    \end{itemize}

  % 19
  \item \textbf{I have tried illegal drugs (for example, marijuana, cocaine, etc.).}
    \begin{itemize} % Replaced non-breaking spaces
      \item True
      \item False
    \end{itemize}

  % 20
  \item \textbf{I always accept others' opinions, even when they don't agree with my own.}
    \begin{itemize} % Replaced non-breaking spaces
      \item True
      \item False
    \end{itemize}

  % 21
  \item \textbf{I take out my bad moods on others now and then.}
    \begin{itemize} % Replaced non-breaking spaces
      \item True
      \item False
    \end{itemize}

  % 22
  \item \textbf{There has been an occasion when I took advantage of someone else.}
    \begin{itemize} % Replaced non-breaking spaces
      \item True
      \item False
    \end{itemize}

  % 23
  \item \textbf{In conversations, I always listen attentively and let others finish their 
  sentences.}
    \begin{itemize} % Replaced non-breaking spaces
      \item True
      \item False
    \end{itemize}

  % 24
  \item \textbf{I never hesitate to help someone in case of an emergency.}
    \begin{itemize} % Replaced non-breaking spaces
      \item True
      \item False
    \end{itemize}

  % 25
  \item \textbf{When I have made a promise, I keep it--no ifs, ands, or buts.}
    \begin{itemize} % Replaced non-breaking spaces
      \item True
      \item False
    \end{itemize}

  % 26
  \item \textbf{I occasionally speak badly of others behind their back.}
    \begin{itemize} % Replaced non-breaking spaces
      \item True
      \item False
    \end{itemize}

  % 27
  \item \textbf{I would never live off other people.}
    \begin{itemize} % Replaced non-breaking spaces
      \item True
      \item False
    \end{itemize}

  % 28
  \item \textbf{I always stay friendly and courteous with other people, even when I am 
  stressed out.}
    \begin{itemize} % Replaced non-breaking spaces
      \item True
      \item False
    \end{itemize}

  % 29
  \item \textbf{During arguments I always stay objective and matter-of-fact.}
    \begin{itemize} % Replaced non-breaking spaces
      \item True
      \item False
    \end{itemize}

  % 30
  \item \textbf{There has been at least one occasion when I failed to return an item that I 
  borrowed.}
    \begin{itemize} % Replaced non-breaking spaces
      \item True
      \item False
    \end{itemize}

  % 31
  \item \textbf{I always eat a healthy diet.}
    \begin{itemize} % Replaced non-breaking spaces
      \item True
      \item False
    \end{itemize}

  % 32
  \item \textbf{Sometimes I only help because I expect something in return.}
    \begin{itemize} % Replaced non-breaking spaces
      \item True
      \item False
    \end{itemize}

  % Additional Likert-scale questions about knowledge/confidence

  % 33
  \item \textbf{I am able to develop programs and services that reflect an 
  understanding of the diversity between and within cultures.}
    \begin{itemize} % Replaced non-breaking spaces
      \item Strongly Disagree
      \item Somewhat Disagree
      \item Neither agree nor disagree
      \item Somewhat Agree
      \item Strongly Agree
    \end{itemize}

  % 34
  \item \textbf{I feel confident about my knowledge and understanding of people with 
  disabilities, needs, traditions, values, family systems, and artistic expressions.}
    \begin{itemize} % Replaced non-breaking spaces
      \item Strongly Disagree
      \item Somewhat Disagree
      \item Neither agree nor disagree
      \item Somewhat Agree
      \item Strongly Agree
    \end{itemize}

  % 35
  \item \textbf{I feel confident about my knowledge and understanding of African American 
  and African history, traditions, values, family systems, and artistic expressions.}
    \begin{itemize} % Replaced non-breaking spaces
      \item Strongly Disagree
      \item Somewhat Disagree
      \item Neither agree nor disagree
      \item Somewhat Agree
      \item Strongly Agree
    \end{itemize}

  % 36
  \item \textbf{I feel confident about my knowledge and understanding of Middle Eastern 
  (South West Asian and North African) history, traditions, values, family systems, 
  and artistic expressions.}
    \begin{itemize} % Replaced non-breaking spaces
      \item Strongly Disagree
      \item Somewhat Disagree
      \item Neither agree nor disagree
      \item Somewhat Agree
      \item Strongly Agree
    \end{itemize}

  % 37
  \item \textbf{I feel confident about my knowledge and understanding of women’s history, 
  traditions, values, family systems, and artistic expressions.}
    \begin{itemize} % Replaced non-breaking spaces
      \item Strongly Disagree
      \item Somewhat Disagree
      \item Neither agree nor disagree
      \item Somewhat Agree
      \item Strongly Agree
    \end{itemize}

  % 38
  \item \textbf{I feel confident about my knowledge and understanding of 
  gay/lesbian/bisexual/transgender history, traditions, values, family systems, 
  and artistic expressions.}
    \begin{itemize} % Replaced non-breaking spaces
      \item Strongly Disagree
      \item Somewhat Disagree
      \item Neither agree nor disagree
      \item Somewhat Agree
      \item Strongly Agree
    \end{itemize}

  % 39
  \item \textbf{I feel confident about my knowledge and understanding of 
  Jewish history, traditions, values, family systems, and artistic expressions.}
    \begin{itemize} % Replaced non-breaking spaces
      \item Strongly Disagree
      \item Somewhat Disagree
      \item Neither agree nor disagree
      \item Somewhat Agree
      \item Strongly Agree
    \end{itemize}

  % 40
  \item \textbf{I am aware of ways in which institutional oppression and the misuse of 
  power constrain the human and legal rights of individuals and groups within 
  American society.}
    \begin{itemize} % Replaced non-breaking spaces
      \item Strongly Disagree
      \item Somewhat Disagree
      \item Neither agree nor disagree
      \item Somewhat Agree
      \item Strongly Agree
    \end{itemize}

  % 41
  \item \textbf{I feel confident about my knowledge and understanding of Native American 
  history, traditions, values, family systems, and artistic expressions.}
    \begin{itemize} % Replaced non-breaking spaces
      \item Strongly Disagree
      \item Somewhat Disagree
      \item Neither agree nor disagree
      \item Somewhat Agree
      \item Strongly Agree
    \end{itemize}

  % 42
  \item \textbf{I have the knowledge to critique and apply culturally competent and 
  social justice approaches to influence assessment, planning, access to resources, 
  intervention, and research.}
    \begin{itemize} % Replaced non-breaking spaces
      \item Strongly Disagree
      \item Somewhat Disagree
      \item Neither agree nor disagree
      \item Somewhat Agree
      \item Strongly Agree
    \end{itemize}

  % 43
  \item \textbf{I feel confident about my knowledge and understanding of Asian and 
  Asian American history, traditions, values, family systems, and artistic expressions.}
    \begin{itemize} % Replaced non-breaking spaces
      \item Strongly Disagree
      \item Somewhat Disagree
      \item Neither agree nor disagree
      \item Somewhat Agree
      \item Strongly Agree
    \end{itemize}

\end{enumerate}
% --- Section for Interviews ---
\section{Interviews}
\label{sec:interviews} % Add label if needed

% --- Subsection for Interview 1 ---
\subsection{Interview 1}
\label{Int1}
\begin{enumerate}[label=\arabic*.]
    \item Tell me about a recent project you worked on where you built an AI system that others would potentially use.
      \begin{enumerate}[label=\alph*.] % Use standard spaces for indent
        \item When did you know you’ve reached the end of the design process for your system?
          \begin{enumerate}[label=\roman*.] % Use standard spaces for indent
            \item When the current assignment is complete.
          \end{enumerate}
        \item When did you know you’ve reached the end of the design process for your Robot/chatbot/AI agent?
        \item What was the overall purpose or objective for your system?
        \item How did you decide the physical characteristics (including UI) of the bot?
          \begin{enumerate}[label=\roman*.]
            \item What do you think would happen if you chose something else?
          \end{enumerate}
        \item How did you decide the behavioral and conversational characteristics of the bot?
          \begin{enumerate}[label=\roman*.]
            \item What do you think would happen if you chose something else?
          \end{enumerate}
      \end{enumerate}

    \item What are the metrics you use for evaluating the performance of your system?
    \item How did you know that your work was effective? (ask if they created a solution to a problem, otherwise skip).
      \begin{enumerate}[label=\alph*.]
        \item Can you tell me about a time when a user interacted with your system and got frustrated or upset?
        \item What are some things that may make a user frustrated or not want to interact with your system?
      \end{enumerate}

    \item Can you walk me through how or why you chose this specific mode of user interaction to reach your objective or solve your problem?

    \item What real-world interactions can you think of that resemble your user interaction?
      \begin{enumerate}[label=\alph*.]
        \item If they say no, ASK - Can you think of any other interaction that might be similar to the interaction you developed?
        \item What are the identities of the people in these interactions?
        \item Explain the interaction for me.
          \begin{enumerate}[label=\roman*.]
            \item What would it look like if one of the people was autistic/ADHD/dyslexic?
            \item How would the interaction change?
          \end{enumerate}
      \end{enumerate}

    \item What behavioral or communicational changes could you make to the bot that would impact the user’s interaction with the bot?
      \begin{enumerate}[label=\alph*.]
        \item How would that be different with a person who is funny?
        \item How would that impact the user’s interactions with folks who are and aren’t funny/don’t use humor in interactions?
      \end{enumerate}

    \item What conclusions would researchers who are building off of your work reach or assume about human interactions, behavior, or AI? What about end-users? What conclusions might they reach?
      \begin{enumerate}[label=\alph*.]
        \item What are the best practices in bot design you have learned in your work?
        \item How do end-users’ interactions differ from those with other bots?
      \end{enumerate}

    \item In your opinion, should we design chatbots, robots, or other AI agents to meet the needs of various community groups?
      \begin{enumerate}[label=\alph*.]
        \item Who are the people most likely to use your chatbot?
        \item What would happen if someone from [specific group] used it?
        \item What would happen if you adjusted the bot to work with multiple user groups? Or specific user groups?
        \item What kind of accessibility features did you consider in your chatbot?
        \item Should we design for people whose mental and neurological abilities may be considered atypical such as folks with autism, ADHD, and dyslexia?
          \begin{enumerate}[label=\roman*.]
            \item Limitations?
          \end{enumerate}
        \item Should we design for different age groups?
      \end{enumerate}

    \item Is there anything else you’d like to say or add in regards to your experience building AI entities?
\end{enumerate}

% --- Subsection for Interview 2 ---
\subsection{Interview 2}
\label{Int2}
\begin{enumerate}[label=\arabic*.]
    \item How would you describe your current professional position?
    \item In your opinion, what experiences from your life greatly impacted or led you to your interest in your current career?
    \item What made you stay or continue to pursue your current career?
    \item Tell me about a time when your personal background overlapped with your current position or your work.
\end{enumerate}

\textbf{Intersectionality Questions}
\begin{enumerate}[label=\arabic*.]
    \item Has an ism (e.g., racism, sexism, ableism, homophobia) ever impacted you during your computer science journey?
      \begin{enumerate}[label=\alph*.]
        \item Has an ism ever impacted you in your life?
        \item If no, ask: At work, do you interact with a number of people who have a similar racial, ethnic, or gender background as you on a regular basis?
      \end{enumerate}
    \item In your opinion, what is intersectionality?
      \begin{enumerate}[label=\alph*.]
        \item Has it ever impacted your work or research? If so, how?
      \end{enumerate}
    \item Tell me about a time when you used or experienced intersectionality in your work/research.
      \begin{enumerate}[label=\alph*.]
        \item If non-tech, let them provide it. If research-related, ask: Do you have any examples related to your current position?
      \end{enumerate}
    \item Tell me about a time when you used or experienced a push for inclusivity in your work/research.
    \item Tell me a time when intersectionality has impacted you at work.
    \item Have you had any training in working with or building tools that support inclusivity?
      \begin{enumerate}[label=\alph*.]
        \item If so, what did the training look like? Can you describe it?
        \item Did it help you?
      \end{enumerate}
    \item Have you had any support in working with or building tools that support inclusivity?
      \begin{enumerate}[label=\alph*.]
        \item If so, what did the support look like? Can you describe it?
        \item Did it help you?
      \end{enumerate}
\end{enumerate}

\textbf{Situational Awareness}
\begin{enumerate}[label=\arabic*.]
    \item Suppose your new colleague shows up wearing headphones to a research conference and continues wearing them even while presenting their work. How would you approach a conversation with them?
      \begin{enumerate}[label=\alph*.]
        \item What if your coworker mentions they’re wearing it because the room is too loud?
      \end{enumerate}
    \item Imagine that your company celebrates its anniversary during Ramadan. Ramadan is a Muslim holiday where Muslims refrain from eating and drinking from sunrise to sunset. This year is their 50th anniversary, and the plan is to have a week full of free buffet-style lunch and brunch banquets over the weekend. However, about three coworkers have expressed concern that they won’t be able to fully participate due to the way things are planned to happen.
      \begin{enumerate}[label=\alph*.]
        \item What is the problem from the coworkers' viewpoint?
        \item How might this problem be solved?
      \end{enumerate}
    \item Suppose your research assistant doodles during meetings, and your supervisor approaches you about the lack of professionalism exhibited by your RA. How would you address this situation?
    \item Max and Alex are working on a research project and are expected to provide feedback to each other to ensure the project stays on track. [Share a conversation transcript.]
      \begin{enumerate}[label=\alph*.]
        \item Why is Alex responding this way?
        \item Why is Max responding this way?
        \item How might you respond as their supervisor?
      \end{enumerate}
\end{enumerate}

% \usepackage{graphicx}

\section{Guidelines Provided Overtime}

\label{app:r1g}

\subsection{Sentence-Level Annotation}
We are annotating text extracted from human-robot interaction literature and social media (Reddit, Twitter). The goal of this study is to classify sentences as either i) implicitly ableist or ii) explicitly ableist iii) implicitly anti-autistic, iv) explicitly anti-autistic, v) not ableist, or vi) needs more context and/or is irrelevant. You will see paragraphs and sentences extracted from scientific literature focusing on robotics for autism published in conferences and journals, and interactions between anonymous social media users. These sentences may or may not contain ableist content.

\subsubsection{Defining Ableism
According to the Center for Disability Rights (emphasis added by us):}
Ableism is a set of beliefs or practices that devalue and discriminate against people with physical, intellectual, or psychiatric disabilities and often rests on the assumption that disabled people need to be ‘fixed’ in one form or the other.

\subsubsection{Neurodiversity vs. Neuroableism}
We center neurodiversity in our work, which means we acknowledge that humans have different ways of thinking, communicating, and experiencing the world, and that these differences are valid forms of human diversity.

The belief that autistic people are “deficient” in certain skills is rooted in eugenicism and dehumanizing research originating in Nazi Germany, and contradict neurodiversity by promoting neuroableism, or the belief that neurotypical people are “normal”, and people with various neurological differences such as autism, ADHD, or dyslexia are “abnormal”. Such understandings may result in researchers studying autism as a “disorder” that needs to be diagnosed, treated, prevented, or cured. This work is ableist because it expects autistic people to be “fixed” by changing their communicational styles and behaviors to adapt to neurotypical social norms.

\subsubsection{Not-Ableist (Round 1-3)}
These sentences may be entirely unrelated to autism or disabilities, or be written by an autistic person reaching out for help and support.
\begin{enumerate}
   
\item Using medical terminology in a more personal manner (e.g. ‘I need therapy’).
 \item Discussions + suggestions from neurodivergent people (community-generated discussions).
\item  General discussions of the medical processes (unrelated to neurodivergence).
\item Example: “I am autistic”, “As an autistic person, I think …”
\end{enumerate}

\subsubsection{Implicitly Ableist (Round 1-3)}
Critical disability theory (CDT) is a framework centering disability which challenges the ableist assumptions present in our society. Using this theory, we define “implicitly ableist” content as:
\begin{enumerate}   
\item Making assumptions about a disabled person’s abilities that would not be made about an able-bodied person. 
\item “Inspiration porn” → looking at disabled people as “inspiration” for able-bodied people
\item Using condescending language such as saying a disabled person “suffers” with their disability 
\item Infantilizing disabled people by portraying them as unable to make their own decisions, or “innocent” and “pure”
\item Example: “She is confined to a wheelchair”, “it must be awful living with bipolar disorder”
\end{enumerate}

\subsubsection{Implicitly Anti-Autistic (Round 3)}

For the purposes of this study, we are applying a critical disability framework and classifying any sentences that describe autism using medical terminology such as an “illness” or “disease”, or focus on clinical applications such as “curing” or “diagnosing” autism as implicitly ableist. These works are considered “implicitly ableist” as they do not use explicitly violent language or slurs, but still may be considered offensive according to critical autism theory.
\begin{itemize}  
    \item Using medical terminology to describe autism (e.g. saying its a “disorder”) E.g., referring to autism as a public health crisis.
    \item Using language that may promote the infantilization or pathologization of autistic people
    \item Using words like “typically developing” to refer to non-autistic people.
    \item Saying autistic people have ‘atypical behavior’.
    \item Ableist jokes using medical terminology (e.g. “It cured every disease, but was it's own cancer.”)
    \item Other anti-autistic language (from the Bottema-Beutel paper)
    Example: “I am so proud of my autistic son for baking this cake”, “even though she is autistic, she is a very good salesperson” 
\end{itemize}

\subsubsection{Explicitly Ableist (Round 1-3)}
Sentences using slurs for disabled people such as: retard, lame, insane, deluded, moron

\subsubsection{Explicitly Anti-Autistic (Round 3)}
This category includes sentences that dehumanize autistic people (e.g. by comparing them to animals or non-living things), containing ableist slurs such as the r-word, other anti-autistic language that promotes negative stereotypes, or expressing negative emotions such as fear, disgust, or hatred toward autistic people. Slurs like the r-word specifically aimed at autistic people: Crazy, Delusional, Deranged, Dumb, Handicap, Idiot, Insane, Maniac, Moron, Madness, Stupid, Wheelchair bound, Differently abled, Handicapable, People with abilities, Special needs	(source) Saying phrases like “autism intensifies” Using “autistic” as an insult (e.g. “that’s so autistic”)

\subsubsection{Unrelated/Needs More Context (Round 1-4)}
This category includes text that is completely unrelated to disabilities, needs more context, and/or contains forms of media other than text. Examples include: tweets with images, ACM CCS categories, etc.
Examples: 

\begin{description}
    \item[Not a sentence: ] “Keywords— Socially assistive robots; Affective robots; ASD; Autism; Developmental ability; Mullen; ADOS”
    \item[Includes non-text media:]
“@RonTerrell http://twitpic.com/3iqv9 - I wish I was there it looks like fun.”
\item[Needs more context:]
“You mean a walk won't cure an abscess in her mouth?” 
\end{description}

\section{Glossary}

An excerpt of our glossary developed as a result of annotators feedback is shown below in Table \ref{tab:glossary}.

\begin{table}[b]
\caption{Glossary of Terms}
\label{tab:glossary}
\centering
\resizebox{\columnwidth}{!}{%
\begin{tabular}{|p{0.19\textwidth}|p{0.8\textwidth}|}
\hline
\multicolumn{1}{|c|}{\textbf{Term}} &
  \multicolumn{1}{c|}{\textbf{Definition}} \\ \hline
\cellcolor[HTML]{F4CCCC}Autism Speaks &
  Controversial group promoting autism `awareness', some autistic people it should be considered a hate group \\ \hline
\cellcolor[HTML]{F4CCCC}ASD &
  Autism Spectrum Disorder \\ \hline
DSM-V &
  The Diagnostic and Statistical Manual of Mental Disorders, 5th Edition published by the American Psychiatric Association (used to diagnose autism) \\ \hline
\cellcolor[HTML]{00FF00}AuDHD &
  Having a combination of autism and ADHD \\ \hline
\cellcolor[HTML]{00FF00}Au, Âû &
  Used by autistic individuals to self-identify as autistic \\ \hline
\cellcolor[HTML]{00FF00}ND &
  Neurodivergent or neurodiverse \\ \hline
\cellcolor[HTML]{00FF00}NT &
  Neurotypical \\ \hline
\cellcolor[HTML]{00FF00}Allistic &
  Non-Autistic \\ \hline
\cellcolor[HTML]{00FF00}Aspie &
  Someone with Aspergers (outdated term, considered offensive unless it is being used to self-identify) \\ \hline
\cellcolor[HTML]{00FF00}Autie &
  Autistic \\ \hline
\cellcolor[HTML]{F4CCCC}Autism mom, dad, or parent &
  A parent or caregiver of an autistic individual \\ \hline
\cellcolor[HTML]{00FF00}Dx &
  Diagnose \\ \hline
\cellcolor[HTML]{F4CCCC}Martyr mom or parent &
  A parent using their child’s autism to gain sympathy, attention, etc. \\ \hline
\cellcolor[HTML]{00FF00}self-Dx &
  Self-diagnose \\ \hline
\cellcolor[HTML]{FFFF00}OT &
  Occupational therapist \\ \hline
\cellcolor[HTML]{00FF00}AA &
  Actually autistic \\ \hline
\cellcolor[HTML]{F4CCCC}ABA &
  Applied Behavioural Analysis, controversial treatment for autism that has been linked to PTSD in autistic individuals \\ \hline
\cellcolor[HTML]{FFFF00}Masking &
  Suppressing behaviors such as stimming to appear neurotypical \\ \hline
\cellcolor[HTML]{FFFF00}SPD &
  Sensory processing disorder \\ \hline
\cellcolor[HTML]{FFFF00}Stimming, stim &
  Repetitive actions such as hand-flapping and singing used for self-soothing \\ \hline
\cellcolor[HTML]{F4CCCC}Functioning label &
  Outdated and offensive way to describe the ‘severity’ of someone’s autism \\ \hline
\cellcolor[HTML]{FFFF00}Echolalia &
  Repeating sounds, words, or phrases, often unintentionally \\ \hline
\cellcolor[HTML]{FFFF00}Co-morbid &
  Having more than one medical condition together \\ \hline
\cellcolor[HTML]{FFFF00}CW/TW &
  Content warning, trigger warning tags. Often used in discussions where the content may be graphic or sensitive in nature \\ \hline
% \cellcolor[HTML]{FFFF00}AFAB/AMAB &
%   Assigned Female at Birth, Assigned Male at Birth, used to describe transgender individuals \\ \hline
% \cellcolor[HTML]{FFFF00}Unalive &
%   To kill someone \\ \hline
% \cellcolor[HTML]{FFFF00}ED, BED, AN, ARFID &
%   Eating Disorders such as: Binge Eating Disorder, Anorexia Nervosa, Avoidant/restrictive food intake disorder \\ \hline
\cellcolor[HTML]{FFFF00}ED &
  Executive dysfunction, such as challenges with planning, time management, organization, and completing tasks often experienced by ND individuals \\ \hline
% \cellcolor[HTML]{FFFF00}EDS &
%   Ehler’s-Danlos Syndrome \\ \hline
\cellcolor[HTML]{FFFF00}IFL/PFL &
  Identity-first language or people-first language. Autistic individuals may prefer IFL (i.e. saying `autistic person' instead of `person with autism' \\ \hline
\cellcolor[HTML]{FFFF00}RSD &
  \textbf{Rejection sensitivity dysphoria, sensitivity to rejection often experienced by ND individuals} \\ \hline
\rowcolor[HTML]{FFFF00} 
2e &
  \cellcolor[HTML]{FFFFFF}Twice exceptional: an ND individual who is also considered `gifted' \\ \hline
\rowcolor[HTML]{FFFF00} 
AAC &
  \cellcolor[HTML]{FFFFFF}Augmentative \& Alternative Communication: ways to communicate besides speaking \\ \hline
\end{tabular}%
}
\end{table}
\chapter{AUTALIC}
\label{AUTALIC}
\section{Guidelines for Responsible Use}
\subsection{Purpose and Scope}
This dataset has been curated to aid in the classification and study of anti-autistic ableist language in a U.S. context using text from Reddit. It aims to support research and educational endeavors focused on understanding, identifying, and mitigating ableist speech directed at autistic individuals, while moving away from mis-classifying any speech related to autism or disability as toxic.

\subsection{Applicability and Cultural Context}

\textbf{U.S.-Specific Results:} The language examples and classification models in this dataset are primarily reflective of usage and cultural nuances in the United States. As a result, the dataset and any models developed from it may not be fully accurate or generalizable for other countries or cultural contexts.

\textbf{Data Curation:} The data included in this has been taken solely from posts and comments on Reddit and may not represent autism discourse on other platforms and in other contexts.

\textbf{Contact for Latest Version:} Language evolves over time. For the most up-to-date version of the dataset, or on more information on when the dataset was last updated, please contact the first author.

\subsection{Access and Security}

\textbf{Password Protection:} The dataset is password-protected to prevent unauthorized or automated scraping (e.g., by bots). While the password is publicly available as of this publication, it may require prior approval in the future as needed to ensure reponsible use.

\textbf{Secure Storage:} Users are expected to maintain secure protocols (e.g., encryption, controlled access) to prevent unauthorized sharing or leaks of the dataset. The dataset may not be shared without consent of the authors. 

\subsection{Permitted Uses}

\textbf{Free Use for Scientific Research:} The dataset is publicly available without charge for legitimate scientific, academic, or educational research purposes, subject to the restrictions outlined below.

\textbf{Academic and Non-Profit Contexts:} Users in academic, research, or non-profit institutions may incorporate the dataset into studies, presentations, or scholarly articles, provided they follow these guidelines and appropriately cite the dataset and its authors.

\subsection{Prohibited or Restricted Uses
Commercial Use:} Commercial use is not authorized without explicit written permission from the dataset authors. If you wish to incorporate the dataset into commercial products or services, you must obtain approval in advance.

\textbf{Automated Content Moderation:} Using the dataset to develop or deploy automated content moderation tools is not authorized without prior approval from the authors. This restriction helps ensure that any moderation system is deployed ethically and with proper considerations for context and language evolution.

\subsection{Ethical Considerations and Privacy}

\textbf{Respect for Individuals and Communities:} Users must handle the dataset with an understanding of the impacts of ableist language on autistic communities. The dataset’s examples are provided solely for research and analysis and  must not be used to perpetuate or normalize ableist attitudes, or to scrutinize or attack any individual annotators or original posters. This work is not intended as an ethical judgment or targeting of individuals, but rather an effort to improve AI alignment with the perspectives of autistic people.

\textbf{Citations and Acknowledgements:} When publishing findings, users should cite this dataset, acknowledging the work of its authors and the communities that provided the materials or data.

\textbf{Compliance with Regulations:} Researchers must comply with all relevant local, national, and international regulations and guidelines relating to data privacy and human subjects research where applicable.

\subsection{How to Request Approval}

\textbf{Commercial or Moderation Use:} If you intend to use the dataset for commercial purposes or automated content moderation, please submit a formal request, detailing:
\begin{itemize}
    \item Project objectives
    \item Potential for data use and distribution
    \item Mechanisms to ensure ethical application and protection of the data
    \item The impact of the project, and its target end-users
\end{itemize}

\textbf{Contact the First Author:} All requests and inquiries should be directed to the first author, as listed in the dataset documentation or project website, available here: [REDACTED FOR PRIVACY]

\subsection{Liability and Disclaimer}
The dataset is provided “as is,” without any guarantees regarding completeness, accuracy, or fitness for a particular purpose, especially outside of the U.S. context, for multi-media posts, or discussions on other platforms outside of Reddit.

\textbf{User Responsibility:} Users bear the responsibility for ensuring their use complies with these guidelines, as well as any applicable laws and ethical standards.

By accessing and using this dataset, you acknowledge that you have read and agreed to these Guidelines for Responsible Use, and that you understand the conditions under which the dataset may be utilized for your research or projects.

\section{Annotator Orientation}
\label{AO}
\subsection{Introduction}
This subsection provides an overview of the annotation orientation session conducted for \textsc{Autalic}. The goal is to ensure annotators understand the history and contemporary examples of anti-autistic ableism, the importance of neurodiversity, and the different ways in which anti-autistic speech may manifest in text. Given the sensitive nature of this work, annotators are advised that they may encounter discussions involving ableist language, violence, self-harm, and suicide mentions.

\subsection{Understanding Anti-Autistic Ableism}
Anti-autistic ableism is the discrimination and devaluation of autistic individuals based on neuronormative standards. A striking example includes cases where caretakers harm autistic individuals due to societal stigma ~\cite{disabilitymemorialDisabilityMourning}. The historical roots of such bias date back to Nazi-era eugenics research, where Hans Asperger categorized autistic individuals as either “useful” or “unfit,” reinforcing a harmful hierarchical perception of autism \cite{thetransmitterEvidenceTies}.

\paragraph{Neuronormativity}
Neuronormativity is the societal belief that neurotypical cognition is the default and that neurodivergence is an abnormality ~\cite{huijg2020neuronormativity}. This belief system marginalizes autistic individuals and contributes to discrimination in various aspects of life, including education, employment, and social interactions.

\paragraph{Deficit-Based Approaches and Their Harms}
Traditional medical models frame autism as a disorder requiring intervention or treatment ~\cite{kapp2013deficit, kapp2019social}. This perspective has led to:
\begin{itemize}
    \item Increased exposure to violence and self-harm risk
    \item Social exclusion and stigmatization
    \item Internalized ableism and lower self-esteem
\end{itemize}

\paragraph{Benevolent Ableism}
Benevolent ableism refers to actions or attitudes that, while seemingly supportive, reinforce autistic individuals as “less than” neurotypicals ~\cite{nario2019hostile}. Examples include organizations like \textit{Autism Speaks}, which promote awareness campaigns that fail to center autistic voices ~\cite{rosenblatt2022autism}. The use of symbols such as the puzzle piece is an example of this issue, as it implies that autism is a mystery to be solved rather than a valid identity.

\subsection{The Neurodiversity Movement}
The neurodiversity paradigm challenges the medical model by recognizing neurological variations as a natural and valid part of human diversity ~\cite{walker2014neurodiversity}. Symbols such as the rainbow infinity sign inspired by the LGBTQ Pride flag have emerged from within the community to counter external narratives that frame autism as a deficit ~\cite{kattari2023infinity}.

\paragraph{Community Perspectives}
Autistic individuals often reclaim language and challenge neuronormative narratives. Important considerations for annotation include:
\begin{itemize}
    \item Identity-first language (e.g., “autistic person” instead of “person with autism”) is preferred by the majority of autistic adults in the United States ~\cite{taboas2023preferences}
    \item Community-adopted terminology such as \textit{Aspie} (a self-identifier used by some autistic individuals)
\end{itemize}

\subsection{Annotation Tasks and Procedures}
In this section, we provide an overview of the annotation task along with video examples of the process.

\paragraph{Common Annotation Challenges}
Annotators should exercise careful judgment when evaluating phrases. For example:
\begin{itemize}
    \item Statements such as \textit{“That’s so autistic”} require contextual interpretation.
    \item The phrase \textit{“This vaccine causes autism”} is categorized as ableist due to its history in promoting autism stigma.
    \item The subtle difference between \textit{“I am not autistic”} and \textit{“At least I am not autistic”} changes the meaning and must be carefully assessed.
\end{itemize}

\subsection{Ethical Considerations and AI Bias}
\paragraph{Challenges in Hate Speech Detection}
Research indicates that many existing AI models misclassify disability-related discourse as toxic, even when the content is neutral or positive ~\cite{narayanan-venkit-etal-2023-automated, venkit-etal-2022-study, gadiraju_offensive_23, gamaartificially}. Specific issues include:
\begin{itemize}
    \item AI models exhibit over-sensitivity to disability-related discussions, frequently labeling them as harmful.
    \item AI models are more confident in detecting ableism when using \textit{person-first language} (e.g., “ableist toward autistic people”) than \textit{identity-first language} (e.g., “anti-autistic”). *
\end{itemize}
*\textit{these are results from our preliminary study}

\paragraph{Project Overview}
This project seeks to mitigate biases in AI hate speech detection by:
\begin{itemize}
    \item Training models using annotations informed by the neurodivergent community.
    \item Ensuring that AI does not misclassify community discourse as hate speech.
    \item Recognizing the distinction between hate speech and reclaimed terminology within the autistic community.
\end{itemize}

\subsection{Resources}
In this section, we provide resources such as our guidelines that contain a glossary to refer to or modify as needed. The terms in the glossary are those commonly used in neurodiversity discourse online.

\subsection{Conclusion}
The annotation orientation session is designed to equip annotators with the necessary knowledge to responsibly and accurately classify anti-autistic hate speech. By following the annotation guidelines and considering the broader socio-historical context, annotators contribute to the development of AI models that better serve neurodivergent individuals. 

\section{Search Keywords}
This list of terms in Table~\ref{tab:searchterms} were used to identify target sentences on Reddit. The number of target sentences containing each term is included.

\begin{table}
    \centering
    \begin{tabular}{cc} 
    \toprule
\bfseries Word & \bfseries Sentence Count \\ \midrule
autis*	& 1,221 \\ 
ASD	& 226 \\ 
disabilit* & 184 \\ 
aspergers & 173 \\ 
ABA	& 167\\ 
neurotyp*	& 158 \\ 
aspie* & 144 \\ 
neurodiver*	& 103 \\ 
AuDHD & 99 \\ 
disable*	& 93 \\ 
autism speaks	& 66 \\ 
stupid*	& 56 \\ 
a11y	& 34 \\ 
NT 	& 27 \\ 
retard*	& 25 \\ 
idiot*	& 18 \\ 
actually autistic	& 13 \\ 
autism intensifies	& 6 \\ 
ND 	      & 5 \\ 
autie* & 2 \\ \bottomrule
% aspergers intensifies	& 0 \\ \hline
    \end{tabular}
    \caption{The keywords included in our Reddit search and the number of sentences containing each term in the \textsc{Autalic} dataset.}
    \label{tab:searchterms}
\end{table}

\section{Glossary (Excerpt)}
Below is an example of terms in our glossary. To view the full glossary, contact the first-author of this paper. This resource was created in collaboration with our annotators to define unfamiliar terms such as slang, medical abbreviations, laws, and other specialized language that they may encounter during the annotation process. In total, it contains 34 words.

\begin{table}[h]
    \centering
    \begin{tabular}{p{2cm} p{5cm}}
        \hline
        \textbf{Term} & \textbf{Definition} \\
        \hline
        AuDHD & A combination of autism and ADHD ~\cite{owensadhd}. \\
        Au, Âû & Used by autistic individuals to self-identify as autistic ~\cite{AU}. \\
        ND & Neurodivergent, neurodiverse, or neurodiversity  ~\cite{Greally_2021}
    \end{tabular}
    \caption{Glossary of specialized terms used during annotation.}
    \label{tab:glossary}
\end{table}

\section{Challenging Cases}
 Here, we show examples of sentences with high disagreements among our annotators, along with their own notes on the aspects of each sentence that made it difficult to classify:
 
\textbf{Example 1}  
\begin{quote}
    “ABA can be very beneficial to low-functioning autists who can't speak, use the bathroom, perform basic tasks, etc.”
\end{quote}
Promotes ABA and the “low-functioning” designation.
\textbf{Example 2}  
\begin{quote}
    “Autism can also lead to social issues like the autistic person saying the wrong thing at the wrong time with absolutely no utility involved.”
\end{quote}
Judgment that the autistic way is socially wrong.

\textbf{Example 3}  
\begin{quote}
    “Autism or ADHD or otherwise, you still have to learn basic etiquette.”
\end{quote}
Specific to cultural context: Implies NT (neurotypical) etiquette is required, but no expectation for NT to learn ways of relating that are natural to autistic people.

\textbf{Example 4}  
\begin{quote}
    “Basically, right after doing so much research, integrating with the autistic community, and accepting ASD as a part of myself, I was back to square one—left feeling like an idiot and immensely confused.”
\end{quote}
Slur against cognitive/intellectual disability, negativity associated with autistic identity, and medicalization of identity.

\section{Self-Agreement Scores}
\begin{figure}
   \centering
   \includegraphics[width=0.95\linewidth]{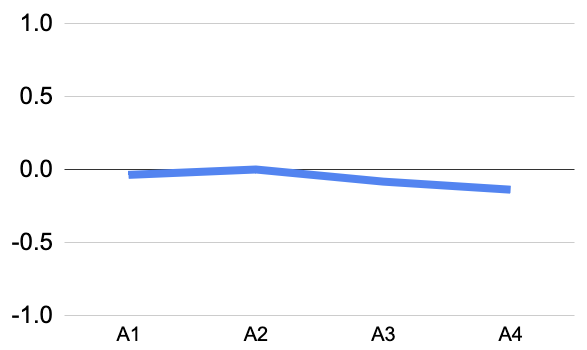}
   \caption{The self-agreement scores among annotators in a preliminary study highlight the difficulty of this task.}
   \label{fig:self-agreement}
\end{figure}

In preliminary studies, we examined different labeling schemes for this task to identify the most efficient and effective annotation strategy. Our experiments revealed high levels of self-disagreement among annotators, as shown in Figure ~\ref{fig:self-agreement}. The observed scores (M = -0.06, SD = 0.06) highlight the difficulty of the task and provide a meaningful baseline for comparison. Notably, our own annotation scores for \textsc{Autalic} were higher (M = 0.21, SD = 0.09), suggesting major improvement.

\section{Annotation Platform}
\begin{figure}[h]
    \centering
    \includegraphics[width=0.99\linewidth]{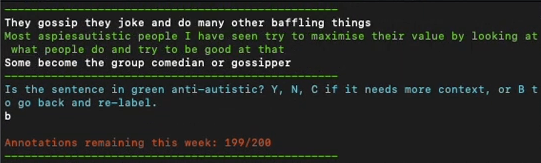}
    \caption{An example of an annotation task on our platform containing the target sentence (green) and contextual sentences (white).}
    \label{fig:platform}
\end{figure}

Figure ~\ref{fig:platform} shows an example of an annotation task on our platform with contextual sentences. 

\end{document}